\pdfoutput=1
\documentclass[11pt,a4paper]{article}
\usepackage{jheppub}
\usepackage[caption=false]{subfig}
\usepackage[export]{adjustbox} 

\newcommand{\beq}{\begin{eqnarray}}
\newcommand{\eeq}{\end{eqnarray}}
\newcommand{\non}{\nonumber\\}

\newcommand{\p}{\partial}

\newcommand{\bphi}{\boldsymbol{\phi}}
\newcommand{\bvarphi}{\boldsymbol{\varphi}}
\newcommand{\bdelta}{\boldsymbol{\delta}}
\newcommand{\bDelta}{\boldsymbol{\Delta}}
\newcommand{\bepsilon}{\boldsymbol{\epsilon}}
\newcommand{\bJ}{\mathbf{J}}
\newcommand{\bw}{\mathbf{w}}
\newcommand{\bx}{\mathbf{x}}
\newcommand{\bn}{\mathbf{n}}
\newcommand{\dc}{\delta{\mkern-1mu}c}
\newcommand{\dphi}{\delta{\mkern-1mu}\phi}
\newcommand{\df}{\delta{\mkern-2.5mu}f}
\newcommand{\ddf}{\delta{\mkern-1mu}\delta{\mkern-2.5mu}f}
\newcommand{\dlambda}{\delta{\mkern-2.5mu}\lambda}
\newcommand{\bdphi}{\bdelta{\mkern-1mu}\bphi}
\newcommand{\dtheta}{\delta{\mkern-1mu}\theta}
\newcommand{\ddtheta}{\delta{\mkern-1mu}\delta{\mkern-1mu}\theta}
\newcommand{\BE}{B{\mkern-1.5mu}E}
\newcommand{\Og}{{\rm O}}
\newcommand{\Lag}{\mathcal{L}}
\renewcommand{\d}{{\mathrm{d}}}
\renewcommand{\i}{\mathrm{i}}

\subheader{\rm\hfill IFUP-TH-2021}
\title{Near-BPS baby Skyrmions with Gaussian tails} 

\author{Sven Bjarke Gudnason$^1$,}
\affiliation{$^1$Institute of Contemporary Mathematics, School of
  Mathematics and Statistics, Henan University, Kaifeng, Henan 475004,
  P.~R.~China}
\emailAdd{gudnason(at)henu.edu.cn}
\author{Marco Barsanti$^{2}$ and}
\emailAdd{marco.barsanti(at)phd.unipi.it}
\author{Stefano Bolognesi$^{2}$}
\affiliation{$^2$Department of Physics ``E. Fermi'', University of
  Pisa and  INFN, Sezione di Pisa, 
  Largo Pontecorvo, 3, Ed. C, 56127 Pisa, Italy}
\emailAdd{stefanobolo(at)gmail.com}

\abstract{
  We consider the baby Skyrme model in a physically motivated limit of
  reaching the restricted or BPS baby Skyrme model, which is a model
  that enjoys area-preserving diffeomorphism invariance. The
  perturbation consists of the kinetic Dirichlet term with a small
  coefficient $\epsilon$ as well as the standard pion mass term, with
  coefficient $\epsilon m_1^2$. The pions remain lighter
  than the soliton for any $\epsilon$ and  therefore the model is
  physically acceptable, even in the $\epsilon \to 0$ limit.
  The version of the BPS baby Skyrme model we use has BPS solutions
  with Gaussian tails. We perform full numerical computations in
  the $\epsilon\to0$ limit and even reach the strict $\epsilon=0$
  case, finding new nontrivial BPS solutions, for which we do not yet
  know the analytic form.  
}

\begin{document}
\maketitle

\section{Introduction}

The Skyrme model \cite{Skyrme:1961vq,Skyrme:1962vh} is an attractive
model for a field-theoretic approach to nuclei and it is the
low-energy effective theory of the Witten-Sakai-Sugimoto model
\cite{Witten:1998zw,Sakai:2004cn}.
Its simplest version overestimates the nuclear binding energies at the
classical level by about an order of magnitude compared with those
observed in experiments.
BPS solitons, on the other hand, saturate a Bogomol'nyi bound which
implies that such solitons have vanishing binding energies.
A BPS Skyrme model was proposed in ref.~\cite{Adam:2010fg} by
Adam--Sanchez-Guillen--Wereszczynski (see also
refs.~\cite{Bonenfant:2010ab,Adam:2010ds,Bonenfant:2012kt,Adam:2013wya})
which is a BPS theory that has BPS solutions for any value of the
topological charge or baryon number, $B$.
Depending on the potential used in this BPS theory, the solutions are
often of a compacton type, that is, the solutions have support on a
compact region of space and has a discontinuous derivative at the
border of the compacton region.
This property can in part be traced back to the fact that the BPS
Skyrme model does not have a kinetic term, but its only derivative
term is of sixth order and is the topological charge-density
squared. 

Real-world nuclei have binding energies at the one-percent level of
their rest mass, and it is expected that a small perturbation of a
BPS model will yield multi-Skyrmion solutions with low binding
energies -- at least at the classical level.
Deforming the BPS Skyrme model is easy enough: a natural deformation
is to try and revert back towards the chiral Lagrangian and include the
kinetic (Dirichlet) term with a small coefficient $\epsilon$
\cite{Gillard:2015eia}, see also ref.~\cite{Bolognesi:2013nja} for a
deformation in holography.
Unfortunately, the numerical solutions become unprecedentedly
difficult in the tiny $\epsilon$ limit, whereas for $\epsilon$ of
order one, there are no severe difficulties with the model.
This fact is due to several factors. The main reason is that the BPS
model has an infinite moduli space, i.e.~of volume-preserving
diffeomorphisms; the theory describes the Skyrmions as an
incompressible fluid \cite{Adam:2014nba}, akin to the liquid-drop
model description of nuclei. 
Indeed this fact about the model is very welcome for phenomenology,
and especially for finite density
applications \cite{Adam:2014dqa,Adam:2015lpa,Adam:2015lra}. 
However, this means that the $\epsilon=0$ theory has infinitely many
solutions, whereas the $0<\epsilon\ll1$ theory picks out a (possibly
finite) subset of these infinitely many solutions.
The mathematical problem of picking out this subset of solutions has
been coined restricted harmonicity by Speight \cite{Speight:2014fqa}.
The idea is that the solution to a near-BPS theory deformed by the
Dirichlet term, whose BPS solutions can have all shapes with a fixed
volume, must be the subset of the solutions that minimize the
Dirichlet energy. It is thus a ``constrained'' harmonicity problem.

Because the near-BPS limit of the BPS Skyrme model is very
challenging, we have in ref.~\cite{Gudnason:2020tps} considered the
analogous problem in the toy model, known as the baby Skyrme model
\cite{Leese:1989gi,Piette:1994jt,Piette:1994ug,Piette:1994mh}.
The baby Skyrme model is ``smaller'' than the Skyrme model in two
senses: 
it is formulated in 2+1 dimensional spacetime instead of 3+1 and it
has the target space $S^2$ instead of $S^3$.
A further similarity with the full Skyrme model, is that it also
possesses a BPS version, which we shall call the BPS baby Skyrme model
or restricted baby Skyrme model.
This BPS version of the model consists, again analogously to the BPS 
Skyrme model in 3+1 dimensions, of the topological charge-density
squared as well as a potential.
Furthermore, the BPS baby Skyrme model also enjoys area-preserving
diffeomorphism invariance for static solutions \cite{Gisiger:1996vb},
and it was in fact discovered earlier in this context, than in the
full 3+1 dimensional Skyrme model.
In our previous paper \cite{Gudnason:2020tps}, we chose to stick with
the most simplistic potential, namely the pion mass term.
This had the consequence that the BPS solutions in that model are
of the compacton type \cite{Adam:2009px,Adam:2010jr,Speight:2010sy}.
Other potentials, however, will provide BPS solutions with either
exponential (Gaussian) tails or power-law tails \cite{Adam:2010jr}.
Baby Skyrmions are also interesting in their own right, and in fact
they are quite similar to the Skyrmions being studied heavily for the
moment in magnetic materials, see ref.~\cite{Fert2017} for a review,
and
refs.~\cite{Barton-Singer:2018dlh,Schroers:2019hhe,Ross:2020hsw,Kuchkin:2020bkg,Ross:2020orc}
for some recent theoretical work. 

The BPS property originally appeared in theories where BPS solitons
were solutions preserving a fraction of supersymmetry. For the Skyrme
model in 3+1 dimensions, this is not the case simply because its
target space is not K\"ahler \cite{Zumino:1979et}.
The first attempts at supersymmetrizing the Skyrme model ended up
with a model with $\mathbb{C}P^1$ target space
\cite{Bergshoeff:1984wb,Freyhult:2003zb}, which is the right target
space for the baby Skyrme model (but not the Skyrme model), although
the bosonic Lagrangian was slightly different than that of the
baby Skyrme model.
The exact supersymmetric version of the baby Skyrme model was later
constructed with $\mathcal{N}=1$ supersymmetry
\cite{Adam:2011hj,Bolognesi:2014ova}, but it does not contain the
bosonic Lagrangian of the restricted baby Skyrme model either.
It turns out that the restricted baby Skyrme model possesses
$\mathcal{N}=2$ supersymmetry \cite{Bolognesi:2014ova,Adam:2013awa} and
interestingly, supersymmetry forbids the presence of the kinetic
Dirichlet term, which exactly corresponds to the $\epsilon\to0$ limit
mentioned above. The baby Skyrmion solitons preserve only a quarter of
supersymmetry (i.e.~one supercharge)
\cite{Nitta:2014pwa,Nitta:2015uba}.
The Skyrme model was successfully supersymmetrized in
ref.~\cite{Gudnason:2015ryh} by complexifying the target space from
SU$(2)$ to SL$(2,\mathbb{C})$, and like the 2-dimensional case,
supersymmetry also forbids the presence of the kinetic Dirichlet term.
This supersymmetric Skyrme model does possess soliton solutions, but
they are not BPS states \cite{Gudnason:2016iex}. 

In this paper, we construct a  BPS sector which is composed by the
Skyrme term (which is also the topological charge-density squared) as
well as a potential that does not give the pion a mass. Specifically,
we will choose the pion mass term squared, which gives the BPS
solitons a Gaussian tail. The deformation of the BPS
sector is given by the kinetic (Dirichlet) term, with coefficient
$\epsilon$, and the normal pion mass term, with coefficient
$\epsilon m_1^2$. This has the nice scaling with $\epsilon$ that keeps
the mass of the perturbative pions constant and equal to $m_1$.
Because of the deformation being in the form of both the kinetic
(Dirichlet) term and a potential term (the standard pion mass term),
the solution of the deformation is not restricted harmonic, but some
generalization which we will call generalized restricted
harmonic (GRH).\footnote{In principle, the deformation \emph{is} restricted
harmonic, but not of the pure BPS solution, but of an
$\epsilon$-dependent BPS solution, where the total potential is used,
including the $\epsilon$-dependent pion mass term, see app.~\ref{app:would-be_BPS_sol}. }
We perform large-scale high-definition brute-force numerical
computations for the topological sectors $Q=2$ and $Q=4$ and are able
to dial $\epsilon$ all the way to $\epsilon=0$ -- finding new
nontrivial BPS solutions, for which we do not know an analytic expression.
We know that for $Q=1$, the restricted harmonic solution, i.e.~the
solution for $\epsilon>0$ is axially symmetric.
We study the axially symmetric baby Skyrmions in the perturbative
scheme, put forward in our previous paper \cite{Gudnason:2020tps} and
find very good agreement between the perturbative results and the full
numerical ODE results.
The main difference here with respect to ref.~\cite{Gudnason:2020tps}
is that the BPS solution here is not a compacton and therefore there
is no need for imposing special cusp conditions at the boundary of the
would-be compacton.
The most stable axially symmetric baby Skyrmion has charge $Q=N=2$ and
therefore it is expected that for large topological charge $Q$, the
most stable solution is made of $\frac{Q}{2}$ almost axially symmetric $N=2$
components -- stitched together in some fashion.
Throughout the paper, we are using the notation that $Q$ denotes the
total charge of the baby Skyrmion, regardless of its shape, and $N$
denotes the charge of an (almost) axially symmetric baby Skyrmion. 
We then turn to calculating the bound states in the perturbative
scheme, using the $\epsilon$ expansion up to $\mathcal{O}(\epsilon^3)$
or next-to-next-to-leading order (N$^2$LO).
To leading order (LO), the correction to the energy is given by
inserting the BPS solution into the deformation terms --  this cannot
explain a bound state, because in order to avoid overcounting, we
simply cut off the tails of the BPS and LO energy
densities, that otherwise would overlap. This does 
explain an attractive force at LO.
In order to understand the separation distance of two $N=2$ (or even
two $N=1$) constituents, we need also a repulsive force at shorter
distances.
In our perturbative scheme, we do find such a force and it is due to
energy accumulation near the middle of the bound state, caused by our
imposed gluing conditions.
Unfortunately, this solution based on the perturbative scheme
does not match with the precise full numerical calculations, neither
for the separation distances nor for the binding energies and we
conclude in this paper that the perturbative scheme fails for solitons
with tails -- as opposed to compactons \cite{Gudnason:2020tps}.

The paper is organized as follows. In sec.~\ref{sec:model}, we
introduce our version of the deformed restricted baby Skyrme model,
its BPS solutions, topological energy bounds and define a physical
length scale for the BPS solutions. In sec.~\ref{sec:perturb}, we
review and modify the perturbative $\epsilon$-expansion scheme
introduced in ref.~\cite{Gudnason:2020tps}. We then test the scheme on
axially symmetric baby Skyrmion solutions. In sec.~\ref{sec:numcalc},
we perform full high-resolution numerical PDE computations and find
nontrivial BPS solutions as well as near-BPS solutions. In
sec.~\ref{sec:int}, we review the long-range interactions of
ref.~\cite{Piette:1994ug} but with the notation of our model.
In sec.~\ref{sec:binding}, we attempt at calculating the binding
energies perturbatively, but discover that
the method fails for solitons with tails.
We conclude the paper with a discussion in sec.~\ref{sec:discussion}.
We have relegated some details of a(n $\epsilon$-dependent) BPS
solution that contains the potential of the deformation Lagrangian to
app.~\ref{app:would-be_BPS_sol}.

\section{The model}\label{sec:model}

The model is based on the restricted (BPS) baby Skyrme model with a
non-BPS deformation in the form of the kinetic term as well as the
standard pion mass term, both with coefficient $\epsilon$,
\beq
\mathcal{L}[\bphi] =
\epsilon\left(\mathcal{L}_2[\bphi] - m_1^2 V_1(\bphi)\right)
+ \mathcal{L}_4[\bphi]
- V_2(\bphi)
+ \frac12\lambda(\bphi\cdot\bphi - 1),
\label{eq:L}
\eeq
where $\bphi=(\phi^1,\phi^2,\phi^3)$ is a real 3-vector,
the kinetic term and the Skyrme terms are given by
\begin{align}
\mathcal{L}_2[\bphi] &= -\frac12 (\p_{\mu}\bphi\cdot\p^{\mu}\bphi),\\
\mathcal{L}_4[\bphi] &= -\frac14 (\p_{\mu}\bphi\cdot\p^{\mu}\bphi) (\p_{\nu}\bphi\cdot\p^{\nu}\bphi)
+\frac14 (\p_{\mu}\bphi\cdot\p_{\nu}\bphi) (\p^{\mu}\bphi\cdot\p^{\nu}\bphi),
\end{align}
respectively, and the two potentials are
\beq
V_p(\bphi) \equiv \frac{1}{p}\left(1 - \phi^3\right)^p,
\label{eq:potp}
\eeq
with $p=1,2$. 
The nonlinear sigma model constraint $\bphi\cdot\bphi=1$ is enforced
by means of the Lagrange multiplier, $\lambda$. In this paper we will use the
mostly positive metric signature.

We have kept the pion mass, $m_1$ as a free parameter, but in order to
avoid clutter, we have scaled away the coefficients in front of the BPS
potential $V_2(\bphi)$ and the Skyrme term by an appropriate choice of
energy and length units (without loss of generality). 

The requirement of finite total energy effectively point compactifies
2-space from $\mathbb{R}^2$ to $\mathbb{R}^2\cup\{\infty\}\simeq S^2$,
i.e.~a 2-sphere.
It would also induce a spontaneous symmetry breaking from $\Og(3)$ to
$\Og(2)$, but we break this symmetry explicitly with the two
potentials.
Therefore the target space is given by $\Og(3)/\Og(2)\simeq S^2$ which
is also a 2-sphere, hence allowing for topologically nontrivial
solutions with degree
\beq
Q = \frac{1}{4\pi}\int\d^2x\; \mathcal{Q}_{12}
  = -\frac{1}{4\pi}\int\d^2x\; \bphi\cdot\p_1\bphi\times\p_2\bphi,
\eeq
which is also equal to the number of baby Skyrmions in
$\mathbb{R}^2$.
The solutions that minimize the static energy
\beq
E = \left.-\int \d{}^2x\;\Lag\right|_{\p_t\bphi=0},
\eeq
are called stable baby Skyrmions, whereas local minima exists which we
shall call meta\-stable baby Skyrmions.

In ref.~\cite{Gudnason:2020tps} we have studied the case with only
$V_1$ as the potential, which in the limit $\epsilon\to 0$ has
solutions of the compacton type \cite{Adam:2010jr}, meaning that the
soliton has only support on a compact domain $D\subset\mathbb{R}^2$,
which could be of the shape of a disc, but not necessarily.
This was the case because the potential $V_2$ was absent.
In fact, any potential with $0<p<2$ would give rise to a
compacton \cite{Adam:2010jr}.
The potential $V_2$ gives rise to solutions with a Gaussian tail,
even in the limit of $\epsilon\to 0$.
We will see this explicitly in the next subsection.

For the analytic calculations, it will be useful to use the following
parametrization
\beq
\bphi =
\frac{1}{1+|\omega|^2}
\begin{pmatrix}
  \omega + \bar{\omega}\\
  -\i(\omega - \bar{\omega})\\
  1-|\omega|^2
\end{pmatrix},
\label{eq:omegacoords}
\eeq
for which $\bphi\cdot\bphi=1$ is manifest.
The Lagrangian components in terms of $\omega$ can be written as  
\begin{align}
  \Lag_2[\omega,\bar{\omega}]
  &= -2\frac{\p_\mu\omega\p^\mu\bar{\omega}}{(1+|\omega|^2)^2},\\
  \Lag_4[\omega,\bar{\omega}]
  &= -2\frac{(\p_\mu\omega\p^\mu\bar{\omega})(\p_\nu\omega\p^\nu\bar{\omega})-(\p_\mu\omega\p_\nu\bar{\omega})(\p^\mu\omega\p^\nu\bar{\omega})}{(1+|\omega|^2)^4},
\end{align}
the potentials as
\beq
V_p(\omega,\bar\omega) =
\frac{1}{p}\left(\frac{2|\omega|^2}{1+|\omega|^2}\right)^p,
\eeq
with $p=1,2$, and the topological charge as
\beq
Q = -\frac{\i}{2\pi}\int d^2x\;
\frac{\epsilon^{ij}\p_i\omega\p_j\bar\omega}{(1+|\omega|^2)^2}.
\eeq
The $\omega$ parametrization will prove useful for finding the BPS
solutions, which we will turn to next, but we will turn back to the
$\bphi$ parametrization for the numerical analysis.

\subsection{BPS solutions}

The model contains a BPS submodel, which is found by sending
$\epsilon$ to zero ($\epsilon\to 0$) in the Lagrangian \eqref{eq:L}: 
\beq
\Lag^{\rm BPS}[\bphi] = \Lag_4[\bphi] - V_2(\bphi)
  + \lambda(\bphi\cdot\bphi - 1).
\label{eq:L_BPS}
\eeq
It contains also a different BPS submodel which includes the potential
$\epsilon m_1^2 V_1$, but the solutions are obviously dependent on
$\epsilon$, see app.~\ref{app:would-be_BPS_sol}.
The BPS equation can be found by performing a so-called Bogomol'nyi
trick (ignoring the Lagrange multiplier term) 
\begin{align}
\mathcal{E}^{\rm BPS}[\bphi] &=
\frac14(\p_i\bphi\cdot\p_i\bphi)(\p_j\bphi\cdot\p_j\bphi)
-\frac14(\p_i\bphi\cdot\p_j\bphi)(\p_i\bphi\cdot\p_j\bphi)
+ \frac12(1-\phi^3)^2\non
&= \frac12(\bphi\cdot\p_1\bphi\times\p_2\bphi)(\bphi\cdot\p_1\bphi\times\p_2\bphi)
+ \frac12(1-\phi^3)^2\non
&= \frac12\left[(\bphi\cdot\p_1\bphi\times\p_2\bphi) \mp (1-\phi^3)\right]^2
\pm (1-\phi^3)(\bphi\cdot\p_1\bphi\times\p_2\bphi).
\label{eq:Bogomolnyi_trick}
\end{align}
The BPS equation is thus
\beq
\mathcal{Q}_{12}
= -\bphi\cdot\p_1\bphi\times\p_2\bphi
= \mp (1-\phi^3),
\label{eq:BPSeq}
\eeq
and when satisfied, the total energy is proportional to the
topological charge or degree of the baby Skyrmion, $Q$, which is the
minimum of the energy in the given topological sector (for given
$Q$). 

We will now change parametrization of $\bphi$ to stereographic
coordinates \eqref{eq:omegacoords} and then the BPS equation reads  
\beq
\frac{\p_1\omega\p_2\bar{\omega} - \p_2\omega\p_1\bar{\omega}}{(1+|\omega|^2)^2}
=\frac{\p_r\omega\p_\theta\bar{\omega} - \p_\theta\omega\p_r\bar{\omega}}{r(1+|\omega|^2)^2}
= \mp\frac{\i|\omega|^2}{1+|\omega|^2}.
\eeq
Inserting the Ansatz $\omega=e^{\i N\theta}\zeta(r)$, we obtain the
equation 
\beq
\frac{\p_r\zeta}{r}
  = -\frac{1}{2N}\zeta(1+\zeta^2),
\label{eq:BPSeq_zeta}
\eeq
where we have chosen the lower sign.
It will now prove convenient to change variables as \cite{Adam:2010jr} 
\beq
1+\zeta^2 &=& \frac{1}{1-\gamma},\label{eq:gamma_var}\\
y &=& \frac12r^2, \label{eq:y_var}
\eeq
for which the differential equation becomes
\beq
\frac{\d\gamma}{\d y} = -\frac{1}{N} \gamma,
\label{eq:diff_gamma}
\eeq
which has the solution \cite{Adam:2010jr}
\beq
\gamma = e^{-\xi^2 - \kappa},
\eeq
where $\xi\equiv\frac{r}{R}$, the characteristic radius is
\beq
R = \sqrt{2N},
\label{eq:radius}
\eeq
and $\kappa$ is an integration constant.
Changing variables back to $\zeta$, we finally obtain the solution
\beq
\zeta = \frac{1}{\sqrt{e^{\xi^2}-1}},
\label{eq:BPSsol}
\eeq
where in order to ensure that the target space is fully covered, we
must make sure that the function is singular at $\xi=0$ and hence we
have set $\kappa=0$.
For large values of $\xi$, we have
\beq
\zeta \simeq \exp\left(-\frac{\xi^2}{2}\right)
+ \mathcal{O}\left(e^{-\frac32\xi^2}\right).
\eeq

It will prove useful to calculate the BPS mass
\begin{align}
M^{\rm BPS} &= \int \d^2x\; \mathcal{E}^{\rm BPS}[\bphi] \non
&= \pm \i4 \int \d^2x\;\frac{|\omega|^2}{(1+|\omega|^2)^{3}}
  \epsilon^{ij}\p_i\omega\p_j\bar\omega \non
&= \pm 16\pi N \int_0^{\infty} \d r\;
  \frac{\zeta^3}{(1+\zeta^2)^{3}}\p_r\zeta \non
&= - 16\pi |N| \int_\infty^0 \d\zeta\;
   \frac{\zeta^3}{(1+\zeta^2)^{3}} \non
&= 4\pi |N|.
\label{eq:BPSmass}
\end{align}
In the fourth line we have chosen the lower sign, corresponding to the
above-found BPS solution with the boundary conditions
$\zeta(0)\to\infty$ and $\zeta(\infty)=0$.

The topological charge of such an axially symmetric
configuration is thus given by
\beq
Q = -2N\int \d r\;\frac{\zeta\p_r\zeta}{(1+\zeta^2)^2} = N.
\eeq
We will call the topological charge $N$ only for axially symmetric
baby Skyrmions, whereas for generically shaped multi-baby Skyrmions we
will denote it by $Q$.

\subsection{Energy bound}

In the previous section, the energy was shown to be bounded from below
by the Bogomol'nyi bound in the BPS limit, $\epsilon=0$,
\beq
E_{4+0} \geq M^{\rm BPS} = 4\pi |Q|.
\label{eq:BPS_bound}
\eeq
There is another BPS limit, which is the double limit
$\epsilon\to\infty$ and $m_1\to0$.
This limit is also bounded from below due to the model becoming the
pure O(3) sigma model (consisting only of the kinetic term)
\cite{Bolognesi:2014ova},
\begin{align}
E_{\epsilon 2} &=
  2\epsilon\int \d^2x\;
  \frac{\p_i\omega\p_i\bar\omega}{(1+|\omega|^2)^2} \non
&\geq 2\epsilon \left|-\i\int \d^2x\;
  \frac{\epsilon^{ij}\p_i\omega\p_j\bar\omega}{(1+|\omega|^2)^2}\right| \non
&\geq 4\pi\epsilon |Q| = M^{\rm lump},
\label{eq:lump_energy_bound}  
\end{align}
where the topological solitons are instead called lumps.

The total energy in the model \eqref{eq:L} in the limit $m_1\to0$
therefore obeys the energy bound
\begin{align}
  E_{\epsilon 2+4+0} &\geq M^{\rm BPS} + M^{\rm lump} \non
  &\geq 4\pi|Q|(\epsilon + 1).
  \label{eq:m1_zero_comp_bound}
\end{align}
This bound is a composite of two limits and it is indeed only saturated
in those two limits: i.e.~in the limit of $\epsilon\to 0$, the model
contains BPS baby Skyrmions and in the limit of $\epsilon\to\infty$,
$m_1\to0$ the model contains BPS lumps \cite{Bolognesi:2014ova}.

Now we consider turning on $m_1$, for which the composite energy bound
\eqref{eq:m1_zero_comp_bound} no longer can be saturated.
It is however possible to calculate a bound similar to that of
eq.~\eqref{eq:BPS_bound}, but instead of including only $V_2$ as the
potential we include both the BPS potential and the deformation
potential, that is $\tilde{m}_1^2V_1+V_2$.
Writing down the would-be BPS mass
\begin{align}
  M^{\rm would-be\ BPS} &= \pm\i4\int \d^2x\;
  \frac{|\omega|\sqrt{\tilde{m}_1^2+(1+\tilde{m}_1^2)|\omega|^2}}{(1+|\omega|^2)^3}
  \epsilon^{ij}\p_i\omega\p_j\bar\omega \non
  &=\pm16\pi N\int_0^\infty\d{r}\;
  \frac{\zeta^2\sqrt{\tilde{m}_1^2+(1+\tilde{m}_1^2)\zeta^2}}{(1+\zeta^2)^3}\p_r\zeta\non
  &=-16\pi|N|\int_\infty^0\d\zeta\;
  \frac{\zeta^2\sqrt{\tilde{m}_1^2+(1+\tilde{m}_1^2)\zeta^2}}{(1+\zeta^2)^3}\non
  &=4\pi|N|\left(\frac12\sqrt{1+\tilde{m}_1^2}(2+\tilde{m}_1^2)
  +\frac{\tilde{m}_1^4}{8}\log\frac{2+\tilde{m}_1^2-2\sqrt{1+\tilde{m}_1^2}}{2+\tilde{m}_1^2+2\sqrt{1+\tilde{m}_1^2}}\right),
  \label{eq:would-be_BPS_bound}
\end{align}
which is the bound for the model \eqref{eq:L} in the limit of
$\epsilon\to0$ and $\epsilon m_1^2=\tilde{m}_1^2$ fixed.
The expression in the parenthesis on the last line of
eq.~\eqref{eq:would-be_BPS_bound} has the limiting value of unity when
$\tilde{m}_1\to0$ tends to zero; it thus coincides with the bound of
eq.~\eqref{eq:BPS_bound} as it must.
The BPS solution corresponding to the above energy bound is given in
app.~\ref{app:would-be_BPS_sol}.

We can now combine the energy bound \eqref{eq:lump_energy_bound} for
the sigma model and for the remaining terms in the Lagrangian
\eqref{eq:L} to obtain the energy bound for the full model as
\begin{align}
  E_{\epsilon(2+0')+4+0} &\geq M^{\rm would-be\ BPS} + M^{\rm lump}\non
  &\geq 4\pi|Q|\left(\epsilon + \frac12\sqrt{1+\epsilon m_1^2}(2+\epsilon m_1^2)
  +\frac{\epsilon^2 m_1^4}{8}\log\frac{2+\epsilon m_1^2-2\sqrt{1+\epsilon m_1^2}}{2+\epsilon m_1^2+2\sqrt{1+\epsilon m_1^2}}\right).
\end{align}
This energy bound can be Taylor expanded in $\epsilon$ as
\begin{equation}
E_{\epsilon(2+0')+4+0}\geq 4\pi|Q|\left[1
+ (1+m_1^2)\epsilon
+ \frac{m_1^4}{8}\left(1 + 2\log\frac{\epsilon m_1^2}{4}\right)\epsilon^2
- \frac{m_1^6\epsilon^3}{8}
+ \mathcal{O}(\epsilon^4)
\right].
\label{bound}
\end{equation}
The BPS bound in the limit $\epsilon\to0$ is clear.
The first order correction to the energy bound, in $\epsilon$, is due
to the mass term and the lump mass.
The higher-order corrections to the energy bound come from the
would-be BPS bound on the Skyrme term together with both potentials,
i.e.~$\tilde{m}_1^2V_1+V_2$.

\subsection{Length scale}

It will prove useful to know the length scale of the baby Skyrmion in
the BPS limit, a.k.a.~its radius.
The short answer is $R=\sqrt{2N}$, which depends only on $N$ (since we
have fixed the coefficient of the BPS potential to be unity, which can
always be done by a rescaling of lengths).

For a more precise estimate of the radius, we can first calculate at
which value of $\zeta$ the fraction $\beta$ of the energy is
contained.
This in turn translates to a radius and hence yields the coefficient
multiplying $R$ for the radius.
The entire BPS energy is contained if $\zeta$ tends to zero in the
integral \eqref{eq:BPSmass},
\beq
-4\int_{\infty}^0\d{\zeta}\frac{\zeta^3}{(1+\zeta^2)^3} = 1.
\eeq
Therefore a fraction $\beta$ of the BPS energy is contained by
\beq
-4\int_{\infty}^{\zeta^\beta}\d{\zeta}\;\frac{\zeta^3}{(1+\zeta^2)^3} = \beta, \qquad
\Rightarrow\qquad
\zeta^\beta = \sqrt{\frac{1}{\beta}+\frac{\sqrt{1-\beta}}{\beta}-1},
\eeq
with $\beta\in(0,1]$.
Inverting now the BPS solution \eqref{eq:BPSsol}, we obtain
\beq
\xi_\beta^2 = -\log\left(1 - \frac{1}{1+(\zeta^\beta)^2}\right)
= -\frac12\log(1-\beta).
\eeq
Using that $\xi\equiv\frac{r}{R}$, we find the radius that contains
the fraction $\beta$ of the total BPS energy reads
\beq
\mathfrak{r}_\beta = \sqrt{-\frac12\log(1-\beta)} R.
\eeq
The radii corresponding to $\beta=0.9,0.95,0.99$ are 
$\mathfrak{r}_{0.9}\simeq1.073R$, $\mathfrak{r}_{0.95}\simeq1.224R$ and
$\mathfrak{r}_{0.99}\simeq1.517R$, respectively.

\section{Perturbation in \texorpdfstring{$\epsilon$}{epsilon}}\label{sec:perturb}

We will now use the framework of perturbation theory around a background
soliton solution, $\bvarphi$, developed in
ref.~\cite{Gudnason:2020tps} where
\beq
\delta\Lag^{\rm BPS}[\bvarphi]=0.
\eeq
The BPS solution in this model is given by eq.~\eqref{eq:BPSsol} in
the $\omega$ parametrization.
Since the numerical calculations will be performed in the $\bphi$
parametrization, it will be useful to write the charge-$N$
axially symmetric baby Skyrmion solution in that parametrization: 
\beq
\bvarphi =
\begin{pmatrix}
  \sin f(r)\cos(N\theta - \alpha)\\
  \sin f(r)\sin(N\theta - \alpha)\\
  \cos f(r)
\end{pmatrix},\qquad
f(r) = \arccos\left(1 - 2e^{-\xi^2}\right),
\label{eq:varphi}
\eeq
with $\xi\equiv r/R$ and $R$ is the characteristic radius
\eqref{eq:radius}, $x+\i y=r e^{\i\theta}$ are the standard polar
coordinates in $\mathbb{R}^2$ and $\alpha$ is a phase modulus.
The phase modulus is an internal parameter of the solution, but for a
single solution is equivalent to a rotation in the plane by
$-\alpha/N$. 

Next we will consider the corrections to the energy order-by-order in
$\epsilon$, in the following sections, starting with the leading
order. 

\subsection{Leading-order correction}\label{sec:LO}

The leading order (LO) correction is the first and linear order in
$\epsilon$, and in contradistinction to the compacton case of
ref.~\cite{Gudnason:2020tps} comes from both the kinetic term as well
as the (perturbative) pion mass term.
Inserting the BPS solution into these terms thus gives
\begin{align}\label{1order}
  \epsilon M^{\rm LO}(N,m_1) &= \epsilon\int \d^2x\;
  \left(-\Lag_2[\bvarphi] + m_1^2V_1(\bvarphi)\right)\non
&= 4\pi\epsilon\int \d r\;\left[
  \frac{r^2\zeta_r^2 + N^2\zeta^2}{r(1+\zeta^2)^2}
  + \frac{r m_1^2\zeta^2}{1+\zeta^2}
  \right]
  \non
& =4\pi\epsilon\left(\frac{\pi^2}{12} + \frac{N^2}{2}\log 2
  + N m_1^2\right),
\end{align}
where in the following we will only consider $N>0$ positive. 
Of course, by including the pion mass in the perturbation, the
leading-order contribution to the energy depends on the mass parameter
$m_1$.
The parenthesis on the last line takes the value $1.169$ for $N=1$ and
$m_1=0$, which is about 17\% above the energy bound for the kinetic
term.
Turning on the pion mass, $m_1=0.5$, increases the value of the
parenthesis to $1.419$.

To this order, the energy reads
\begin{align}
\label{0+1order}
  E(\epsilon,N,m_1) &= M^{\rm BPS}(N) + \epsilon M^{\rm LO}(N,m_1) \non
  &= 4\pi N + \frac{\pi^3\epsilon}{3} + 2\pi\epsilon N^2\log 2 +
  4\pi N \epsilon m_1^2.
\end{align}
Note that this is strictly above the BPS bound \eqref{bound}
considered at the linear level in $\epsilon$.
Considering now the energy per $N$ as a function of $N$, we can 
determine which solution has the lowest energy per baby Skyrmion
(nucleon). 
We have
\beq
\frac{\d}{\d N}\left(\frac{E(\epsilon,N,m_1)}{N}\right) =
-\frac{\pi\epsilon}{3N^2}
\left(\pi^2 - 6N^2\log 2\right) = 0,
\eeq
with solution 
\beq
N_\star = \frac{\pi}{\sqrt{6\log 2}}
\approx 1.541,
\label{eq:Nstar}
\eeq
which is the $N$ with the minimum energy per $N$.
We notice that the formal minimum of the energy per $N$, to leading
order in $\epsilon$, does not depend on $m_1$. 

\begin{figure}[!htp]
  \begin{center}
    \includegraphics[width=0.5\linewidth]{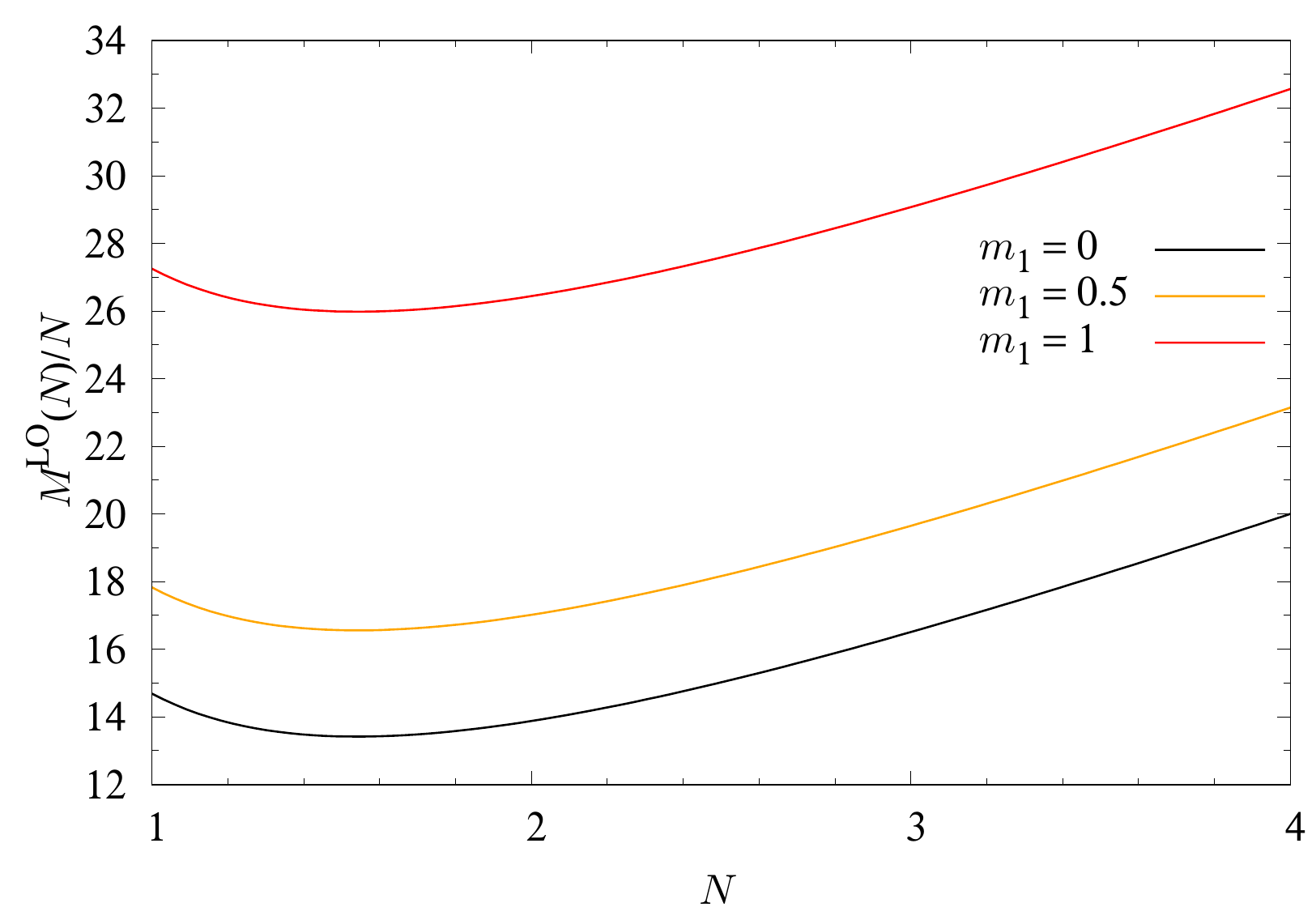}
    \caption{Leading-order energy per nucleon (baby Skyrmion) divided
      by $\epsilon$: $\frac{M^{\rm LO}}{N}$ as a function of $N$ for
      $m_1=0,0.5,1$ (from bottom to top).
    } 
    \label{fig:LON}
  \end{center}
\end{figure}
In fig.~\ref{fig:LON} is shown the leading-order energy correction,
divided by $\epsilon N$. This is the energy correction per nucleon, to
be multiplied by $\epsilon$.
We can see from the figure that the stable axially symmetric baby
Skyrmion, to leading order in $\epsilon$, will have $N=2$.
For example, two separated 1-Skyrmions will have a higher energy than
an axially symmetric 2-Skyrmion and an axially symmetric 4-Skyrmion
will have higher energy than two separated 2-Skyrmions.
Also an axially symmetric 3-Skyrmion will have higher energy than
three 1-Skyrmions or a 1-Skyrmion and a 2-Skyrmion.

\subsection{NLO and \texorpdfstring{N$^2$LO}{N2LO} corrections}\label{sec:NLO}

The next corrections are the next-to-leading order (NLO) and
next-to-next-to-leading order (N$^2$LO) corrections to the energy
which are of order $\epsilon^2$ and $\epsilon^3$, respectively, and
they will be calculated by introducing a linear perturbation around
the background field
\beq
\bphi = \bvarphi + \bdphi,
\eeq
where $\bvarphi=(\varphi^1,\varphi^2,\varphi^3)$ is the background
solution around which $\bdphi=(\dphi^1,\dphi^2,\dphi^3)$ is a small
perturbation.
We assume that $\bdphi$ is of order $\epsilon$, although this is not
clear \emph{a priori}; we have to check it \emph{a posteriori}.

In order to capture the NLO and N$^2$LO corrections, we have to
calculate the variation of the model \eqref{eq:L} up to third order in
the fields (assuming that $\bdphi=\mathcal{O}(\epsilon)$)
\begin{align}
\Lag^{\rm perturb}[\bvarphi,\bdphi] &=
\left.\frac{\p\Lag}{\p\phi^a}\right|\dphi^a
+\left.\frac12\frac{\p^2\Lag}{\p\phi^a\p\phi^b}\right|\dphi^a\dphi^b
+\left.\frac{\p\Lag}{\p\p_\mu\phi^a}\right|\p_\mu\dphi^a
\non
&\phantom{=\ }
+\left.\frac12\frac{\p^2\Lag}{\p\p_\mu\phi^a\p\p_\nu\phi^b}\right|\p_\mu\dphi^a\p_\nu\dphi^b
+\left.\frac16\frac{\p^3\Lag}{\p\p_\mu\phi^a\p\p_\nu\phi^b\p\p_\rho\phi^c}\right|\p_\mu\dphi^a\p_\nu\dphi^b\p_\rho\dphi^c\non
&=
\epsilon m_1^2 \dphi^3
+\dlambda\bvarphi\cdot\bdphi
-\frac12\left(\dphi^3\right)^2
+\frac{\lambda_0 + \dlambda}{2}\bdphi^2
-\epsilon J_a^\mu\p_\mu\dphi^a
\non
&\phantom{=\ }
-\frac12 V_{a b}^{\mu\nu}\p_\mu\dphi^a\p_\nu\dphi^b
-\frac16\Gamma_{a b c}^{\mu\nu\rho}\p_\mu\dphi^a\p_\nu\dphi^b\p_\rho\dphi^c,
\label{eq:Lperturb}
\end{align}
where the vertical bar ``\ $|$\ '' denotes evaluation on the
background by setting $\bphi=\bvarphi$ (to the left of the bar) and we
have defined the following quantities 
\begin{align}
J_a^\mu &\equiv \left.-\frac{\p\Lag_2}{\p\p_\mu\phi^a}\right|
= \p^\mu\varphi^a, \\
V_{a b}^{\mu\nu} &\equiv
\left.-\frac{\p^2\Lag}{\p\p_\mu\phi^a\p\p_\nu\phi^b}\right|
\equiv V_{0 a b}^{\mu\nu} + \epsilon V_{1 a b}^{\mu\nu}, \non
V_{0 a b}^{\mu\nu} &\equiv (\p_\rho\bvarphi\cdot\p^\rho\bvarphi)\eta^{\mu\nu}\delta^{a b}
+2\p^\mu\varphi^a\p^\nu\varphi^b
-\p^\mu\bvarphi\cdot\p^\nu\bvarphi\delta^{a b}
-\p_\rho\varphi^a\p^\rho\varphi^b\eta^{\mu\nu} 
-\p^\mu\varphi^b\p^\nu\varphi^a, \non
V_{1 a b}^{\mu\nu} &\equiv \eta^{\mu\nu}\delta^{a b},\\
\Gamma_{a b c}^{\mu\nu\rho} &\equiv
\left.-\frac{\p^3\Lag}{\p\p_\mu\phi^a\p\p_\nu\phi^b\p\p_\rho\phi^c}\right|\non
&=
\eta^{\mu\nu}\big(2\p^\rho\varphi^c\delta^{a b}
-\p^\rho\varphi^a\delta^{b c}
-\p^\rho\varphi^b\delta^{c a}\big)
+\eta^{\nu\rho}\big(2\p^\mu\varphi^a\delta^{b c}
-\p^\mu\varphi^b\delta^{c a}
-\p^\mu\varphi^c\delta^{a b}\big) \non
&\phantom{=\ }
+\eta^{\mu\rho}\big(2\p^\nu\varphi^b\delta^{c a}
-\p^\nu\varphi^c\delta^{a b}
-\p^\nu\varphi^a\delta^{b c}\big),
\end{align}
and the Lagrange multiplier for the background fields, $\lambda_0$, is
given by
\begin{align}
\lambda_0 &=
-(\bvarphi\cdot\p^2\bvarphi)
  (\p_\mu\bvarphi\cdot\p^\mu\bvarphi)
+(\bvarphi\cdot\p_\mu\p_\nu\bvarphi)
  (\p^\mu\bvarphi\cdot\p^\nu\bvarphi)
-(1 - \varphi^3)\varphi^3.
\end{align}
The Lagrange multiplier $\lambda$ has been replaced by
the expansion $\lambda_0+\dlambda$, where $\lambda_0$ enforces the sigma
model constraint for the background solution $\bvarphi$ and $\dlambda$
makes sure the total field $\bphi=\bvarphi+\bdphi$ remains inside the
O(4) group, i.e.~it preserves the constraint
$\bphi\cdot\bphi=1+\mathcal{O}(\epsilon^4)$.
Varying $\dlambda$ yields
\beq
\left(\frac12\bdphi^2 + \bvarphi\cdot\bdphi\right) = 0,
\label{eq:dlambda_constraint}
\eeq
which has the solution \cite{Piette:1994ug,Gudnason:2020tps}
\beq
\bdphi = \bDelta\times\bvarphi
+ \frac12\bDelta\times(\bDelta\times\bvarphi).
\label{eq:Delta_form}
\eeq
This form of the perturbation satisfies the constraint
\eqref{eq:dlambda_constraint} up to $\mathcal{O}(\bDelta^4)$ and since
$\bDelta$ will turn out to be of order $\epsilon$, that makes it of
order $\mathcal{O}(\epsilon^4)$.
Notice that the perturbative contribution to the energy from
the terms multiplying $\dlambda$ is exactly the bracket in
eq.~\eqref{eq:dlambda_constraint} and hence is of order
$\mathcal{O}(\epsilon^4)$, which we can safely disregard.
Therefore, we do not need to find the explicit expression for
$\dlambda$ to this order in perturbation theory; its job was to
produce eq.~\eqref{eq:Delta_form}.

We will now substitute the form of the variation \eqref{eq:Delta_form}
into the perturbation Lagrangian \eqref{eq:Lperturb}:
\begin{align}
\Lag^{\rm perturb}[\bvarphi,\bDelta] &=
\Lag_2^{\rm perturb}[\bvarphi,\bDelta]
+\Lag_3^{\rm perturb}[\bvarphi,\bDelta]
+\mathcal{O}(\epsilon^4),\\
\Lag_3^{\rm perturb}[\bvarphi,\bDelta] &=
\Lag_{3,{\rm quad}}^{\rm perturb}[\bvarphi,\bDelta]
+\Lag_{3,{\rm cubic}}^{\rm perturb}[\bvarphi,\bDelta],
\end{align}
where the NLO Lagrangian is given by
\begin{align}
\Lag_2^{\rm perturb}[\bvarphi,\bDelta] &=
\epsilon m_1^2\left(\Delta^1\varphi^2-\Delta^2\varphi^1\right)
-\frac12\left(\Delta^1\varphi^2-\Delta^2\varphi^1\right)^2
+\frac{\lambda_0}{2}\left(\bDelta^2 - (\bDelta\cdot\bvarphi)^2\right)
\label{eq:L2perturb}
\\&\phantom{=\ }
-\epsilon\bJ^\mu\cdot\p_\mu\bDelta\times\bvarphi
-\epsilon\bJ^\mu\cdot\bDelta\times\p_\mu\bvarphi
-\frac12V_{0ab}^{\mu\nu}\varepsilon^{acd}\p_\mu(\Delta^c\varphi^d)\varepsilon^{bef}\p_\nu(\Delta^e\varphi^f),\nonumber
\end{align}
and the N$^2$LO Lagrangians read
\begin{align}
\Lag_{3,{\rm quad}}^{\rm perturb}[\bvarphi,\bDelta] &=
\frac12\epsilon m_1^2\left((\bDelta\cdot\bvarphi)\Delta^3 - \bDelta^2\varphi^3\right)\non
&\phantom{=\ }
-\frac\epsilon2(\bvarphi\cdot\bDelta)(\bJ_\mu\cdot\p^\mu\bDelta)
-\frac\epsilon2(\bJ_\mu\cdot\bDelta)(\p^\mu\bvarphi\cdot\bDelta)
-\frac\epsilon2(\bJ_\mu\cdot\bDelta)(\bvarphi\cdot\p^\mu\bDelta)\non
&\phantom{=\ }
+\frac\epsilon2(\bJ_\mu\cdot\p^\mu\bvarphi)\bDelta^2
+\epsilon(\bJ_\mu\cdot\bvarphi)(\bDelta\cdot\p^\mu\bDelta)
\non
&\phantom{=\ }
-\frac\epsilon2V_{1ab}^{\mu\nu}\varepsilon^{acd}\p_\mu(\Delta^c\varphi^d)\varepsilon^{bef}\p_\nu(\Delta^e\varphi^f),\label{eq:L3perturb_quad}\\
\Lag_{3,{\rm cubic}}^{\rm perturb}[\bvarphi,\bDelta] &=
-\frac12\left(\Delta^1\varphi^2-\Delta^2\varphi^1\right)\left((\bDelta\cdot\bvarphi)\Delta^3
- \bDelta^2\varphi^3\right) \non
&\phantom{=\ }
-\frac12V_{0ab}^{\mu\nu}\varepsilon^{acd}\p_\mu(\Delta^c\varphi^d)\p_\nu\left(\Delta^b(\bDelta\cdot\bvarphi)- \bDelta^2\varphi^b\right) \non
&\phantom{=\ }
-\frac16\Gamma_{a b c}^{\mu\nu\rho}
  \varepsilon^{a d e}\p_\mu(\Delta^d\varphi^e)
  \varepsilon^{b f g}\p_\nu(\Delta^f\varphi^g)
  \varepsilon^{c h i}\p_\rho(\Delta^h\varphi^i),
\label{eq:L3perturb_cubic}
\end{align}
which is the complete perturbation Lagrangian to third order in
$\epsilon$.

In principle, we should solve this nonlinear problem for $\bDelta$,
which however is almost as difficult as the original problem, without
introducing the perturbation theory on top of the soliton background.
Therefore, we will linearize the above Lagrangian, i.e.~we will only
use the linear and quadratic parts in $\bDelta$
(i.e.~eqs.~\eqref{eq:L2perturb} and \eqref{eq:L3perturb_quad}) to
determine its equation of motion.

One may then wonder why go to the third order in $\epsilon$ if it
results in cubic terms in $\bDelta$, that we anyway will discard once
we turn to solving the equation of motion for the perturbation.
The answer is that we need the last term of
eq.~\eqref{eq:L3perturb_quad}, as it gives a term $\epsilon\p^2$ in the
equation of motion, whereas the last term in eq.~\eqref{eq:L2perturb}
will give a term $f(\varphi,\p_\mu\varphi)\p^2$ in the equation of
motion for $\bDelta$.
Although we are preparing for the study of the limit of very small
$\epsilon$ (say about $1/100$), it is still much larger than the
exponentially suppressed operator coming from
eq.~\eqref{eq:L2perturb}.
Therefore, neglecting it will yield an incorrect behavior at distances
of the order of the Skyrmion size away from the Skyrmion (dependent on
$\epsilon$ of course).

The linearized static equation of motion for $\bDelta$ can thus be
written down
\begin{equation}
  X \Delta_{i i}^a
  +X^{a b} \Delta_{i i}^b
  +X_{i j}^{a b} \Delta_{ij}^b
  +X_i^{a b} \Delta_{i}^b
  +\Lambda\Delta^a
  +\Lambda^{a b}\Delta^b
  = -\epsilon\varepsilon^{a b c}\varphi^b\varphi_{i i}^c
  -\epsilon m_1^2\varepsilon^{a b 3}\varphi^b,
  \label{eq:Xlineq}
\end{equation}
where we have defined the quantities
\begin{align}
  X &\equiv \epsilon,\\
  X^{a b} &\equiv
  -\epsilon\varphi^a\varphi^b
  +\varphi_j^a\varphi_j^b,\\
  X_{ij}^{ab} &\equiv -\varphi_i^a\varphi_j^b,\\
  X_i^{a b} &\equiv
  -2\epsilon\varphi_i^a\varphi^b
  -\varphi_{ij}^a\varphi_j^b
  +\varphi_{jj}^a\varphi_i^b
  -2\varphi_i^a\varphi_{jj}^b
  +2\varphi_j^a\varphi_{ij}^b
  +(\bvarphi_j\cdot\bvarphi_j)(\varphi^a\varphi_i^b-\varphi_i^a\varphi^b)\non
  &\phantom{=\ }
  -(\bvarphi_i\cdot\bvarphi_j)(\varphi^a\varphi_j^b-\varphi_j^a\varphi^b),\\
  \Lambda &\equiv -(1-\varphi^3)\varphi^3
  -\epsilon m_1^2\varphi^3,\\
  \Lambda^{a b} &\equiv
  -\frac\epsilon2\varphi_{ii}^a\varphi^b
  +\frac\epsilon2\varphi^a\varphi_{ii}^b
  +(1-\varphi^3)\varphi^3\varphi^a\varphi^b
  -(\bvarphi_{ij}\cdot\bvarphi_j)\varphi_i^a\varphi^b
  +(\bvarphi_{ii}\cdot\bvarphi_j)\varphi_j^a\varphi^b \non
  &\phantom{=\ }
  +(\bvarphi_i\cdot\bvarphi_j)\varphi_{ij}^a\varphi^b 
  -(\bvarphi_j\cdot\bvarphi_j)\varphi_{ii}^a\varphi^b
  +(\bvarphi_i\cdot\bvarphi_j)^2\varphi^a\varphi^b
  -(\bvarphi_i\cdot\bvarphi_i)^2\varphi^a\varphi^b
  -\varepsilon^{a c 3}\varepsilon^{b d 3}\varphi^c\varphi^d \non
  &\phantom{=\ }
  +\frac\epsilon2 m_1^2(\varphi^a\delta^{b3} + \delta^{a3}\varphi^b).
  \label{eq:Xtensors}  
\end{align}

It will now prove instructive to consider the equation of motion for
the fluctuation $\bDelta$ at an asymptotic distance from the BPS
baby Skyrmion background.
In that case, the background solution \eqref{eq:varphi} can be
approximated as
\beq
\bvarphi =
\begin{pmatrix}
  2e^{-\frac{\xi^2}{2}}\sqrt{1-e^{-\xi^2}}\cos(N\theta-\alpha)\\
  2e^{-\frac{\xi^2}{2}}\sqrt{1-e^{-\xi^2}}\sin(N\theta-\alpha)\\
  1-2e^{-\xi^2}
\end{pmatrix}
\simeq
\begin{pmatrix}
  2e^{-\frac{\xi^2}{2}}\cos(N\theta-\alpha)\\
  2e^{-\frac{\xi^2}{2}}\sin(N\theta-\alpha)\\
  1-2e^{-\xi^2}
\end{pmatrix}
+\mathcal{O}(e^{-\frac{3}{2}\xi^2}).
\eeq
Inserting this approximation into eq.~\eqref{eq:Xlineq}, we can
write the equation of motion for the fluctuation, to second order in
$e^{-\frac{\xi^2}{2}}$ as:
\begin{align}
&\epsilon\left(\p^2\Delta^a - \p^2\Delta^3\delta^{3a}\right)
-\epsilon m_1^2\left(\Delta^a - \Delta^3\delta^{3a}\right) \non
&\mathop-2\epsilon e^{-\frac{\xi^2}{2}}
\left(w^a \delta^{b3} + \delta^{a3} w^b\right)
\p^2\Delta^b
+\frac{4\epsilon}{R^2} e^{-\frac{\xi^2}{2}} w^ax^i\p_i\Delta^3
+\frac{4\epsilon N}{\xi^2R^2} e^{-\frac{\xi^2}{2}} \widehat{w}^a \epsilon_{ij}x^j\p_i\Delta^3 \non
&\mathop+\frac{\epsilon}{R^2}\left(\xi^2 - \frac{N^2}{\xi^2} - 2\right)
e^{-\frac{\xi^2}{2}}\left(-w^a\delta^{b3} + \delta^{a3}w^b\right)\Delta^b
+\epsilon m_1^2 e^{-\frac{\xi^2}{2}}
\left(w^a\delta^{b3} + \delta^{a3}w^b\right)\Delta^b \non
&
\mathop-4\epsilon e^{-\xi^2}(w^a w^b - \delta^{a3}\delta^{b3})\p^2\Delta^b
-\frac{8\epsilon}{R^2} e^{-\xi^2}x^i\p_i\Delta^3\delta^{a3}\non
&\mathop+\frac{8\epsilon}{R^2} e^{-\xi^2}\left(x^iw^aw^b
  + \frac{N}{\xi^2}\varepsilon_{ij}x^j\widehat{w}^a w^b\right)\p_i\Delta^b +2\epsilon m_1^2 e^{-\xi^2} (\Delta^a - \Delta^3\delta^{a3})
\non
&
\mathop+\frac{4\xi^2}{R^2}e^{-\xi^2}w^a w^b \p^2\Delta^b
+\frac{4N^2R^2}{\xi^2}\widehat{w}^a\widehat{w}^b e^{-\xi^2} \p^2\Delta^b \non
&\mathop-\frac{4}{R^4}e^{-\xi^2}
\left(w^ax^i + \frac{N}{\xi^2}\widehat{w}^a\epsilon_{ik}x^k\right)
\left(w^bx^j + \frac{N}{\xi^2}\widehat{w}^b\epsilon_{jl}x^l\right)
\p_i\p_j\Delta^b \non
&\mathop-\frac{4N}{\xi^2R^4}e^{-\xi^2}\left[(3\xi^2+1)w^a\widehat{w}^b
  - (3\xi^2-1)\widehat{w}^aw^b\right]\epsilon_{ij}x^j\p_i\Delta^b\non
&\mathop-\frac{4}{R^4}e^{-\xi^2}\left[
  \left(1+\frac{N^2}{\xi^2}\right)w^aw^b
  + N^2\left(\frac{1}{\xi^2} + \frac{1}{\xi^4}\right)\widehat{w}^a\widehat{w}^b
  \right]x^i\p_i\Delta^b\non
&\mathop-4e^{-\xi^2}\widehat{w}^a\widehat{w}^b\Delta^b
-2e^{-\xi^2}(\Delta^a - \Delta^3\delta^{a3})
\non
&\quad = \,
-\frac{2\epsilon}{R^2}\left(\xi^2 - \frac{N^2}{\xi^2} - 2 - m_1^2R^2\right)
e^{-\frac{\xi^2}{2}} \widehat{w}^a,
\label{eq:asymptotic}
\end{align}
where we have defined
\beq
\bw =
\begin{pmatrix}
  \cos(N\theta - \alpha)\\
  \sin(N\theta - \alpha)\\
  0
\end{pmatrix}, \qquad
\widehat{\bw} =
\begin{pmatrix}
  -\sin(N\theta - \alpha)\\
  \cos(N\theta - \alpha)\\
  0
\end{pmatrix}.
\eeq
Clearly, only the first line of eq.~\eqref{eq:asymptotic} is not
exponentially suppressed and therefore guarantees the propagation of
the fluctuation $\bDelta$ in the asymptotic regime (away from the
background baby Skyrmion).
Now, importantly, all the terms with a factor of $\epsilon$ on the
left-hand side of eq.~\eqref{eq:asymptotic} come from the
$\mathcal{O}(\epsilon^3)$ Lagrangian \eqref{eq:L3perturb_quad} and
therefore had we only gone to second order in the $\epsilon$
expansion, the linearized equation for the fluctuation would have
looked like this:
\begin{align}
\frac{4\xi^2}{R^2}e^{-\xi^2}w^a w^b \p^2\Delta^b
-\frac{4}{R^4}e^{-\xi^2} w^a w^b x^i x^j \p_i\p_j\Delta^b 
-\frac{12N}{R^4}e^{-\xi^2}\left[w^a\widehat{w}^b -
  \widehat{w}^aw^b\right]\epsilon_{ij}x^j\p_i\Delta^b \non
-\frac{4}{R^4}e^{-\xi^2} w^a w^b x^i\p_i\Delta^b
-4e^{-\xi^2}\widehat{w}^a\widehat{w}^b\Delta^b
-2e^{-\xi^2}(\Delta^a - \Delta^3\delta^{a3})
=
-\frac{2\epsilon}{R^2}\xi^2 e^{-\frac{\xi^2}{2}} \widehat{w}^a,
\label{eq:asymptotic_second_order}
\end{align}
where we have kept only the terms with the highest powers of $\xi$ on
both sides.
To this order, the equation of motion for the fluctuation is badly
behaved, because it schematically takes the form
\beq
\mathcal{O}^{a b}\Delta^b =
-\frac{2\epsilon}{R^2}\xi^2 e^{\frac{\xi^2}{2}} \widehat{w}^a, \qquad
\xi\equiv\frac{r}{R},
\eeq
whose right-hand side diverges exponentially.
We expect of course that the fluctuations go to zero at asymptotic
distances, and this will also be the case for the third-order in
$\epsilon$ equation of motion \eqref{eq:asymptotic}, because of the
terms $\epsilon (\p^2 - m_1^2) \Delta^a$, which have the expected
exponential falloff compatible with the boundary conditions.

We have assumed that the perturbation field is proportional
to $\epsilon$. In order to attempt to address this point, let us
consider (for simplicity) a 
few of the terms of the third-order in $\epsilon$ equation of motion
\eqref{eq:asymptotic} for the fluctuation:
\begin{align}
\epsilon\left(\p^2\Delta^a - \p^2\Delta^3\delta^{3a}\right)
-\epsilon m_1^2\left(\Delta^a - \Delta^3\delta^{3a}\right) 
+\frac{4\xi^2}{R^2}e^{-\xi^2}w^a w^b \p^2\Delta^b
-4e^{-\xi^2}\widehat{w}^a\widehat{w}^b\Delta^b \non
-2e^{-\xi^2}(\Delta^a - \Delta^3\delta^{a3}) = 0.
\end{align}
Since $\bw$ and $\widehat{\bw}$ are of order one, we can estimate at
which distance from the background baby Skyrmion the third order terms
become dominant:
\beq
\epsilon \gg \frac{4\xi^2}{R^2}e^{-\xi^2},
\label{eq:epsilon_vs_xi}
\eeq
where $\xi\equiv \frac{r}{R}$ and $R=\sqrt{2N}$.
The smaller values of $\epsilon$, the larger distances are needed
before the third-order terms take over.
Since the exponential of $-r^2$ quickly becomes infinitesimally small,
the tail of the perturbation at asymptotic distances will take the
form
\beq
\bDelta =
\begin{pmatrix}
  c_1\\
  c_2\\
  0
\end{pmatrix}
e^{-m_1 r },
\label{eq:tail}
\eeq
where $c_{1,2}$ are constants.
Notice that this is seemingly independent of $\epsilon$.
All the $\epsilon$-dependence is contained at distances
\beq
\epsilon \lesssim \frac{4\xi^2}{R^2}e^{-\xi^2},
\label{eq:smaller_distances}
\eeq
for which the $\epsilon$-dependence becomes quite complicated.
Indirectly, the dependence on $\epsilon$ in eq.~\eqref{eq:tail} is
possessed by $c_{1,2}$ by gluing it together with the solution at
distances given by eq.~\eqref{eq:smaller_distances}.
Importantly, in the limit of $\epsilon=0$, $c_{1,2}=0$
because there is no tail correction in the BPS limit.
To leading order, the coefficients must behave like
$c_{1,2}\propto\epsilon^p$, with $p\in\mathbb{Z}_{>0}$ a positive
integer. 
We have, however, not been able to prove rigorously that $p=1$
(i.e.~it could be larger than one).

It is worthwhile to compare the situation of the setting at hand -- a
baby Skyrmion having a Gaussian tail in the BPS limit and where
the $\epsilon$ perturbation of the model entails both the Dirichlet
kinetic term as well as the standard pion mass term, see
eq.~\eqref{eq:L} -- with the compacton case studied in
ref.~\cite{Gudnason:2020tps}.

In the compacton case, as the name suggests, the BPS limit of the baby
Skyrmion is a compacton (hence no tail in the BPS limit) and the
$\epsilon$ perturbation in that case was done solely by adding the
Dirichlet kinetic term (with coefficient $\epsilon$).
Let us recapitulate the situation in the compacton case of
ref.~\cite{Gudnason:2020tps}.
The tail of the perturbation in that case became non-dynamic unless we
went to the third order in $\epsilon$ (which would prevent the
compactons of knowing of one another and hence prevent the calculation
of binding energies).
Going to the third order in the $\epsilon$ expansion gave a dynamic
tail to the perturbation, which however was nonanalytic in $\epsilon$.
The nonanalycity was due to the fact that the (inverse) propagator had
the form $\epsilon\p^2-m^2$ and the tail thus took the form
$e^{-\frac{m r}{\sqrt{\epsilon}}}$, which when Taylor expanded in $\epsilon$
vanishes at any finite order.

In this case of a baby Skyrmion with a Gaussian tail in the BPS
limit, the situation draws some similarities to the compacton case,
but nevertheless is dramatically different.
The similar property of the $\epsilon$ expansion, is that we still
have to go to the third order in $\epsilon$, because otherwise -- as
demonstrated above -- the equations of motion for the perturbation
become ill-defined at asymptotic distances or at least incompatible
with suitable boundary conditions.
A huge difference is that the baby Skyrmions, in the BPS limit,
themselves have Gaussian tails, and therefore already ``feel'' each
other when two or more of them are placed at a finite distance from
each other.
The BPS solution is thus nontrivial for any separation distance.
If we now turn on a small but finite $\epsilon$ according to the
Lagrangian at hand \eqref{eq:L}, the perturbation field $\Delta$ adds
an exponential tail to the background solution.
The field configuration now flows to the nearest GRH solution in field
space. 
The solution to the generalized restricted harmonicity is, however,
extremely difficult compared to the compacton case, which is nearly
trivial. That is, axially symmetric baby Skyrmions placed at distances
such that their compacton regions do not overlap -- for details, see
ref.~\cite{Gudnason:2020tps}.
Now, since the background fields (meaning the fields of the background
BPS baby Skyrmion) tend to the vacuum exponentially (or rather
like the Gaussian), the governing
equations of motion for the perturbation are simply
\beq
\epsilon\left(\p^2\Delta^a - \p^2\Delta^3\delta^{3a}\right)
-\epsilon m_1^2\left(\Delta^a - \Delta^3\delta^{3a}\right) = 0,
\label{eq:very_asymptotic}
\eeq
and thus the tail of the perturbation (not the BPS solution) is
exponential too, see eq.~\eqref{eq:tail}, but in stark
contradistinction to the compacton case, it is independent of
$\epsilon$ (it is nevertheless ``glued together'' with a solution
closer to the baby Skyrmion that \emph{is} dependent on $\epsilon$ and
hence a suppression of the tail is still in effect, but we shall check
this \emph{a posteriori}).
Of course, this is not an accident, but a consequence of the
construction of this more elaborate (and aimed at being more physical)
model, compared with the compacton case of
ref.~\cite{Gudnason:2020tps}. 
We include the Dirichlet kinetic term with a coefficient $\epsilon$,
because we want small binding energies (that is a small perturbation
added to the BPS model).
Then we include the mass term, also with coefficient $\epsilon$, in
order to prevent that the pion mass becomes unrealistically large in
the small-$\epsilon$ limit. 
Another feature -- also by construction -- is that the potential in
the BPS sector ($V_2$) does not give rise to a pion mass and therefore
the coefficient of the potential is not restricted by the value of the 
pion mass.

The picture that forms, which we will elaborate on in a later
section, is that at asymptotically large distances, the governing
equation for the perturbation is eq.~\eqref{eq:very_asymptotic} and
the tails of two or more baby Skyrmions attract each other (in the
attractive channels) with a force that is at least cubic in $\epsilon$,
but the mass of the fluctuations is independent of $\epsilon$.
The pion mass can thus be set to any realistic value\footnote{Of
  course, we are working with the 2-dimensional baby Skyrme model, but the
  aim is to test this framework in two dimensions before attempting at
  addressing the 3-dimensional model.}.

A word of caution is that we calculate the perturbation using the
linearized equation of motion that contains some (crucial) terms
\eqref{eq:L3perturb_quad} at $\mathcal{O}(\epsilon^3)$, but not the
cubic terms \eqref{eq:L3perturb_cubic} at the same order.
As for the calculation of the energy, we will compare the energy
calculated at NLO and N$^2$LO to the exact numerical solutions in the
axially symmetric case in the next section.

The static energy of the perturbation is simply given by
\beq
\mathcal{E}^{\rm perturb}[\bvarphi,\bDelta]
= \left.-\Lag^{\rm perturb}[\bvarphi,\bDelta]\right|_{\p_0=0}.
\eeq

A comment in store about the perturbation, $\bDelta$, is that the
entire static energy vanishes for $\bDelta\propto\bvarphi$.
However, since the expression \eqref{eq:Delta_form} for $\bdphi$ is
nonlinear in $\bDelta$, a few cross terms survive if we take
$\bDelta=\dc\bvarphi+\bDelta_{\bot}$, with 
$\bDelta_{\bot}\cdot\bvarphi=0$ a perpendicular perturbation. 
The cross terms only give rise to a linear first-order PDE for $\dc$,
which must vanish once subject to the boundary conditions
$\lim_{|x|\to\infty}\dc=0$ and $\dc(0)=0$, where the origin is at each
background baby Skyrmion center.
This justifies setting $\bDelta=\bDelta_{\bot}$. 
For more details, see ref.~\cite{Gudnason:2020tps}.

It will now prove useful to specialize to the case of the background
BPS baby Skyrmion solution for $\bvarphi$, using transverse
perturbations for $\bDelta=\bDelta_{\bot}$ and switching to polar
coordinates in the plane, for which the static perturbation energy
reads 
\beq
\mathcal{E}^{\rm perturb}[f,\df,\dtheta] =
\mathcal{E}_2^{\rm perturb}[f,\df,\dtheta]
+\mathcal{E}_{3,{\rm quad}}^{\rm perturb}[f,\df,\dtheta]
+\mathcal{E}_{3,{\rm cubic}}^{\rm perturb}[f,\df,\dtheta],
\label{eq:Eperturb_df_dtheta}
\eeq
with
\begin{align}
&\mathcal{E}_2^{\rm perturb}[f,\df,\dtheta] =
\left(\epsilon m_1^2\sin f+\frac{\epsilon N^2}{2r^2}\sin(2f)\right)\df
+\epsilon f_r\df_r 
+\frac{N^2}{2r^2}\sin^2(f)\df_r^2
+\frac{f_r^2}{2r^2}\dtheta_\theta^2\non
&\phantom{=\ }
+\frac12\left(\cos f - \cos(2f) + \frac{N^2}{r^2}\cos(2f)f_r^2\right)\df^2
+\left(\cos f\sin^2\left(\frac{f}{2}\right) - \frac{N^2}{2r^2}\sin^2(f)f_r^2\right)\dtheta^2\non
&\phantom{=\ }
+\frac{N}{r^2}\sin(f)f_r
  \left[2\df_r\dtheta_\theta - \df_\theta\dtheta_r\right]
+\frac{N^2}{2r^2}\sin(2f)f_r
  \left[2\df_r\df + \dtheta_r\dtheta\right]
+\frac{N}{r^2}\cos(f)f_r^2\df\dtheta_\theta,
\end{align}
for the NLO terms,
\begin{align}
\mathcal{E}_{3,{\rm quad}}^{\rm perturb}[f,\df,\dtheta] &=
\frac\epsilon2\left(\df_r^2+\frac{\df_\theta^2}{r^2}\right)
+\frac{\epsilon}{2}\left(m_1^2\cos f + \frac{N^2}{r^2}\cos(2f)\right)\df^2
+\frac\epsilon2\left(\dtheta_r^2 + \frac{\dtheta_\theta^2}{r^2}\right)\non
&\phantom{=\ }
+\frac\epsilon2\left(m_1^2\cos f - f_r^2 + \frac{N^2}{r^2}\cos^2f\right)\dtheta^2
-\frac{\epsilon N}{r^2}\cos(f)\left(\df_\theta\dtheta - \df\dtheta_\theta\right),
\end{align}
for the N$^2$LO terms quadratic in $\bDelta_{\bot}$ and
\begin{align}
&\mathcal{E}_{3,{\rm cubic}}^{\rm perturb}[f,\df,\dtheta] =
\frac{1}{2r^2}\left(
  N\sin(f)\df_r
  +f_r\dtheta_\theta
  +N\cos(f)f_r\df
  \right)\times\non&
\left[
  2N\cos(f)\df_r\df
  +2\df_r\dtheta_\theta
  +2N\cos(f)\dtheta_r\dtheta
  -2\df_\theta\dtheta_r
  -N\sin(f)f_r\df^2
  -N\sin(f)f_r\dtheta^2
\right]\non
&\phantom{=\ }
+\frac14\sin2f\left(\df^2+\dtheta^2\right)\df,
\end{align}
for the N$^2$LO terms cubic in $\bDelta_{\bot}$, and we have defined
the transverse perturbations $\bDelta_{\bot}$ as
\beq
\bDelta_\bot \equiv
\begin{pmatrix}
  -\sin N\theta\\
  \cos N\theta\\
  0
\end{pmatrix}\df -
\begin{pmatrix}
  \cos f\cos N\theta\\
  \cos f\sin N\theta\\
  -\sin f
\end{pmatrix}\dtheta,
\label{eq:Delta_transverse}
\eeq
and the background BPS baby Skyrmion solution is given by 
\beq
f = f^{\rm BPS}
= \arccos\left(1 - 2e^{-\frac{r^2}{R^2}}\right),
\eeq
with $R=\sqrt{2N}$, (eq.~\eqref{eq:radius}).

The corresponding static equations of motion read
\begin{align}
  X_{rr}
  \begin{pmatrix}
    \df_{rr}\\
    \dtheta_{rr}
  \end{pmatrix}
  +\frac{1}{r}X_r
  \begin{pmatrix}
    \df_r\\
    \dtheta_r
  \end{pmatrix}
  +\frac{1}{r^2}X_{\theta\theta}
  \begin{pmatrix}
    \df_{\theta\theta}\\
    \dtheta_{\theta\theta}
  \end{pmatrix}
  +\frac{1}{r}
  \begin{pmatrix}
    0 & X_\theta^{\dtheta}\\
    X_\theta^{\df} & 0
  \end{pmatrix}
  \begin{pmatrix}
    \df_\theta\\
    \dtheta_\theta
  \end{pmatrix}
  +\frac{X_{r\theta}}{r}
  \begin{pmatrix}
    0 & 1\\
    1 & 0
  \end{pmatrix}
  \begin{pmatrix}
    \df_{r\theta}\\
    \dtheta_{r\theta}
  \end{pmatrix}\non
  +
  \begin{pmatrix}
    \Lambda^{\df} & 0\\
    0 & \Lambda^{\dtheta}
  \end{pmatrix}
  \begin{pmatrix}
    \df\\
    \dtheta
  \end{pmatrix}
  =
  \begin{pmatrix}
    -\epsilon\left(f_{rr} + \frac{1}{r}f_r - \frac{N^2}{2r^2}\sin(2f)
    -m_1^2\sin f\right)\\
    0
  \end{pmatrix},
  \label{eq:Xpde_polar}
\end{align}
where we have defined the matrices
\begin{align}
  X_{rr} &=
  \begin{pmatrix}
    \epsilon + \frac{N^2}{r^2}\sin^2f & 0\\
    0 & \epsilon
  \end{pmatrix},\\
  X_r &=
  \begin{pmatrix}
    \epsilon - \frac{N^2}{r^2}\sin^2f + \frac{N^2}{r}\sin(2f)f_r & 0\\
    0 & \epsilon
  \end{pmatrix},\\
  X_{\theta\theta} &=
  \begin{pmatrix}
    \epsilon & 0\\
    0 & \epsilon + f_r^2
  \end{pmatrix},
\end{align}
as well as the functions
\begin{align}
  X_\theta^{\df} &= \frac{2\epsilon N}{r}\cos f
    -\frac{N}{r}\sin(f)f_{rr}
    +\frac{N}{r^2}\sin(f)f_r,\\
  X_\theta^{\dtheta} &= -\frac{2\epsilon N}{r}\cos f
    +\frac{2N}{r}\sin(f)f_{rr}
    +\frac{N}{r}\cos(f)f_r^2
    -\frac{2N}{r^2}\sin(f)f_r,\\
  X_{r\theta} &= \frac{N}{r}\sin(f)f_r,\\
  \Lambda^{\df} &=
  -\epsilon m_1^2\cos f
  -\frac{\epsilon N^2}{r^2}\cos(2f)
  +\frac{N^2}{r^2}\sin(2f)f_{rr}
  +\frac{N^2}{r^2}\cos(2f)f_r^2
  -\frac{N^2}{r^3}\sin(2f)f_r
  -\cos f\non&\phantom{=\ }
  +\cos 2f,\\
  \Lambda^{\dtheta} &=
  \epsilon f_r^2
  -\frac{\epsilon N^2}{r^2}\cos^2(f)
  +\frac{N^2}{2r^2}\sin(2f)f_{rr}
  +\frac{N^2}{r^2}\cos^2(f)f_r^2
  -\frac{N^2}{2r^3}\sin(2f)f_r
  -m^2\cos f.
\end{align}
Firstly, the source term, i.e.~the right-hand side of
eq.~\eqref{eq:Xpde_polar} only exists for the upper equation, that is
for the equation of motion for $\df$.
Secondly, the mixing between the upper and the lower equations only
appears in terms involving a $\theta$ derivative and a mixed $r$ and
$\theta$ derivative. 
Thus, $\dtheta$ is a homogeneous source-free equation of motion,
unless $\df$ has nontrivial $\theta$ dependence.
Therefore, if we restrict to axially symmetric background BPS
solutions and turn on only axially symmetric perturbations,
$\df=\df(r)$, then $\dtheta$ decouples and is trivially satisfied
($\dtheta=0$ everywhere).

The equation of motion for the perturbation of axially symmetric baby
Skyrmions thus reduces to
\begin{align}
\left(\epsilon + \frac{N^2}{r^2}\sin^2f\right) \df_{rr}
+\frac1r\left(\epsilon - \frac{N^2}{r^2}\sin^2f + \frac{N^2}{r}\sin(2f)f_r\right) \df_r
+\Lambda^{\df}\df \qquad\non
= -\epsilon\left(f_{rr} + \frac{1}{r}f_r - \frac{N^2}{2r^2}\sin2f
-m_1^2\sin f\right),
\label{eq:axially_sym_EOM_df}
\end{align}
and the corresponding static energy for the perturbation is
\begin{align}
\mathcal{E}_2^{\rm perturb}[f,\df] &=
\left(\epsilon m_1^2\sin f + \frac{\epsilon N^2}{2r^2}\sin(2f)\right)\df
+\epsilon f_r\df_r  
+\frac{N^2}{2r^2}\sin^2(f)\df_r^2\non
&\phantom{=\ }  
+\frac12\left(\cos f - \cos(2f) + \frac{N^2}{r^2}\cos(2f)f_r^2\right)\df^2
+\frac{N^2}{r^2}\sin(2f)f_r\df_r\df,
\label{eq:E2axial}
\end{align}
for the NLO terms,
\begin{align}
\mathcal{E}_{3,{\rm quad}}^{\rm perturb}[f,\df] &=
\frac\epsilon2\df_r^2
+\frac{\epsilon}{2}\left(m_1^2\cos f + \frac{N^2}{r^2}\cos(2f)\right)\df^2,
\label{eq:E3axialquad}
\end{align}
for the N$^2$LO terms quadratic in $\df$ and
\begin{align}
\mathcal{E}_{3,{\rm cubic}}^{\rm perturb}[f,\df] &=
\frac{N^2}{2r^2}\sin(2f)\df_r^2\df
+\frac{N^2}{r^2}\left(1 - \frac32\sin^2f\right)f_r\df_r\df^2
-\frac{N^2}{4r^2}\sin(2f)f_r^2\df^3 \non
&\phantom{=\ }
\mathop+\frac14\sin(2f)\df^3,
\label{eq:E3axialcubic}
\end{align}
for the N$^2$LO terms cubic in $\df$.

\subsection{Axially symmetric solutions}\label{sec:axial}

In order to verify the accuracy of our perturbative scheme, we start
with axially symmetric baby Skyrmions.
As was shown in ref.~\cite{Gudnason:2020tps}, the perturbation can in
this case be written as
\begin{align}
\bphi &= \bvarphi + \bDelta_\bot\times\bvarphi + \frac12\bDelta_\bot\times(\bDelta_\bot\times\bvarphi)\non
&\simeq
\begin{pmatrix}
  \sin(f+\df)\cos N\theta\\
  \sin(f+\df)\sin N\theta\\
  \cos(f+\df)
\end{pmatrix}
+ \mathcal{O}(\df^3),
\label{eq:axial_sym_ansatz}
\end{align}
where we have used eqs.~\eqref{eq:Delta_form},
\eqref{eq:Delta_transverse} and set $\dtheta=0$.
It is thus clear that $\df$ is indeed an additive correction to the
BPS background profile function $f$ in the axially symmetric case.

\begin{figure}[!htp]
  \begin{center}
    \mbox{\includegraphics[width=0.49\linewidth]{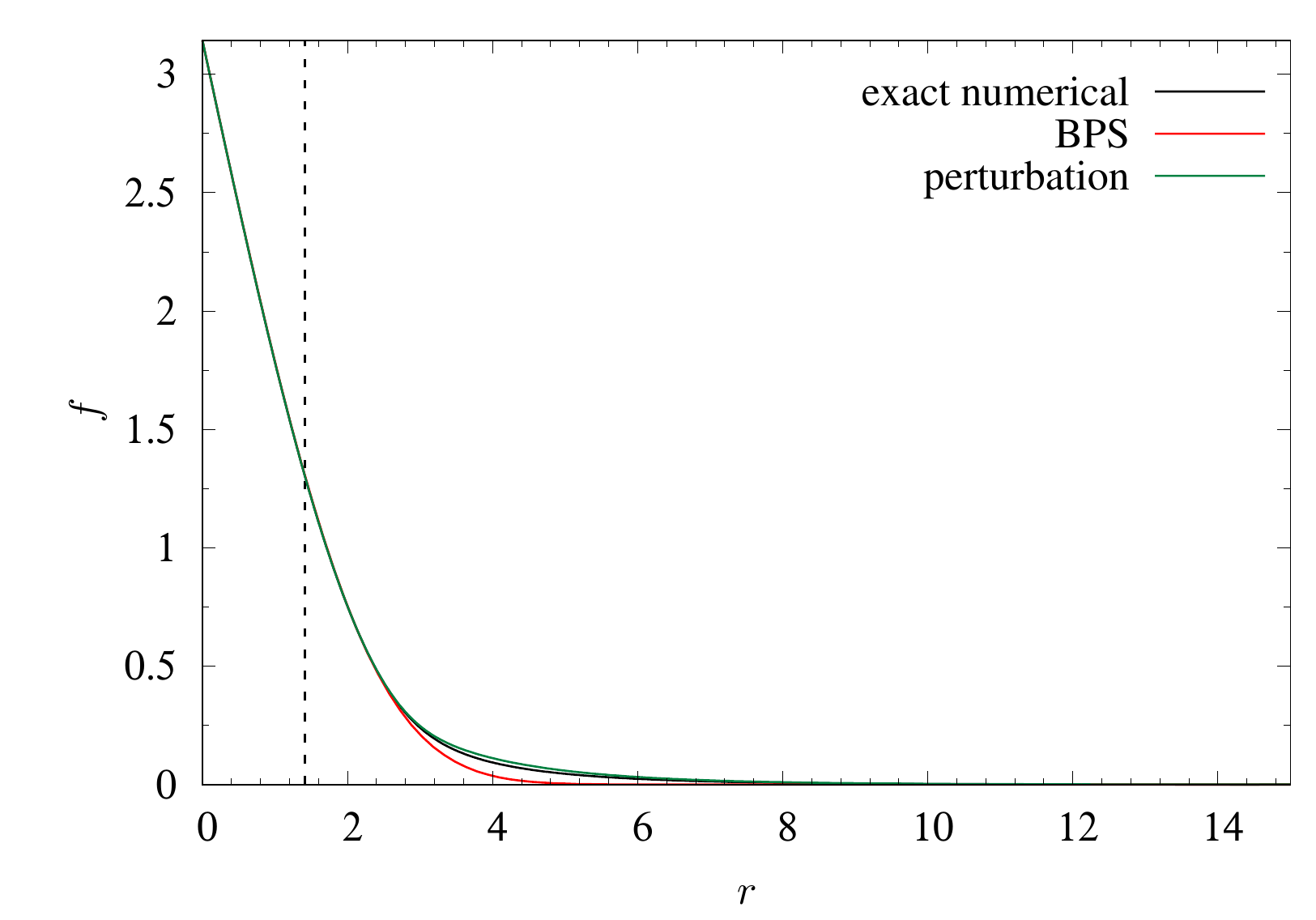}
      \includegraphics[width=0.49\linewidth]{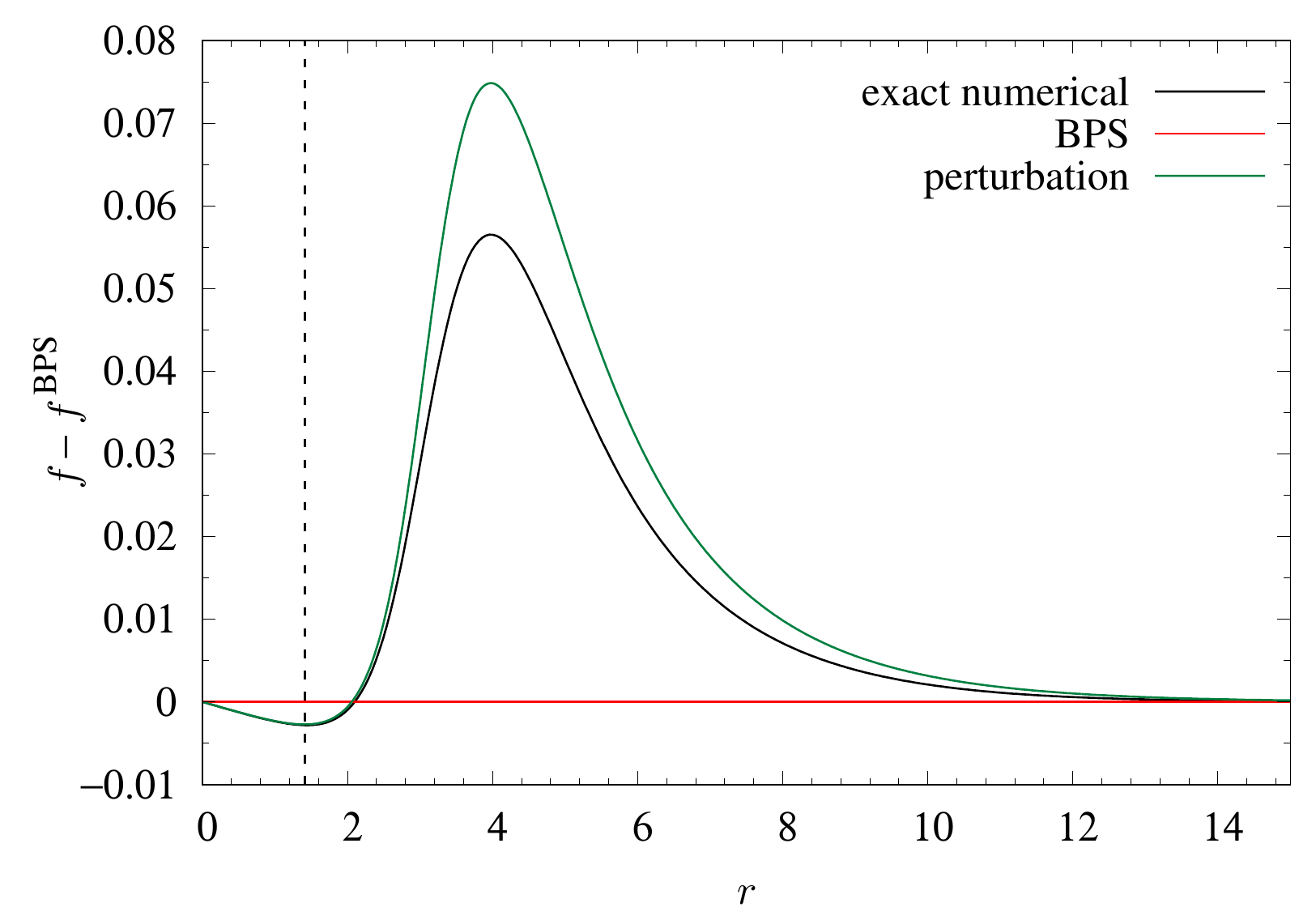}}
    \mbox{\includegraphics[width=0.49\linewidth]{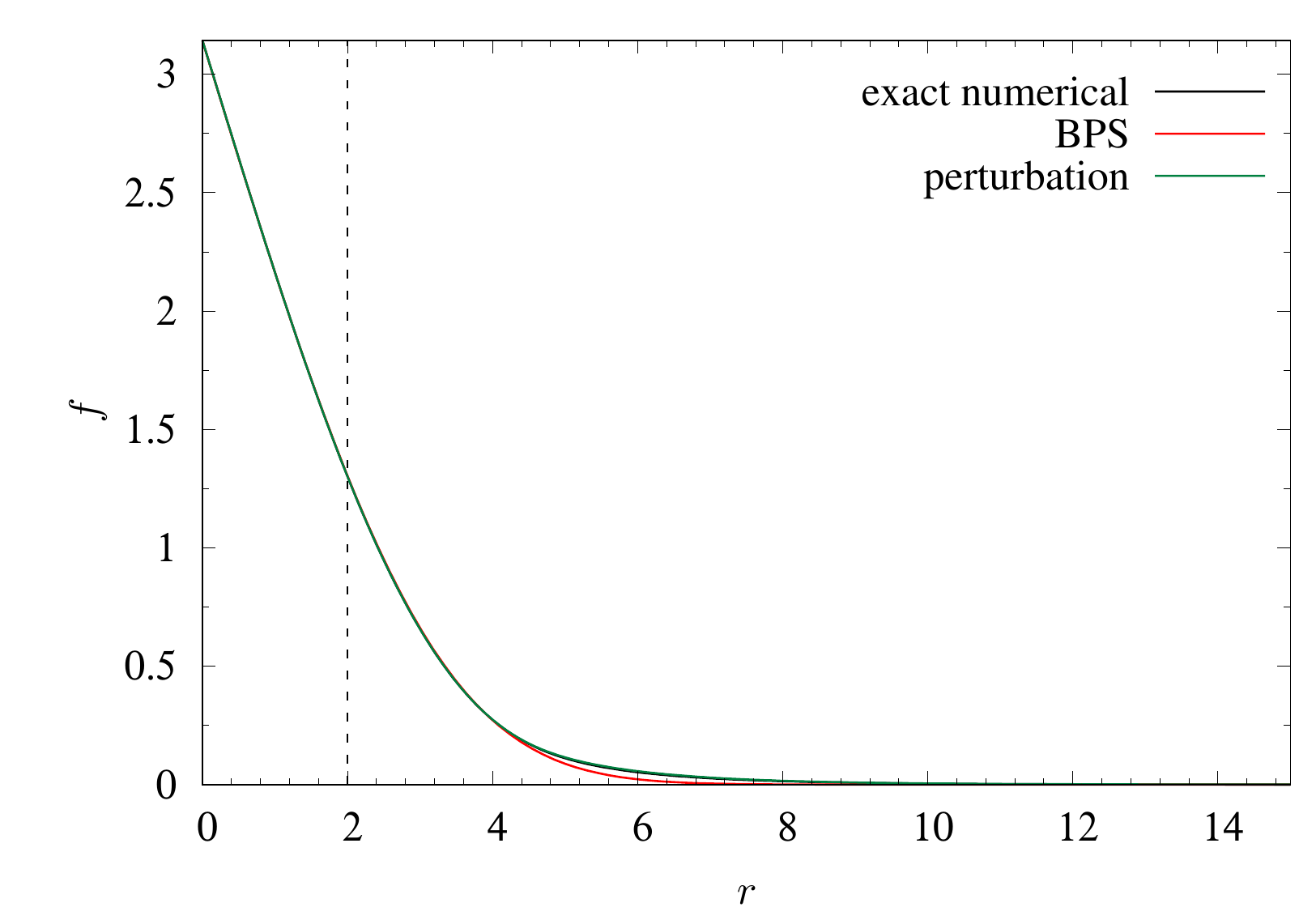}
      \includegraphics[width=0.49\linewidth]{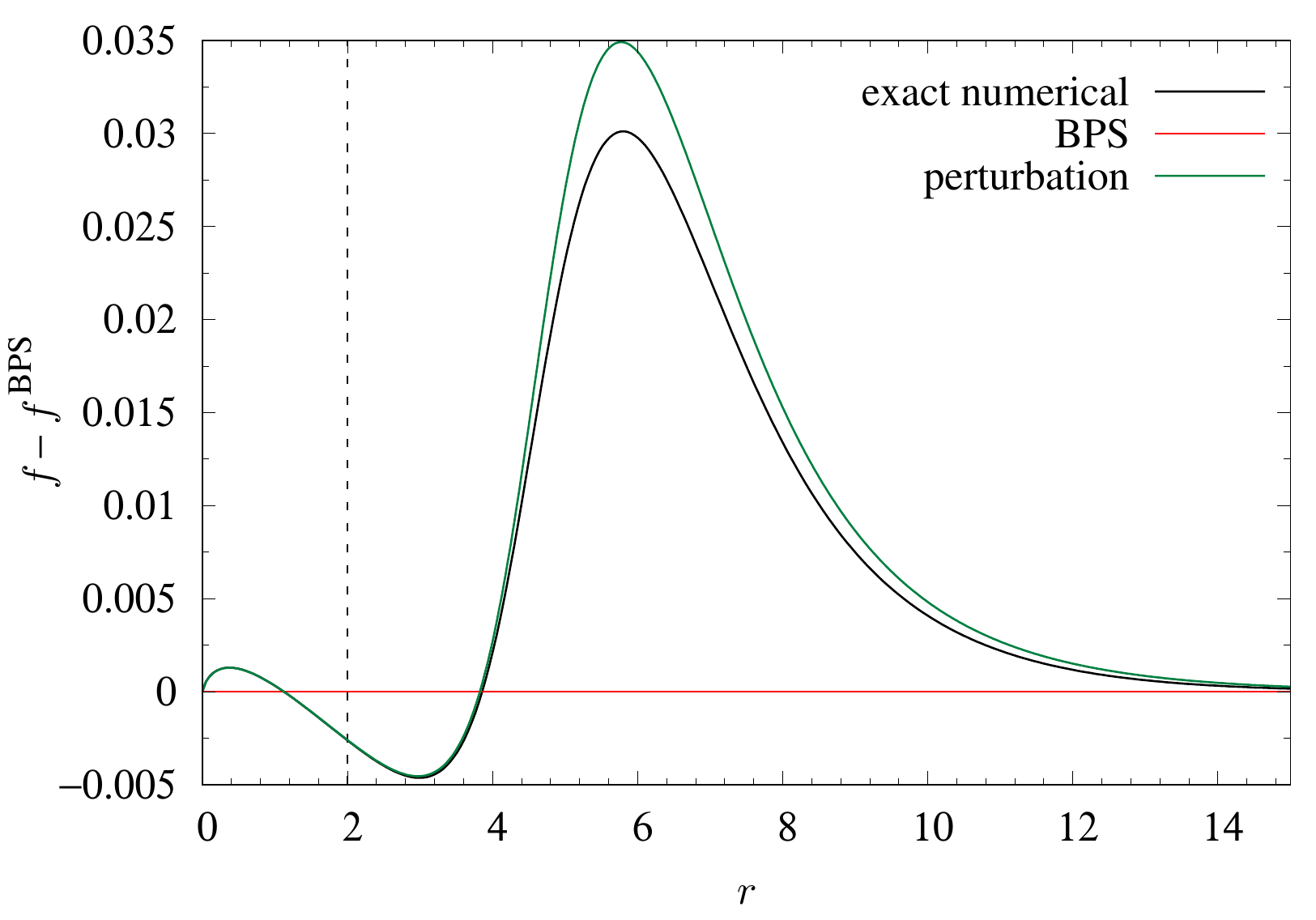}}
    \mbox{\includegraphics[width=0.49\linewidth]{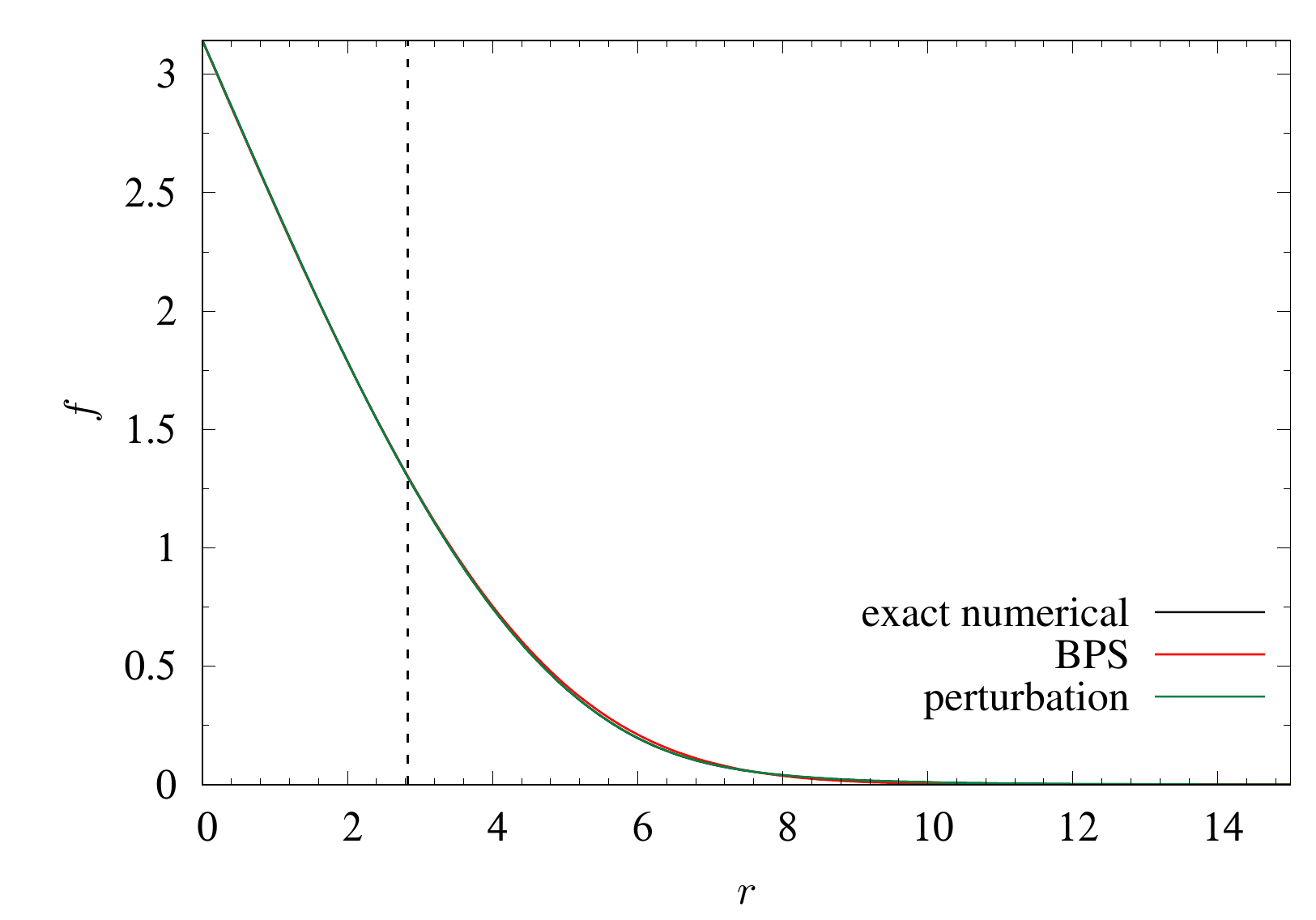}
      \includegraphics[width=0.49\linewidth]{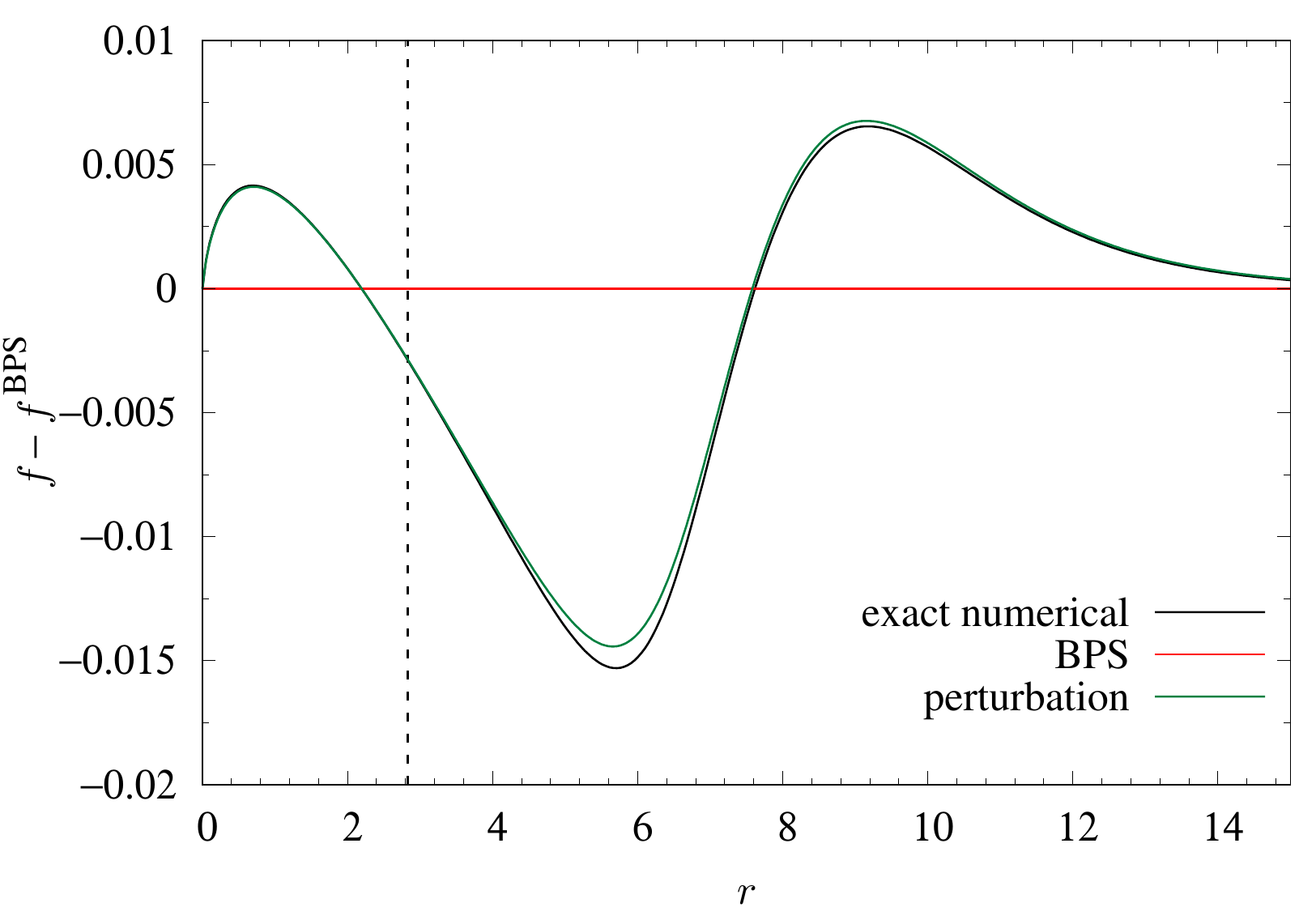}}
    \caption{The profile function $f=\arccos(\phi^3)$ for $N=1,2,4$
      baby Skyrmions with $\epsilon=0.01$ and $m_1=0.5$. The BPS
      background profile function is shown as a red solid line.
      The perturbation (dark-green solid line) gets closer to the true 
      solution (black solid line), obtained by numerical calculation,
      as $N$ gets larger. 
      (a) profile functions. (b) profile function with the BPS one
      subtracted off.
      The dashed black vertical line marks $R=\sqrt{2N}$ on each
      panel, which is the characteristic length scale of the baby
      Skyrmion. 
    } 
    \label{fig:lpprofile}
  \end{center}
\end{figure}
Fig.~\ref{fig:lpprofile} shows the profile function
$f=\arccos(\phi^3)$ for the $N=1,2,4$ axially symmetric baby Skyrmions
with $\epsilon=0.01$ and $m_1=0.5$ obtained by three methods: exact
numerical calculation,  BPS approximation ($\epsilon=0$) and 
$\epsilon$ expansion, denoted as ``perturbation''.
In the right-hand side panel, the BPS profile is subtracted off of all
the profiles, so the differences between the solutions are clearly
visible.
The characteristic radius, $R=\sqrt{2N}$ (see eq.~\eqref{eq:radius})
of the baby Skyrmion is shown with a vertical dashed black line on
each panel of the figure. 

Notice that the perturbation qualitatively captures the correct
behavior of the exact solution everywhere.
For $r\lesssim 1.5R$ the perturbative correction to the BPS solution
matches the exact solution extremely well, see the right-hand side
panels of fig.~\ref{fig:lpprofile} (the characteristic radius
$R=\sqrt{2N}$ is marked with a vertical dashed black line on each
graph) and this is the radius within which 99\% of the energy of the
BPS baby Skyrmion is contained. 
However, for $r\gtrsim 2R$ the perturbative correction overshoots
(undershoots) the exact solution for $N=1,2$ ($N=4$), which indicates
the level of precision of the perturbative method at the linear order
in the perturbation field (corresponding to second and some third order
terms in $\epsilon$).

One could think that it would be necessary to include all terms at the
third order in the $\epsilon$ expansion, hence making the problem a
nonlinear one.
However, as the discrepancy between the perturbative solution and the
exact solution appears only at distances $r\gtrsim 2R$, the
exponential suppression of the tail of the BPS background solution
makes the perturbative contribution to the energy subdominant in that
(asymptotic) region; in fact, less than 1\% of the energy density of
the BPS background baby Skyrmion resides beyond the distance of
$\sim1.5R$ from the center of the soliton. 
In order to demonstrate that this statement is true, we now turn to
calculating the perturbative corrections to the energy of the
axially symmetric baby Skyrmions.

\begin{figure}[!ht]
  \begin{center}
    \mbox{\subfloat{\includegraphics[width=0.33\linewidth]{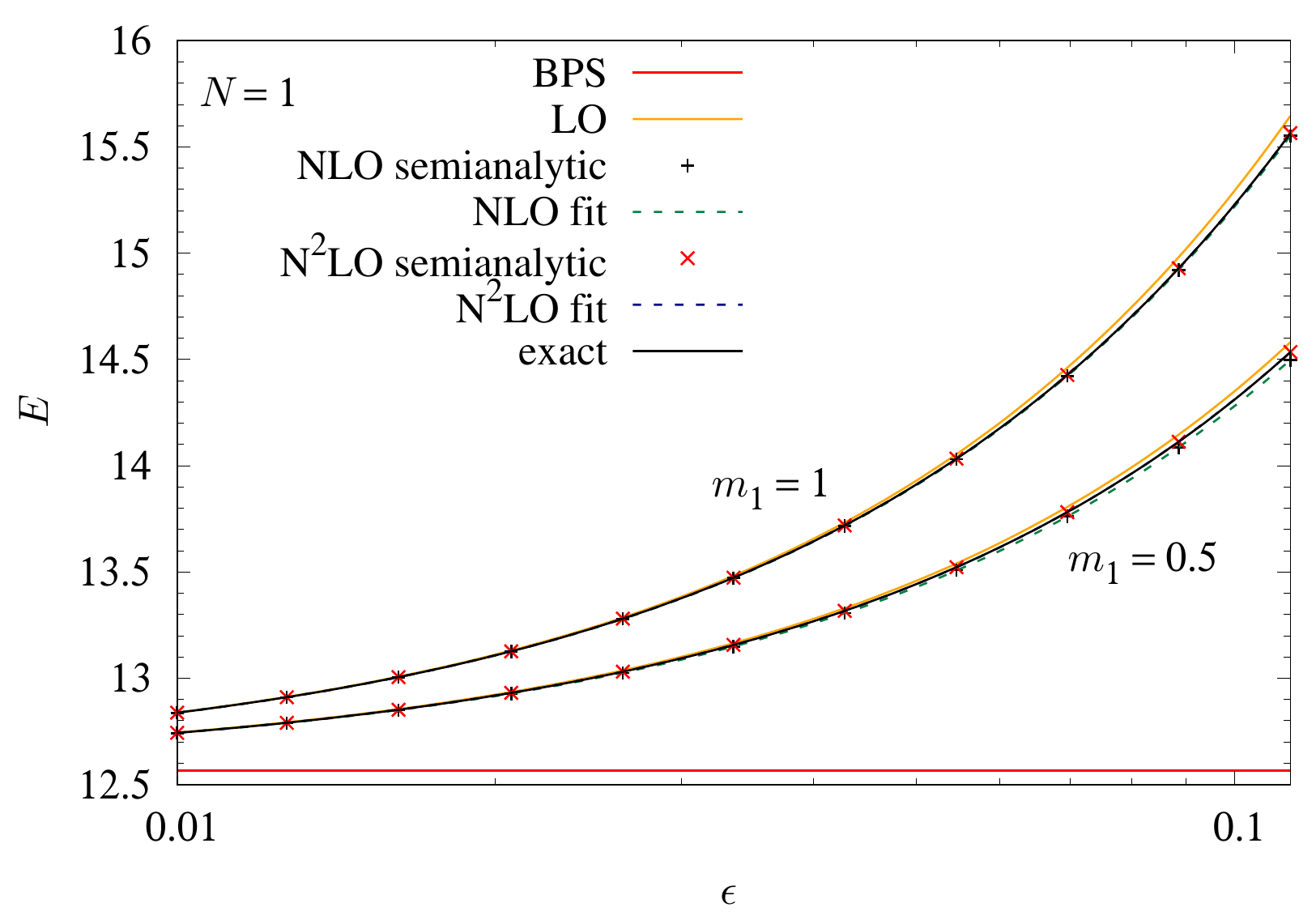}}
      \subfloat{\includegraphics[width=0.33\linewidth]{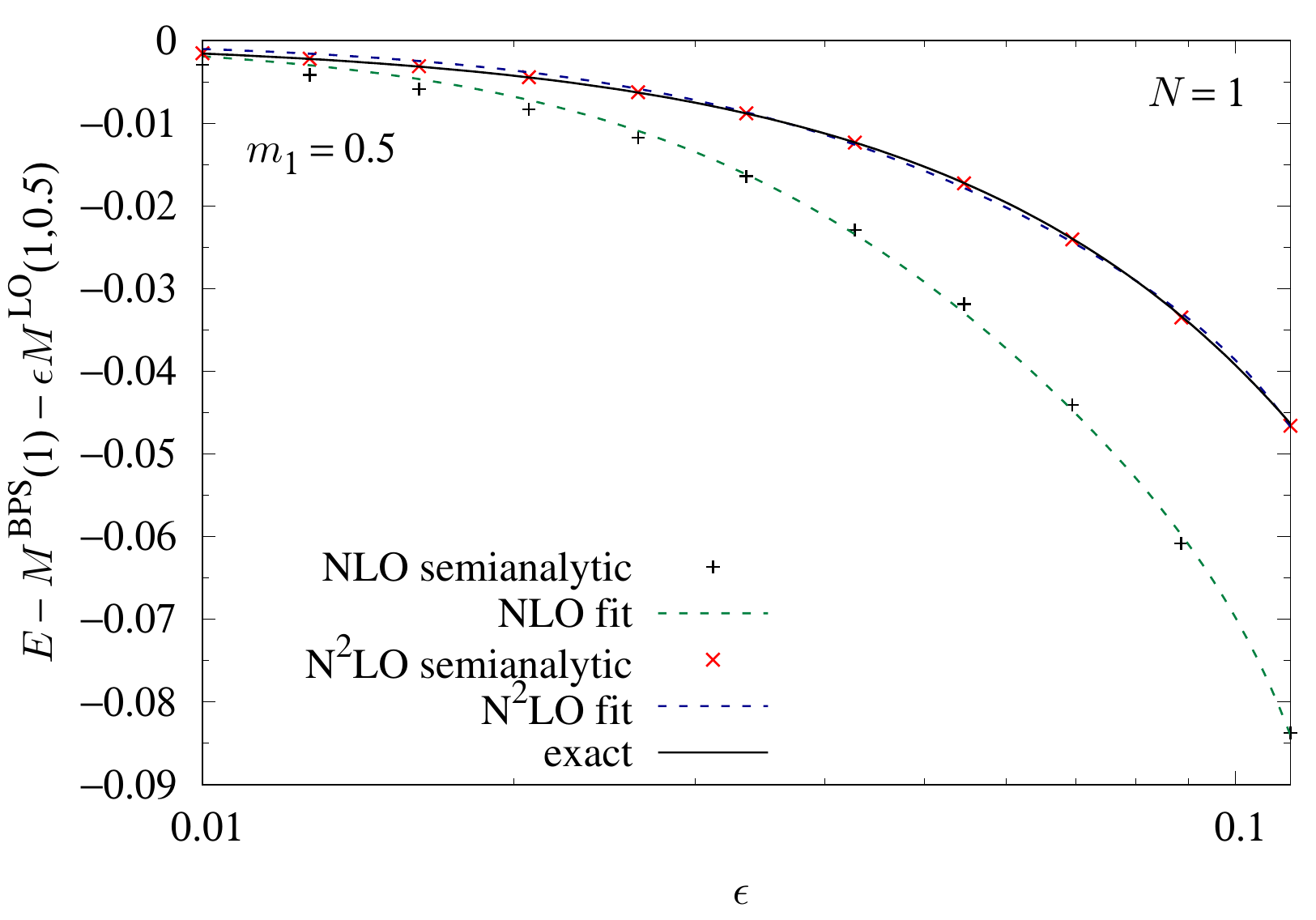}}
      \subfloat{\includegraphics[width=0.33\linewidth]{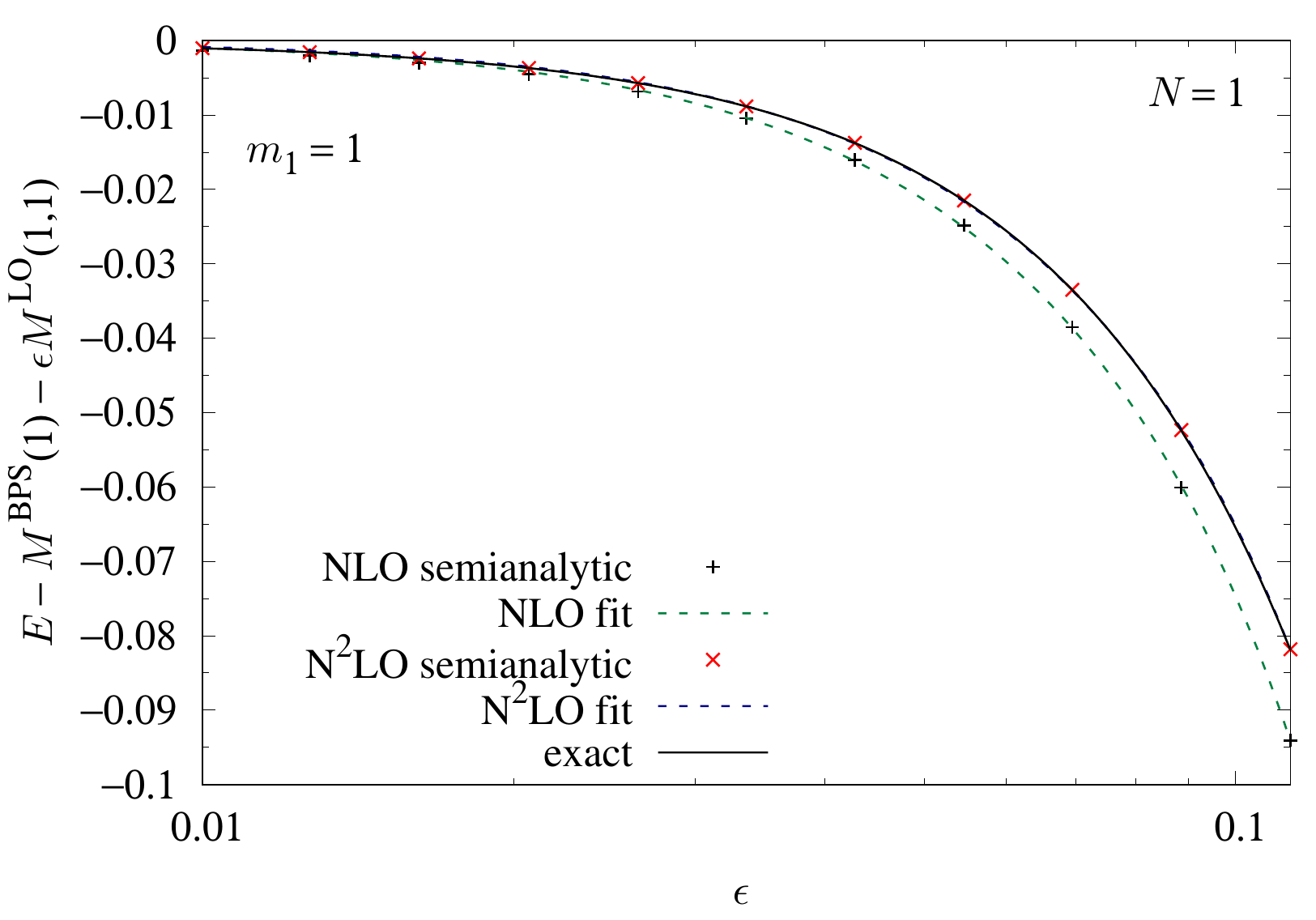}}}
    \mbox{\subfloat{\includegraphics[width=0.33\linewidth]{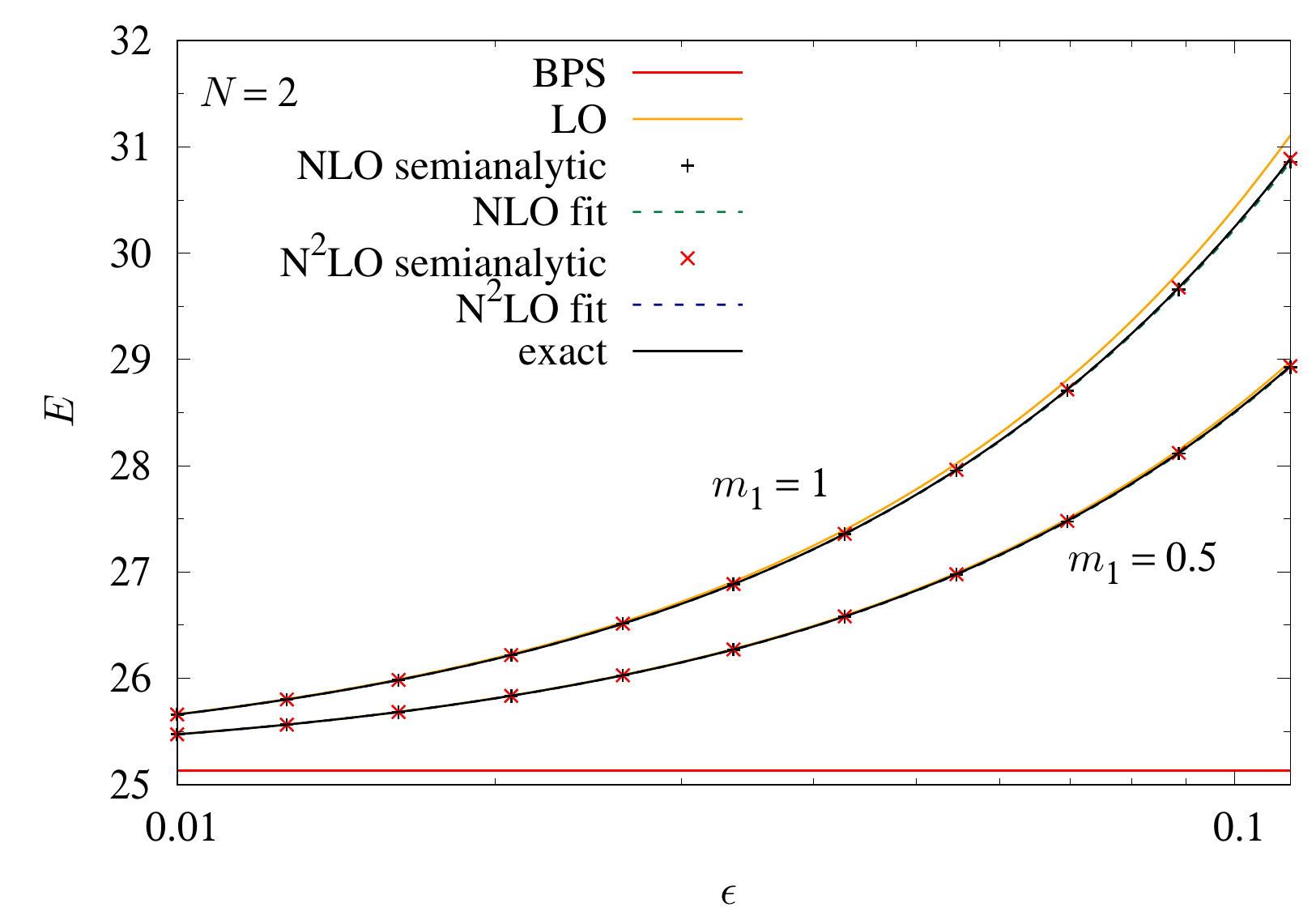}}
      \subfloat{\includegraphics[width=0.33\linewidth]{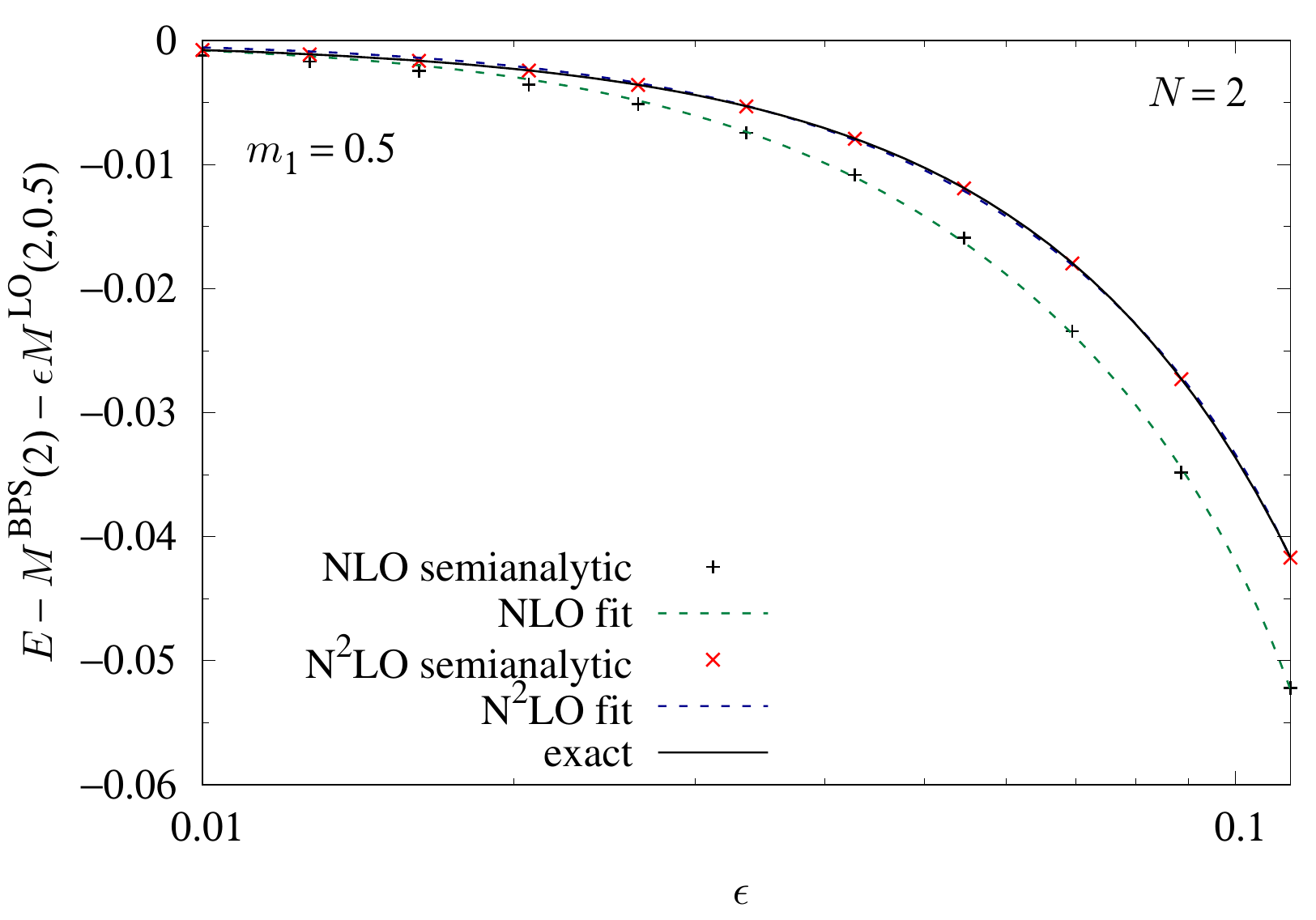}}
      \subfloat{\includegraphics[width=0.33\linewidth]{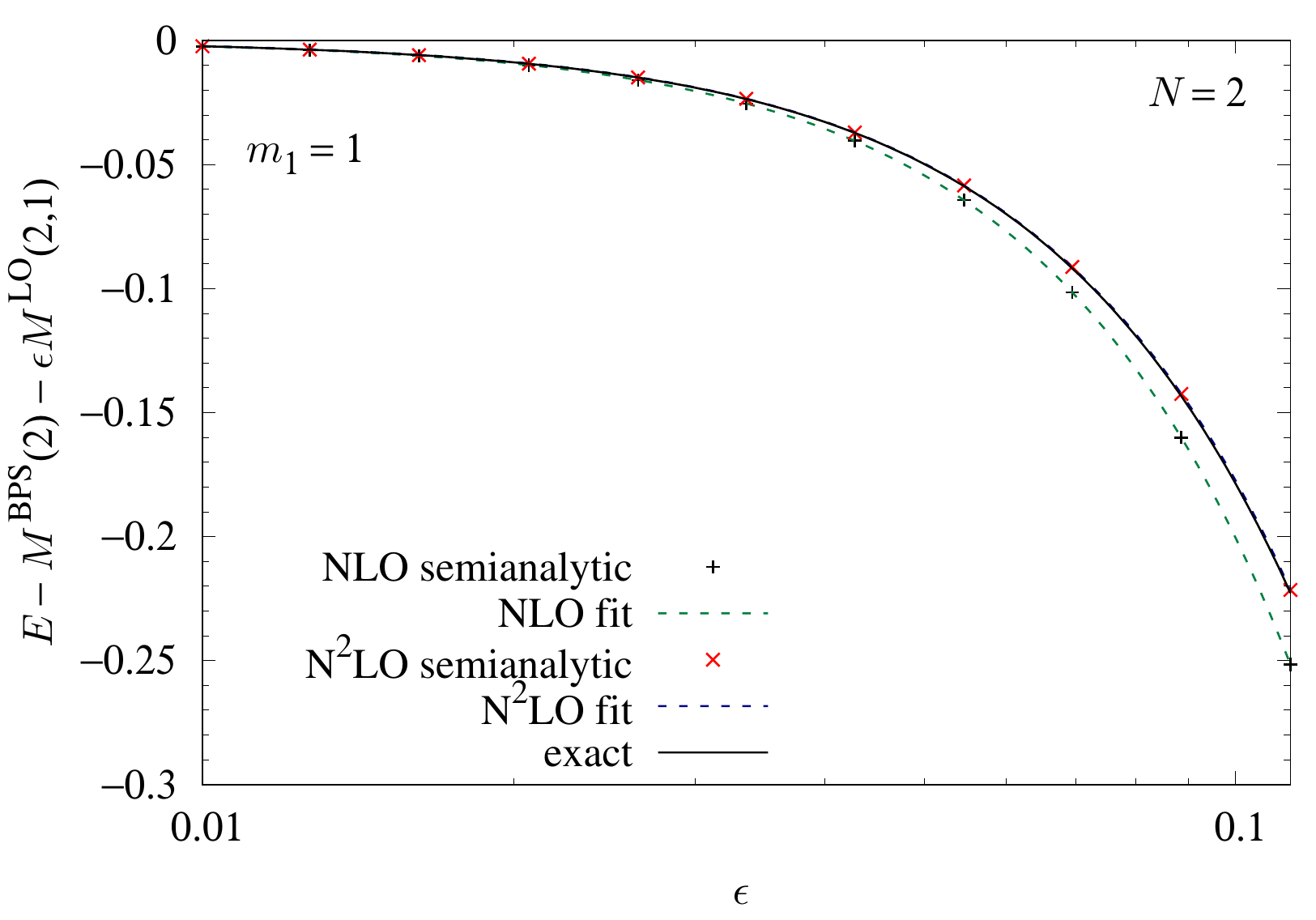}}}
    \mbox{\subfloat{\includegraphics[width=0.33\linewidth]{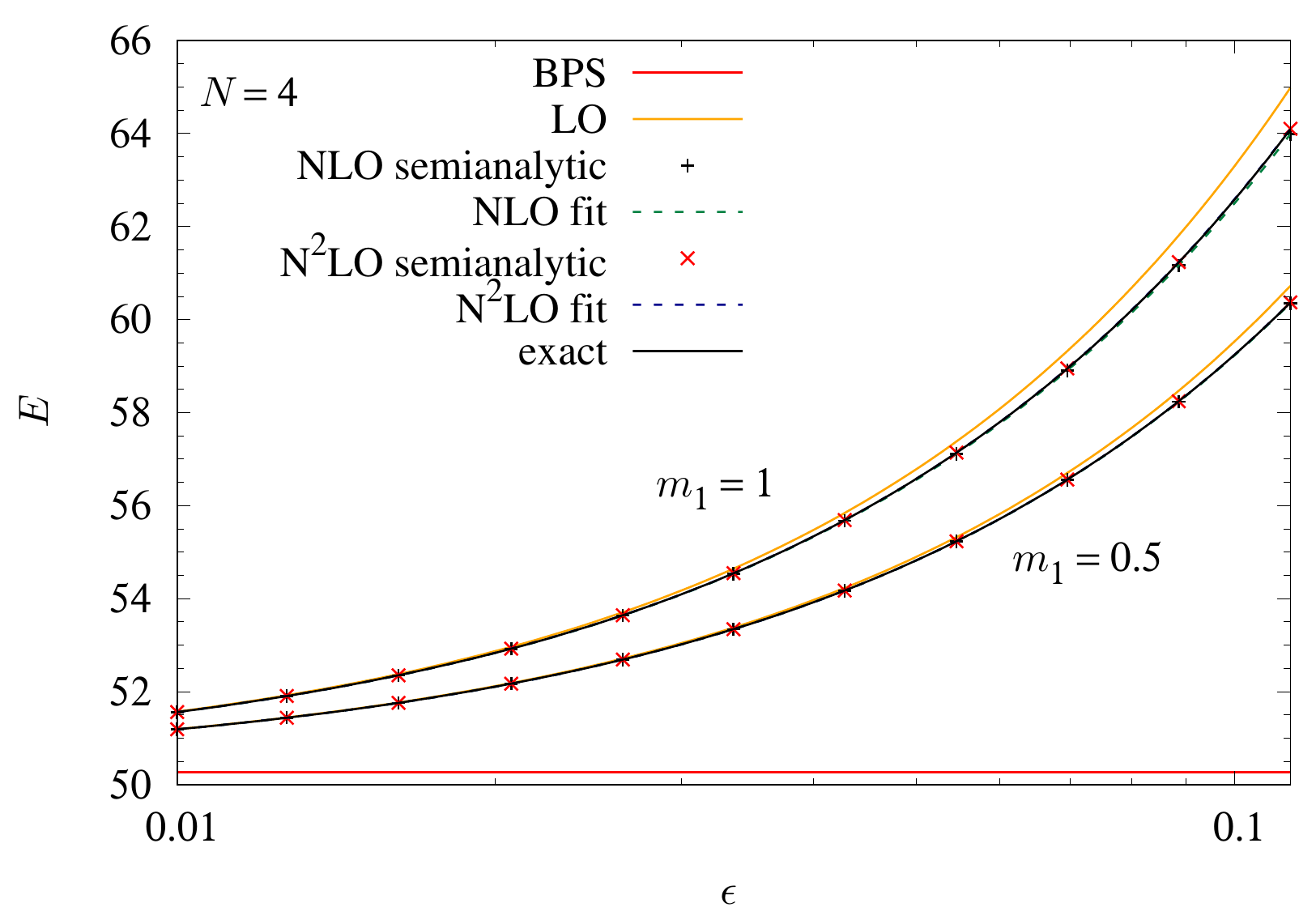}}
      \subfloat{\includegraphics[width=0.33\linewidth]{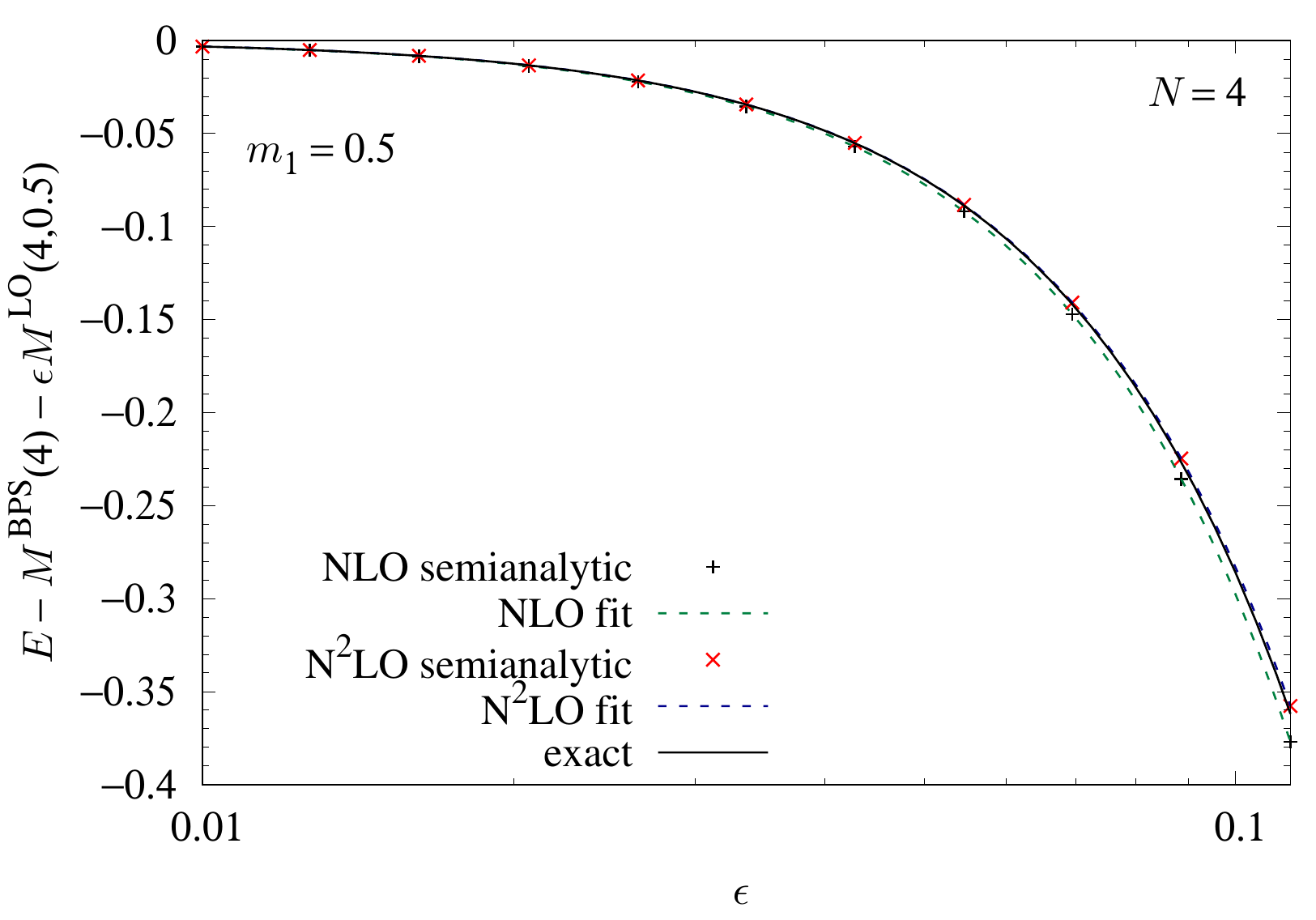}}
      \subfloat{\includegraphics[width=0.33\linewidth]{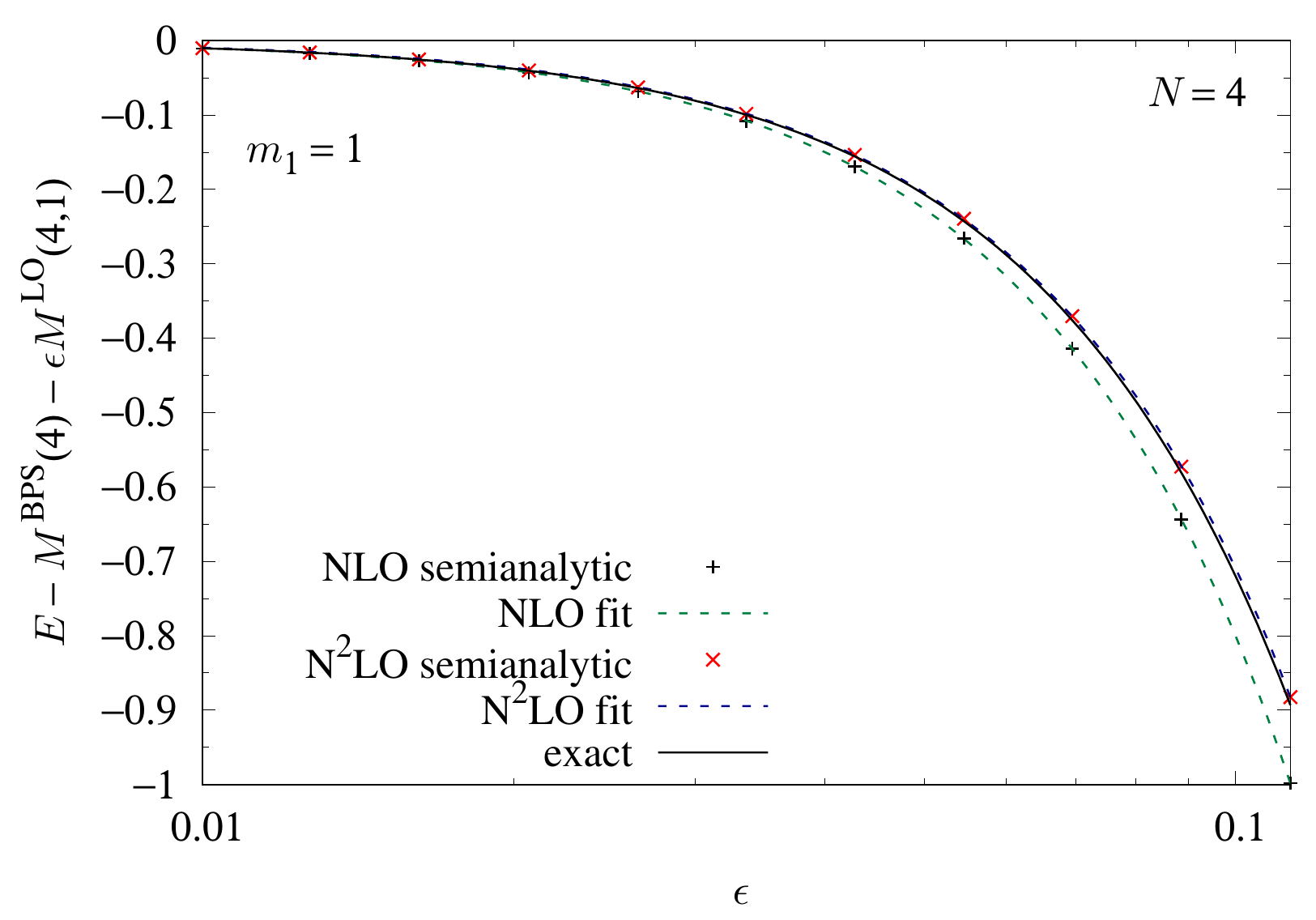}}}
    \caption{The energy (mass) of the baby Skyrmions in the
      perturbative $\epsilon$-expansion scheme, as a function of
      $\epsilon$ (on a log-scale): The red line shows the BPS bound,
      the orange line is the leading-order (LO) correction, the black
      pluses are the next-to-leading order (NLO) corrections in
      $\epsilon$ calculated using the linear perturbation, the
      green dashed line is a fit to the NLO points, the red crosses
      are the next-to-next-to-leading order (N$^2$LO) corrections in
      $\epsilon$ calculated using the same linear perturbation, the
      dark-blue dashed line is a fit to the N$^2$LO points and the
      black line is the exact energy calculated using the full
      (nonlinear) equations of motion. The left columns show the total
      energy, $E$, the middle and right columns show the energy with
      the BPS and the LO correction subtracted off for $m_1=0.5$ and
      $m_1=1$, respectively, to better see the differences between the 
      exact, NLO and N$^2$LO corrections.
    } 
    \label{fig:perturbative_energy_axial}
  \end{center}
\end{figure}

The perturbative corrections to the energy of the $N=1,2,4$
axially symmetric baby Skyrmions for $m_1=0.5,1$ are shown in 
fig.~\ref{fig:perturbative_energy_axial} with the left-hand side
panels showing the exact energies (solid black lines), the BPS
energies (solid red lines), the NLO corrections (black pluses), the
N$^2$LO corrections (red crosses).
Both the NLO (i.e.~$\mathcal{O}(\epsilon^2)$) corrections
\eqref{eq:E2axial} and the N$^2$LO (i.e.~$\mathcal{O}(\epsilon^3)$)
corrections \eqref{eq:E3axialquad}+\eqref{eq:E3axialcubic} are
calculated using the perturbation, which is a solution to the
linearized equation of motion \eqref{eq:axially_sym_EOM_df}, which
contains all terms of the second-order-in-$\epsilon$ Lagrangian
\eqref{eq:L2perturb} and the quadratic terms of the
third-order-in-$\epsilon$ Lagrangian \eqref{eq:L3perturb_quad}. 
The green dashed line shows a fit to the NLO corrections and the
dark-blue dashed line for the N$^2$LO corrections.
The middle and right panels of the figure show the same information,
for $m_1=0.5$ and $m_1=1$, respectively, but with the BPS and LO
corrections subtracted off, so as to better see the differences
between the NLO corrections, the N$^2$LO corrections and the exact
energies. 

First of all, we can see from the total energies, shown in the
left-hand side panels of fig.~\ref{fig:perturbative_energy_axial},
that the precision of the perturbative scheme is extremely good.
More precisely, the LO correction overshoots the exact result
slightly, but the NLO correction is negative and undershoots the exact
result. 
The N$^2$LO correction is then positive and comes extremely close to the 
exact energies.
In fact, by a close inspection of the figures, it can be seen that the
red crosses, corresponding to the perturbative N$^2$LO corrections,
fit the exact energies (black solid lines) better than the N$^2$LO
fits (dark-blue dashed lines). 
Although negligibly small, we expect the remaining discrepancy between
the N$^2$LO corrected energies and the exact energies to be due to the
approximation of linearizing the equation of motion for the
perturbation and for truncating the $\epsilon$ expansion at the third
order. 

The fits shown in fig.~\ref{fig:perturbative_energy_axial} contain no
linear term in $\epsilon$ for the NLO corrections and no linear and 
quadratic terms in $\epsilon$ for the N$^2$LO corrections.
Nevertheless they describe the perturbative calculations very well and
this is thus the \emph{a posteriori} confirmation that $\bdphi$ is of
order $\epsilon$ (and possibly containing higher-order corrections
too). 
It should be fairly convincing that the LO corrections, which are
linear in $\epsilon$ are orders of magnitude larger than the NLO and
N$^2$LO corrections, that thus do not contain a linear term. 

Because we now have a dependence on $m_1$ in this model, compared with
the compacton case of ref.~\cite{Gudnason:2020tps}, we have calculated
the perturbative corrections for two different values of $m_1$, see
fig.~\ref{fig:perturbative_energy_axial}.
Since the physical case has a nonvanishing pion mass, we have chosen
two nonvanishing values, namely $m_1=0.5$ and $m_1=1$.

We will now use the fits to the NLO and N$^2$LO results in
fig.~\ref{fig:perturbative_energy_axial} to write an approximate
expression for the N$^2$LO energy:
\begin{align}
  E(\epsilon,N,m_1) &=
  M^{\rm BPS}(N)
  + \epsilon M^{\rm LO}(N,m_1)
  + \epsilon^2 M^{\rm NLO}(N,m_1)
  + \epsilon^3 M^{{\rm N}^2{\rm LO}}(N,m_1)\non
  &\phantom{=\ }
  + \epsilon^4 M^{{\rm residual},4}(N,m_1)\non
  &= 4\pi N
  + \epsilon\left(\frac{\pi^3}{3} + 2\pi N^2\log 2\right)
  + \epsilon m_1^2(4\pi N)\non
  &\phantom{=\ }
  + \epsilon^2\left(-34.87 + 28.16 N - 5.434 N^2\right)
  + \epsilon^2 m_1^2\left(23.75 - 17.87 N - 2.660 N^2\right) \non
  &\phantom{=\ }
  + \epsilon^3\left(318.1 - 182.9 N + 20.74 N^2\right)
  + \epsilon^3 m_1^2\left(-218.9 + 75.03 N + 26.12 N^2\right)\non
  &\phantom{=\ }
  + \epsilon^4\left(-1253 + 681.1 N - 72.10 N^2\right)
  + \epsilon^4 m_1^2\left(820.1 - 218.5 N - 99.01 N^2\right),
  \label{eq:E_N2LO_compacton_fit}
\end{align}
where we have assumed that the dependence on $m_1$ is quadratic.
This fit should work reasonably well, at least for $m_1\in[0.5,1]$.
The reason for including a fourth-order term in $\epsilon$ in the fit
is to allow for some higher-order (residual) behavior of the
perturbation. 
The higher-order result for the energies in turn gives a correction
to $N_\star$ of eq.~\eqref{eq:Nstar}, i.e.:
\begin{align}
  N_\star &= 1.541
  +\epsilon(-1.638 + 2.240 m_1^2)
  +\epsilon^2(17.12 - 16.75 m_1^2 - 0.2606 m_1^4)\non
  &\phantom{=\ }
  +\epsilon^3(-32.18 + 22.19 m_1^2 + 0.6709 m_1^4 + 1.215 m_1^6)
  + \mathcal{O}(\epsilon^4).
\end{align}

\begin{figure}[!ht]
  \begin{center}
    \mbox{\subfloat[]{\includegraphics[width=0.5\linewidth]{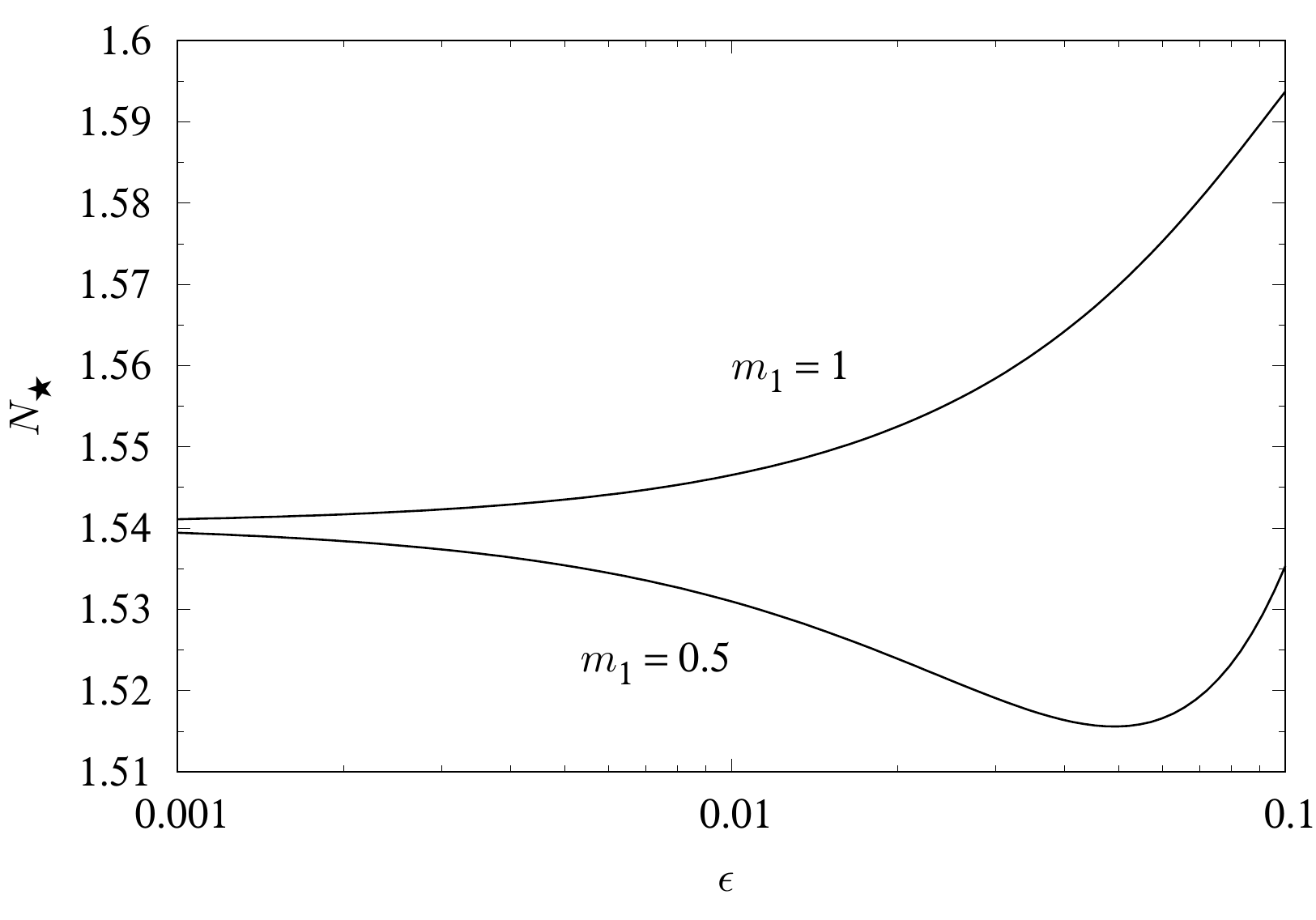}}
      \subfloat[]{\includegraphics[width=0.5\linewidth]{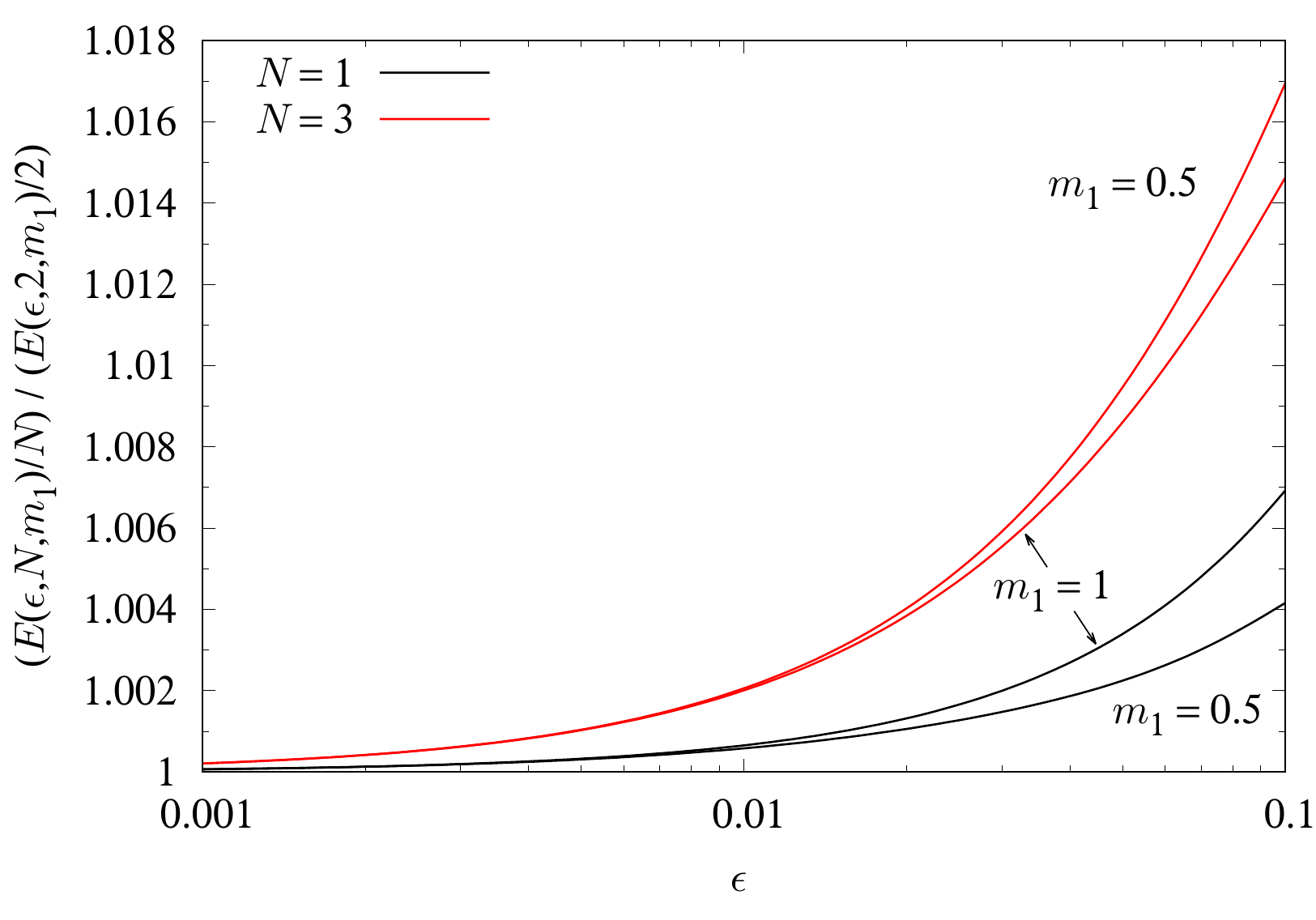}}}
    \caption{(a) The perturbatively corrected critical value,
      $N_\star$, of the topological charge $N$ of the
      axially symmetric baby Skyrmion, as a function of $\epsilon$. This 
      critical value formally corresponds to the smallest energy per
      $N$.
      (b) The ratio of the energy of the $N$-Skyrmion per $N$ and the
      energy of the 2-Skyrmion divided by 2, for $N=1,3$ and
      $m_1=0.5,1$, as functions of $\epsilon$.
    }
    \label{fig:Nstar_ENratio}
  \end{center}
\end{figure}

The perturbatively corrected $N_\star$ is shown in
fig.~\ref{fig:Nstar_ENratio}(a) as a function of $\epsilon$ for
$m_1=0.5,1$. 
It is seen from the figure that for $\epsilon$ sufficiently small
(i.e.~$\epsilon<0.1$) the perturbative corrections to $N_\star$ are
too small to decisively make another topological charge sector
the stable one.
In order to check this explicitly, we plot in
fig.~\ref{fig:Nstar_ENratio}(b) the ratio of the energy per
topological charge for $N=1,3$ to the same quantity with $N=2$; that
is $\frac{E(\epsilon,N,m_1)}{N}\frac{2}{E(\epsilon,2,m_1)}$, for
$m_1=0.5,1$ as functions of $\epsilon$.
We can see that there is only a mild dependency on $m_1$ and all
values in the plot are greater than unity:
Hence, the $N=2$ axially symmetric baby Skyrmion remains the stable
solution.
Increasing $\epsilon$ in the range shown in the figure actually makes
the 2-Skyrmion more stable, as can be seen from 
fig.~\ref{fig:Nstar_ENratio}(b).

\section{Numerical calculations}\label{sec:numcalc}

We will now present numerical solutions to the full nonlinear
equations of motion. It is computationally a very expensive task and
because the static baby Skyrmions require only 2-dimensional PDEs, we
are able to use refined enough square grids for the computations.
The numerical method used is the so-called arrested Newton flow on
grids with about $4096^2$ lattice points and lattice spacing down to
about $\sim0.0038$.
The numerical derivatives are approximated using a finite-difference
method on a fourth-order 5-point stencil.
The numerical accuracy is about $10^{-6}$ or better.
The code is an adaptation of the CUDA C code used in
ref.~\cite{Gudnason:2020tps}, which is executed on a GPU cluster.

In this section, we explore numerical solutions in the topological
sector $Q=4$ and $Q=2$ and include the axially symmetric solutions,
which can be directly compared to the very precise (exact) numerical
ODE calculations.
This gives another estimate on our numerical precision.
It also confirms in which part of the parameter space, the
axially symmetric solutions exist and which type of solution has the
lowest energy in the given topological sector.

Our first and main aim of this numerical exercise, is to confirm that
the lowest energy baby Skyrmion solution -- for small but finite
$0<\epsilon\ll 1$ -- is made of two $N=2$ baby Skyrmions sitting at
some unknown separation distance from one another.
This is supported by the LO energy correction, which in the
axially symmetric case is confirmed to be the dominant correction to
the energy in the small-$\epsilon$ limit. 

\begin{figure}[!t]
  \begin{center}
    \includegraphics[width=\linewidth]{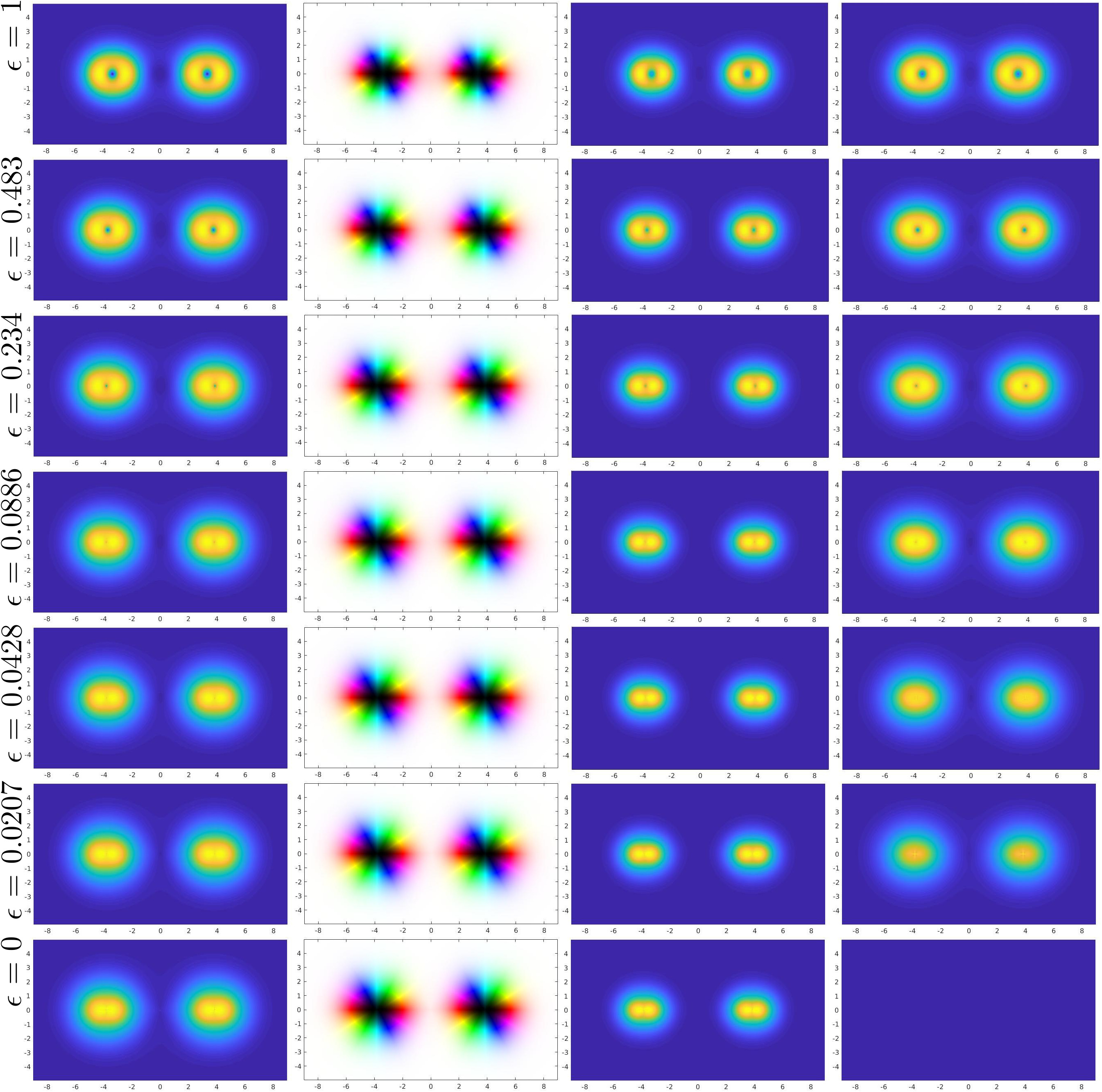}
  \caption{The $Q=2+2$ baby Skyrmion for various values of $\epsilon$;
    this is the stable $Q=4$ baby Skyrmion solution.
    The columns show the topological charge density, the pion vector
    orientation using a color scheme described in the text, the total
    energy density and the perturbation energy density
    $\epsilon(-\Lag_2+m_1^2V_1)$.
    In this figure $m_1=0.5$. 
  }
  \label{fig:N=2+2}
  \end{center}
\end{figure}

\begin{figure}[!t]
  \begin{center}
    \includegraphics[width=\linewidth]{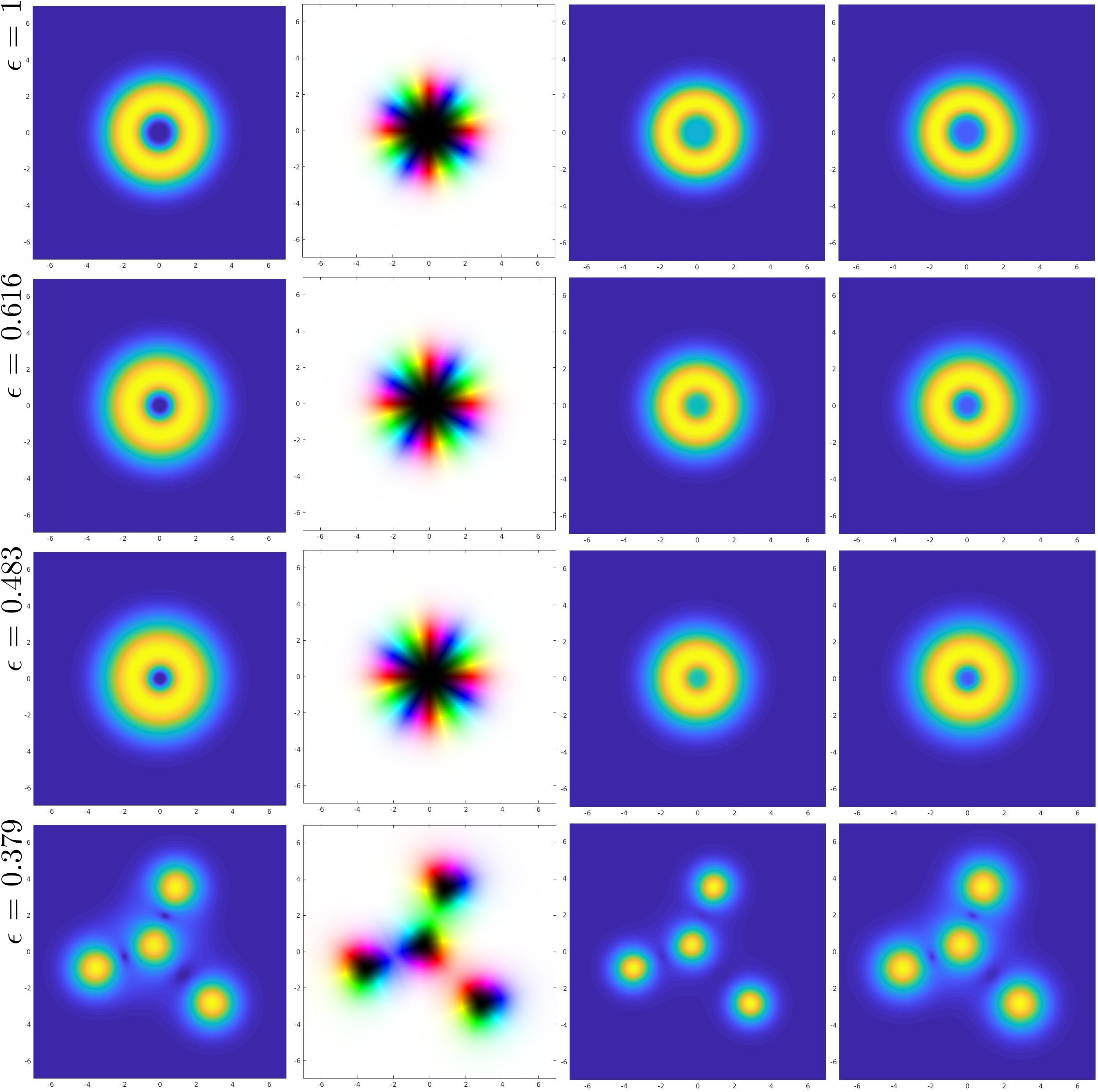}
    \caption{The $Q=N=4$ baby Skyrmion for various values of $\epsilon$.
    The columns show the topological charge density, the pion vector
    orientation using a color scheme described in the text, the total
    energy density and the perturbation energy density
    $\epsilon(-\Lag_2+m_1^2V_1)$. 
    The last row (i.e.~for $\epsilon=0.379$) shows that the solution
    has decayed into a lower-energy configuration with
    almost-triangular symmetry instead of axial symmetry.
    In this figure $m_1=0.5$. 
  }
  \label{fig:N=4}
  \end{center}
\end{figure}

\begin{figure}[!p]
  \begin{center}
    \includegraphics[width=\linewidth]{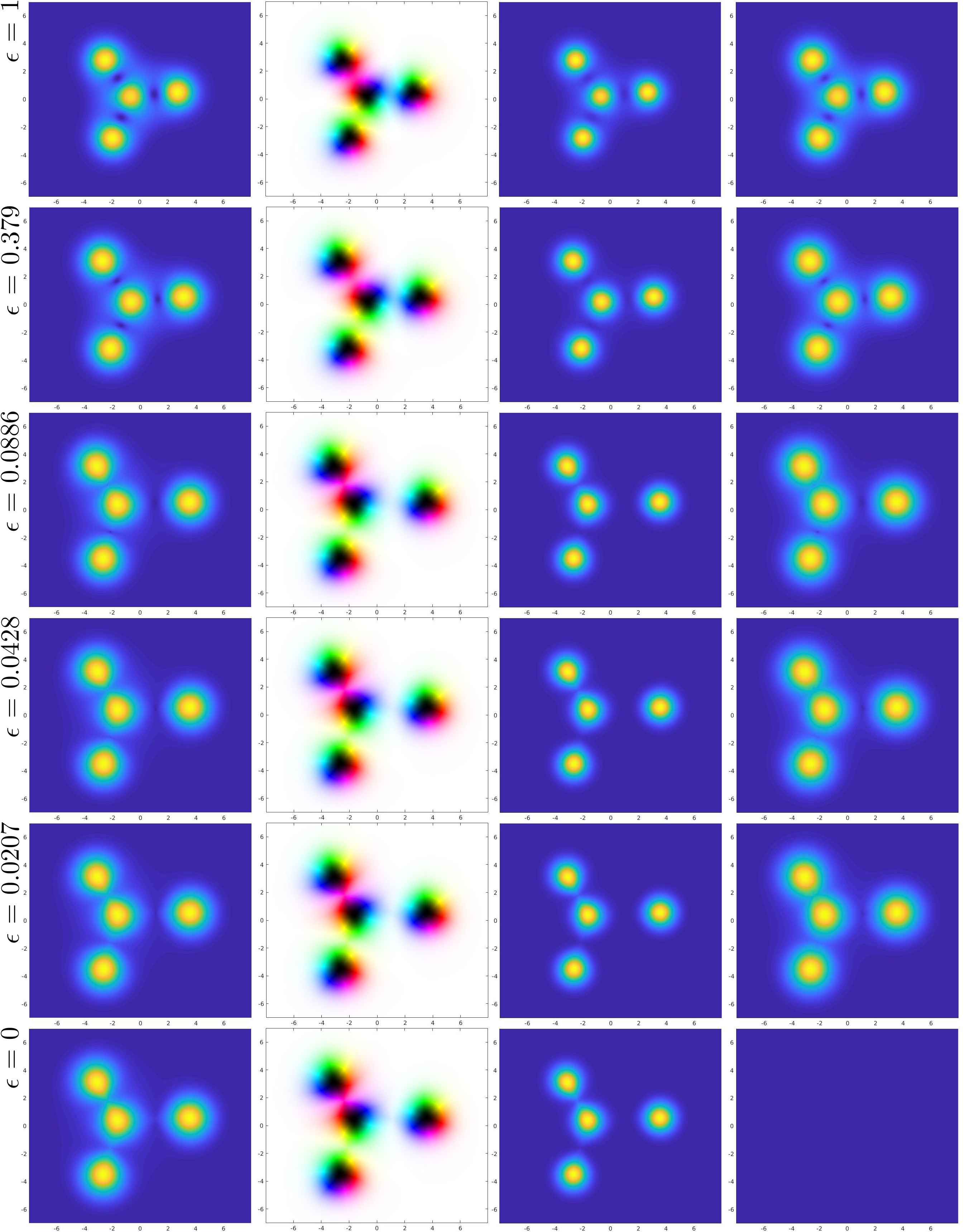}
    \caption{The $Q=1+1+1+1$ baby Skyrmion with (almost) triangular
      symmetry, for various values of $\epsilon$. 
      The columns show the topological charge density, the pion vector
      orientation using a color scheme described in the text, the total
      energy density and the perturbation energy density
      $\epsilon(-\Lag_2+m_1^2V_1)$. 
      In this figure $m_1=0.5$. 
    }
    \label{fig:N=1+1+1+1_triangle}
  \end{center}
\end{figure}

\begin{figure}[!t]
  \begin{center}
    \includegraphics[width=\linewidth]{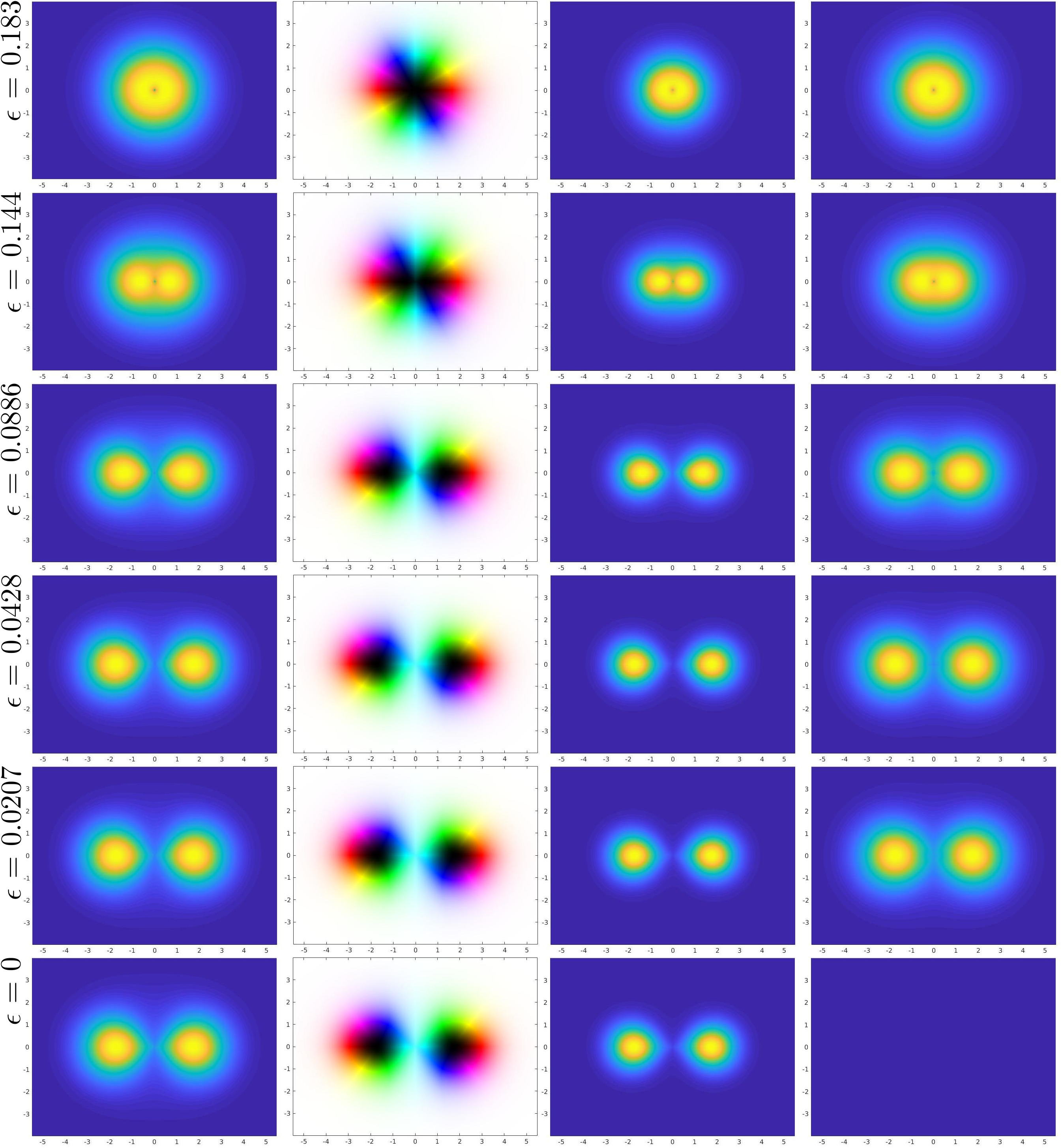}
    \caption{The $Q=1+1$ baby Skyrmion for various values of
      $\epsilon$. 
      The columns show the topological charge density, the pion vector
      orientation using a color scheme described in the text, the total
      energy density and the perturbation energy density
      $\epsilon(-\Lag_2+m_1^2V_1)$.
      The first row (i.e.~for $\epsilon=0.183$) shows that the
      solution has decayed into the axially symmetric $N=2$ solution.
      In this figure $m_1=0.5$. 
    }
  \label{fig:N=1+1}
  \end{center}
\end{figure}

\begin{figure}[!p]
  \begin{center}
    \includegraphics[width=0.98\linewidth]{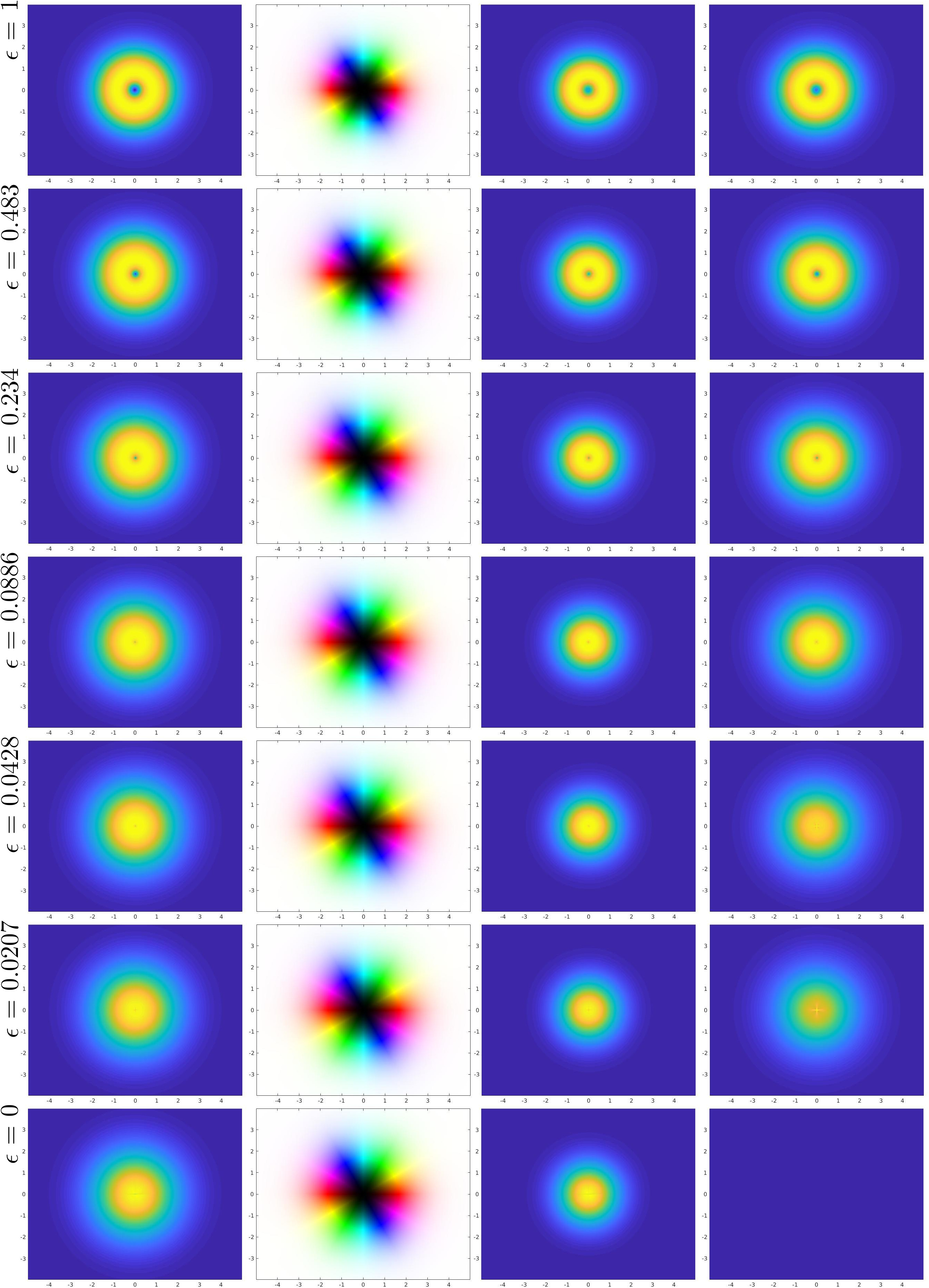}
    \caption{The $Q=N=2$ baby Skyrmion for various values of
      $\epsilon$.
      The columns show the topological charge density, the pion vector
      orientation using a color scheme described in the text, the total
      energy density and the perturbation energy density
      $\epsilon(-\Lag_2+m_1^2V_1)$.
      In this figure $m_1=0.5$.
    }
  \label{fig:N=2}
  \end{center}
\end{figure}

\begin{figure}[!ht]
  \begin{center}
    \mbox{\includegraphics[width=0.49\linewidth]{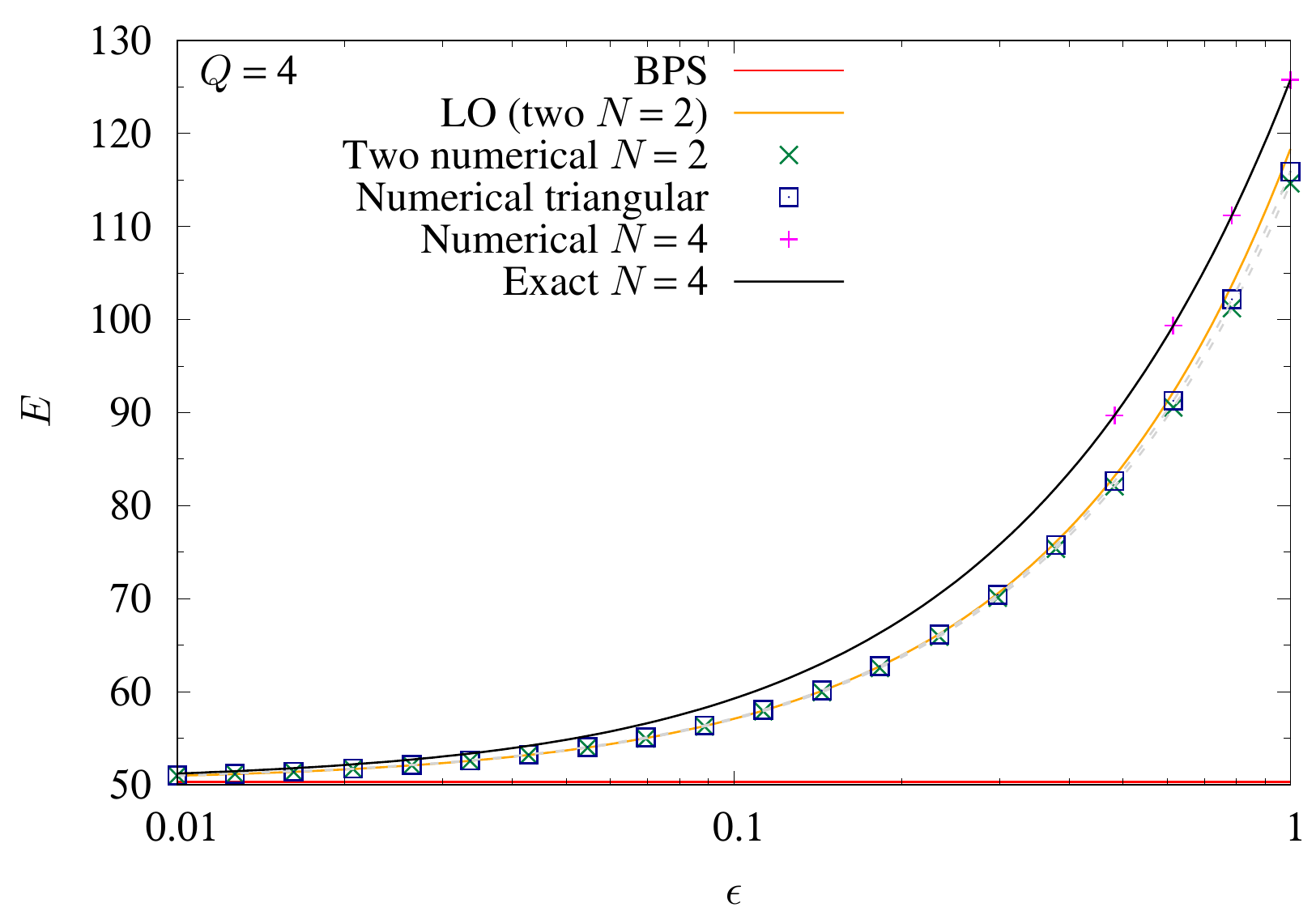}
      \includegraphics[width=0.49\linewidth]{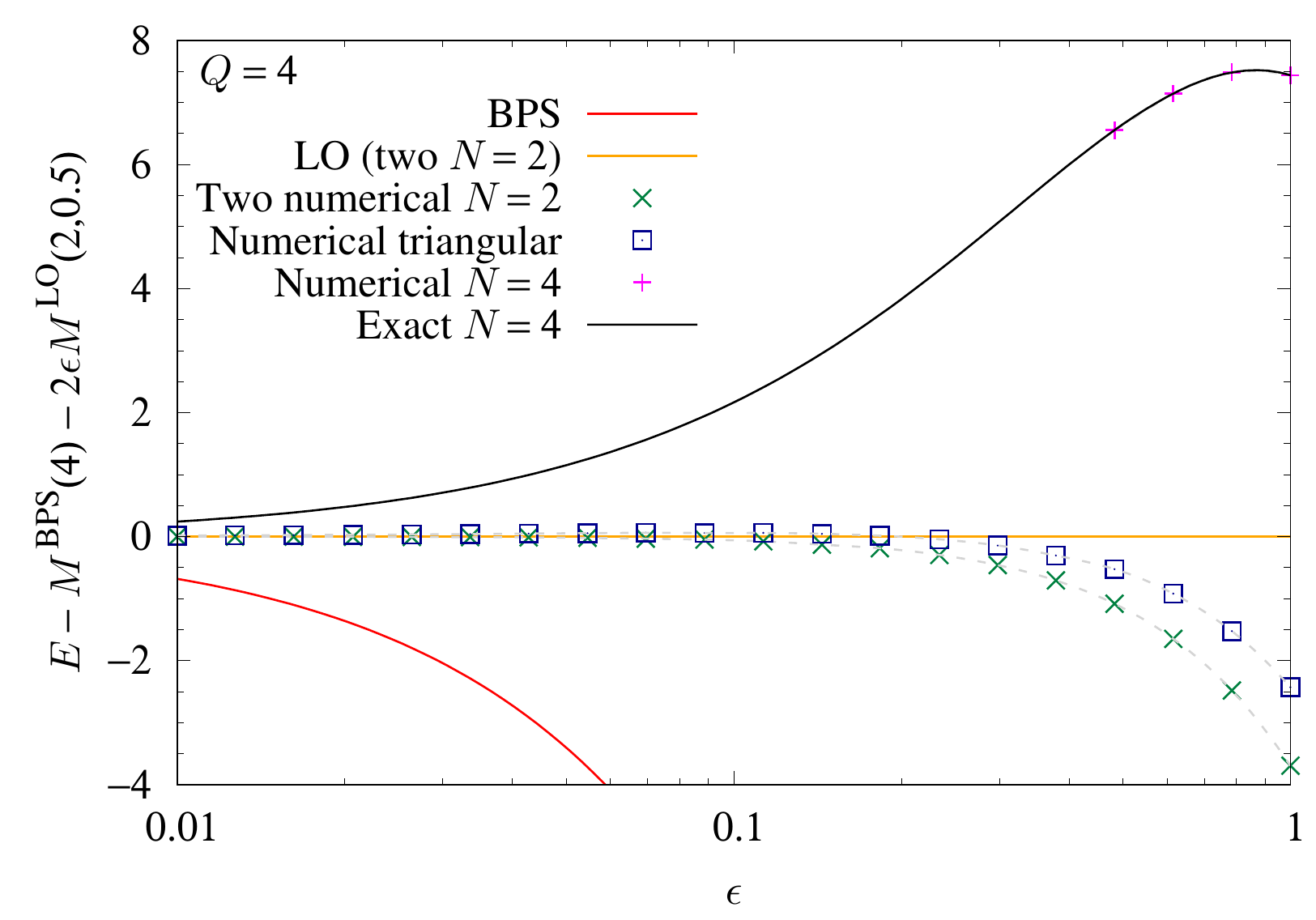}}
    \caption{Left panel: Energies of the baby Skyrmion solutions in
      the $Q=4$ sector. The red line shows the BPS bound, the orange
      line is the LO correction for two $N=2$ baby Skyrmions.
      The numerical solutions are: the two $N=2$ Skyrmions
      side-by-side (green crosses), the triangularly symmetric baby
      Skyrmion (blue dotted squares), and the axially symmetric $N=4$
      Skyrmion (magenta pluses). The numerical $N=4$ solution is
      compared to the exact solution (using ODE computations) shown
      with a black solid line. 
      The right-hand side panel shows the same energies, but with the
      BPS and the LO correction subtracted off.
      Around $\epsilon\lesssim0.4$ the axially symmetric $N=4$
      soliton becomes unstable and decays to the triangularly
      symmetric one -- which is just metastable.
    } 
    \label{fig:NumEnQ4}
  \end{center}
\end{figure}

\begin{figure}[!ht]
  \begin{center}
    \mbox{\includegraphics[width=0.49\linewidth]{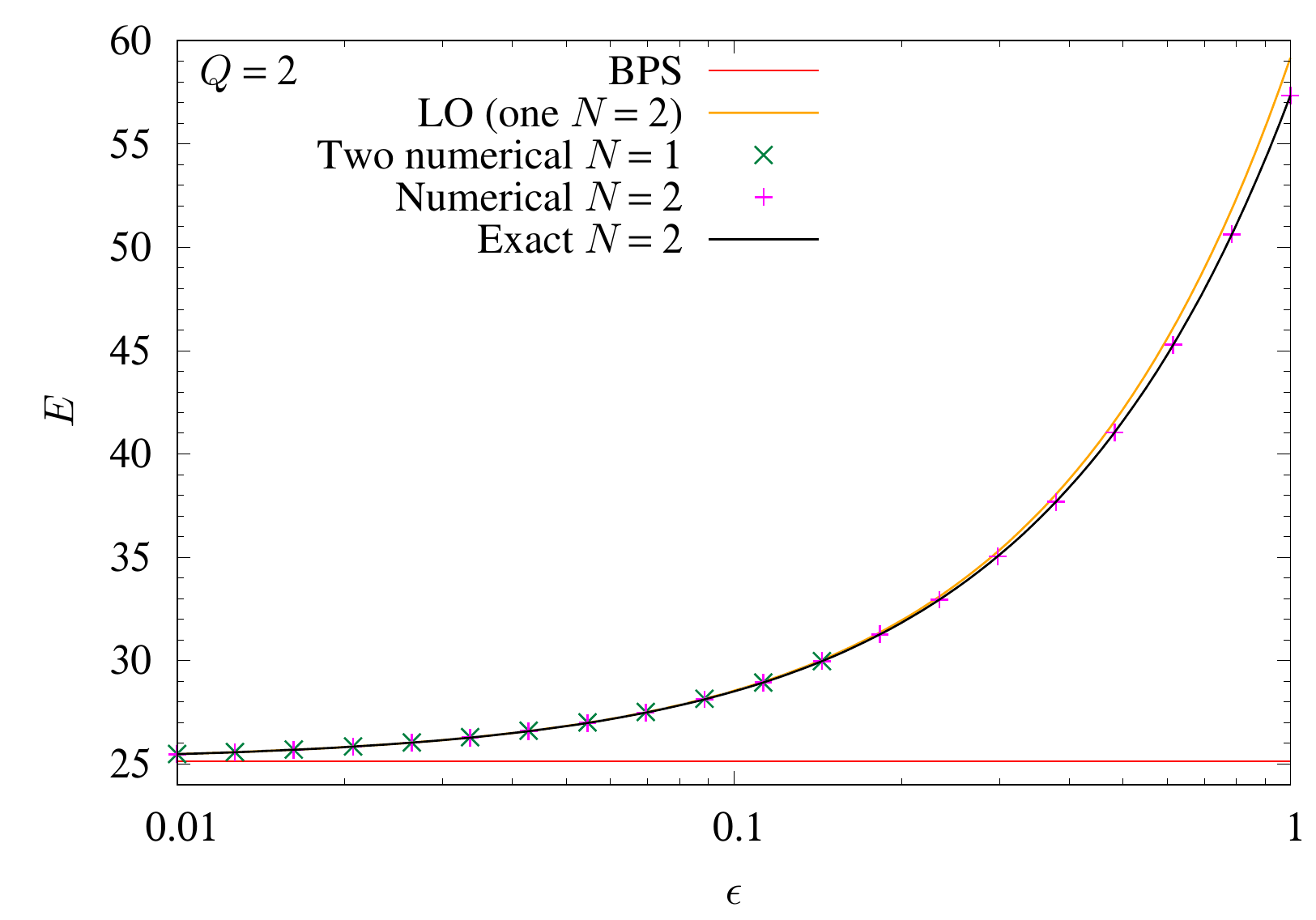}
      \includegraphics[width=0.49\linewidth]{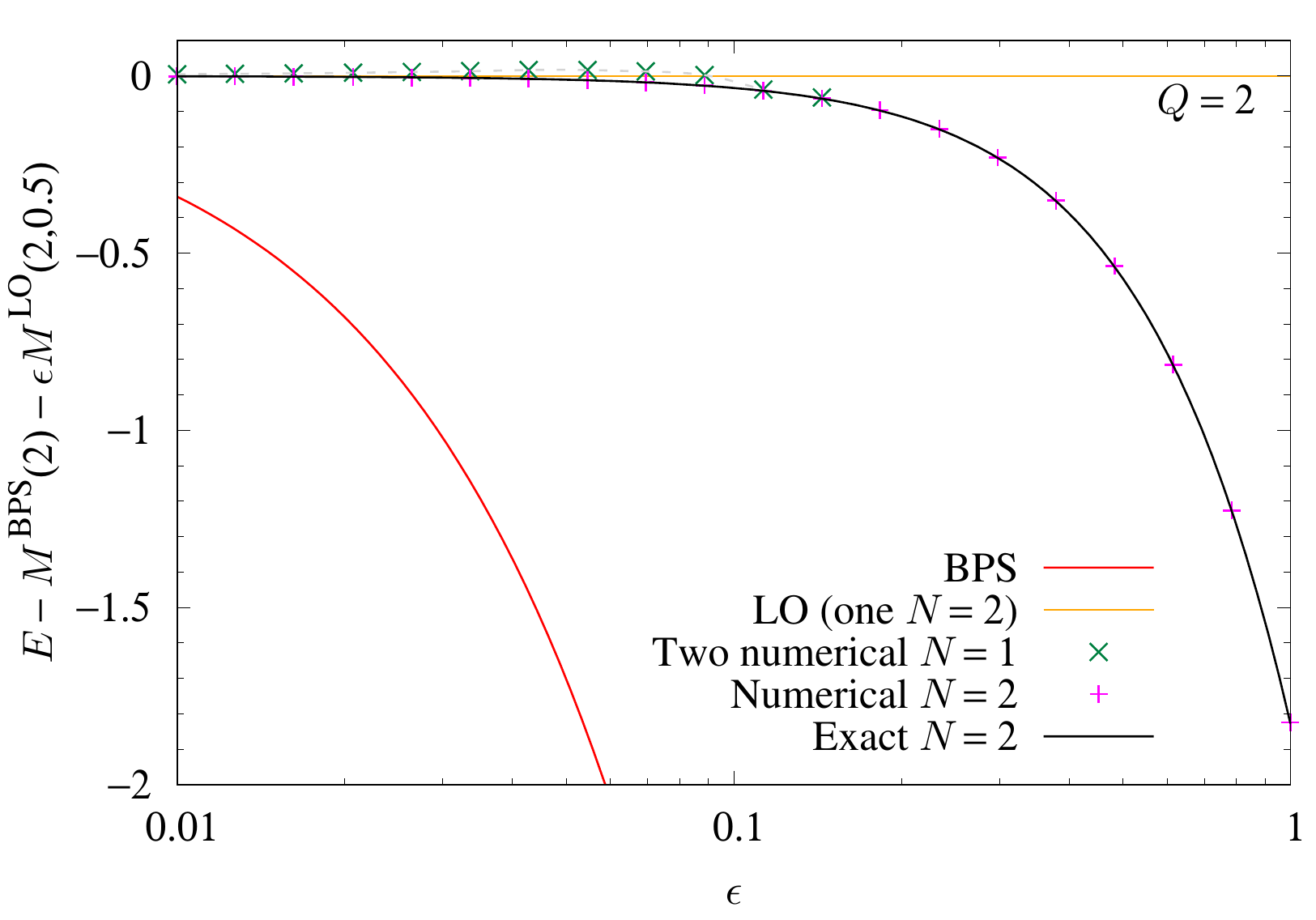}}
    \caption{Left panel: Energies of the baby Skyrmion solutions in
      the $Q=2$ sector. The red line shows the BPS bound, the orange
      line is the LO correction for one $N=2$ baby Skyrmion.
      The numerical solutions are: the two $N=1$ Skyrmions
      side-by-side (green crosses), and the axially symmetric $N=2$
      Skyrmion (magenta pluses). The numerical $N=2$ solution is
      compared to the exact solution (using ODE computations) shown
      with a black solid line.
      The right-hand side panel shows the same energies, but with the
      BPS and the LO correction subtracted off.
      Around $\epsilon\gtrsim0.15$ the two $N=1$ baby Skyrmions
      side-by-side become unstable and decay to the axially symmetric
      $N=2$ solution -- which is the stable solution.
    }
    \label{fig:NumEnQ2}
  \end{center}
\end{figure}

We begin with two $N=2$ baby Skyrmions placed side-by-side, which thus
is in the topological charge sector $Q=4$.
The numerical results are shown in fig.~\ref{fig:N=2+2} for
representative values of $\epsilon$.
This figure and the following four are organized into four columns
displaying the topological charge density, the baby Skyrmion
orientation in O(3) space, the total energy density and finally the
perturbation part of the energy density $\epsilon(-\Lag_2+m_1^2V_1)$.
The figures in columns 1, 3 and 4 are shown with a color scheme
normalized to the content on each graph.
The figures in column 2 show the pion vector of the baby Skyrmion in
the following sense: $\phi^3=1$ corresponds to the vacuum and is shown
with white, $\phi^3=-1$ is the anti-vacuum and is shown with black,
$\arg(\phi^1+\i\phi^2)=0$ is shown with red,
$\arg(\phi^1+\i\phi^2)=\frac{2\pi}{3}$ is shown with green and finally
$\arg(\phi^1+\i\phi^2)=\frac{4\pi}{3}$ is shown with blue.

The two $N=2$ baby Skyrmions sitting side-by-side exist for all values
of $\epsilon$ (in our scanned range), even $\epsilon=1$ and
$\epsilon=0$. 
Starting with $\epsilon=1$, the solution is made of two ring-like baby
Skyrmions, that however are prolonged along the axis that joins them.
Upon lowering $\epsilon$, the greatest effect appears to be on the
distribution of the energy and topological charge within each
(approximately) axially symmetric $N=2$ baby Skyrmion.
The separation distance between them is affected only mildly.
More precisely, the separation distance between the two $N=2$
Skyrmions is seen to increase slightly for $\epsilon$ decreasing from 
$\epsilon=1$ to $\epsilon\sim0.379$ and below that, the separation
distance appears to be constant in the limit of $\epsilon\to0$.
This situation is quite different from the compacton case of
ref.~\cite{Gudnason:2020tps}, where the limit of $\epsilon\to0$ yields
two perfectly axially symmetric compactons touching each other at a
mathematical point with the pion vectors anti-aligned, so the point
they touch has the same pion orientation on both sides of said point.
In this case, it appears that the limit of $\epsilon\to0$ yields a
nontrivial GRH solution, where neither of the two
$N=2$ Skyrmions are axially symmetric.
This is obviously because of their Gaussian tails in the BPS limit
(i.e.~in the limit of $\epsilon\to0$) that makes it impossible to
place two perfectly axially symmetric BPS solutions next to each other
at any finite distance.
Although we are able to lower $\epsilon$ all the way down to
$\epsilon=0$ with extreme resolution and in the same time taking into
account the tails up to radii $r\sim15$, the two $N=2$ baby Skyrmions
do not tend to perfectly axially symmetric solutions, see that
last row of fig.~\ref{fig:N=2+2} (i.e.~for $\epsilon=0$).
The BPS solution shown in the last row of the figure is evidently the
BPS solution closest to the limiting sequence of solutions for
$0<\epsilon\ll1$, but we should point out that the moduli space for
$\epsilon=0$ is infinite dimensional and hence this is just one
solution out of infinitely many BPS solutions that can take any shape
as long as the volume is preserved.\footnote{For a definition of the
  volume of the soliton, see refs.~\cite{Adam:2015zhc,Adam:2017ouo}. } 
A slight polarization of the constituent $N=2$ Skyrmions -- stretching
the baby Skyrmions along the axis that joins them -- persists in the
limit of $\epsilon\to0$, see fig.~\ref{fig:N=2+2}.
This configuration in the topological charge sector $Q=4$, consisting
of two $N=2$ Skyrmions is the stable solution (i.e.~with the smallest
energy per $Q$) for all values of $\epsilon$ studied in this paper. 

Apart from the global minimum of the energy functional, which for
$Q=4$ sector we claim is attained by two $N=2$ baby Skyrmions with a
small separation distance, there exists also local minima --
viz.~metastable states.
For this reason, we have tried with several good guesses as initial
conditions and observed what they flow to under the numerical
minimization of the energy by means of the arrested Newton flow
algorithm.

As $N_\star\sim 1.5$ and the energy contains a term that grows
quadratically with $N$, it is expected that the $N=4$
axially symmetric baby Skyrmion is just a metastable state, if it
exists at all. 
An explicit check shows that the $N=4$ solution exists for
$\epsilon=1$, but it decays to a(n almost) triangularly symmetric
arrangement of four $N=1$ baby Skyrmions for $\epsilon\approx0.4$, see
fig.~\ref{fig:N=4}. 
This is analogous to what happens in the compacton case
\cite{Gudnason:2020tps}.
This solution is also a metastable state and in particular for large
$\epsilon\sim1$, two $N=1$ baby Skyrmions are unstable and quickly
combine into an axially symmetric $N=2$ solution, see below.
However, the almost triangularly symmetric configuration enjoys
metastability due to the fact that the center $N=1$ Skyrmion is
confused by 3 attractive channels and cannot decide which other $N=1$
to combine into an $N=2$ with.

We increase $\epsilon$ for the (almost) triangularly symmetric
configuration of four $N=1$ Skyrmions from $\epsilon=0.379$ all the
way up to $\epsilon=1$, see fig.~\ref{fig:N=1+1+1+1_triangle}.
The arrangement, although metastable, turns out to remain, even for
$\epsilon=1$ and in fact it becomes perfectly triangularly symmetric
for the larger values of $\epsilon$ due to larger attraction between
the four $N=1$ constituents.
We also decrease $\epsilon$ all the way down to $\epsilon=0$ and find that
it becomes less triangularly symmetric for the smaller values of
$\epsilon$ -- although we do not know exactly why.
The attractive force between the $N=1$'s become very weak and we can
see from the second-last row of fig.~\ref{fig:N=1+1+1+1_triangle} that
it would be fairly easy to knock off the $N=1$ baby Skyrmion on the 
right-hand side of the solution.
In that sense, the model will be able to describe weakly bound nuclear
clusters.
Of course, in two dimensions it makes little sense to try and compare
with actual nuclear physics knowledge.
In the last row of the figure, where $\epsilon=0$, the position of the
constituents are just moduli and can be moved freely, of course. 

We could in principle search for further arrangements of four $N=1$
baby Skyrmions, but we already know from the leading order calculation
of the energy in $\epsilon$, that the energetically preferred
constituents are the axially symmetric $N=2$ baby Skyrmions.
We will leave such a search for more clusters to future work as they
will not be relevant in the small-$\epsilon$ limit.

We will now determine the phase diagram in the $Q=4$ topological
charge sector.
To this end, we plot the energies of the baby Skyrmions displayed in
figs.~\ref{fig:N=2+2}, \ref{fig:N=4} and \ref{fig:N=1+1+1+1_triangle}
in fig.~\ref{fig:NumEnQ4} (with exception of the $\epsilon=0$ ones,
due to the logarithmic scale of the ordinate).
Starting with the least stable solution, the axially symmetric $N=4$
baby Skyrmion (magenta pluses) only exists for $\epsilon\gtrsim0.4$
and then decays to the (almost) triangularly symmetric solution (blue
dotted squares), which in turn is only slightly higher in energy
compared with the lowest-energy state -- the two $N=2$ baby Skyrmions
side-by-side.
Indeed it is the globally stable solution in the entire range of
$\epsilon$ considered here.
We also note that the LO energy is actually an extremely good
approximation to that of the solutions made of two $N=2$ Skyrmions or
four $N=1$ Skyrmions for $\epsilon\lesssim0.1$.

We now turn to the $Q=2$ topological charge sector, for which the
combinatoric possibilities obviously are more limited.
Of course, from sections \ref{sec:LO} and \ref{sec:axial} we know that
the $N=1$ baby Skyrmion has a larger mass per $N$ than the $N=2$ one, both of
them with axial symmetry. 
Nevertheless for a full understanding of the phase diagram in the
$Q=2$ sector, we need to study where the two $N=1$ baby Skyrmions
side-by-side are metastable and where they are unstable to collapse
into an axially symmetric $N=2$ baby Skyrmion, see
fig.~\ref{fig:N=1+1}. 
With the intuition from the compacton case of
ref.~\cite{Gudnason:2020tps}, we expect the two $N=1$'s side-by-side
to exist only for small values of $\epsilon$.
The critical value of $\epsilon$ above which the two $N=1$'s
side-by-side are unstable is roughly
$\epsilon_{\rm crit}\approx0.15$. 
In the small-$\epsilon$ limit, the two $N=1$ baby Skyrmions
side-by-side appear to be two nearly axially symmetric baby Skyrmions
vaguely connected in the attractive channel.
Because of the Gaussian tail in the BPS limit, the two $N=1$ baby
Skyrmions cannot recover axial symmetry in the $\epsilon\to0$ limit --
in sharp contradistinction to the compacton case of
ref.~\cite{Gudnason:2020tps}.

We do not need to solve the full PDEs to obtain the energies for the
stable (in the $Q=2$ sector) axially symmetric $N=2$ baby Skyrmions.
Nevertheless, we perform these calculations as a check of our
numerical accuracy.
The solutions are shown in fig.~\ref{fig:N=2}.

Finally we plot the energies of the two types of solution in the
topological charge $Q=2$ sector in fig.~\ref{fig:NumEnQ2}.
In the left-hand side panel of the figure, we can see that both types
of solution are very close in energy and are actually described quite
well by the LO correction to the energy for $\epsilon\lesssim0.1$.
In the right-hand side panel, we show the same energies but with the
BPS and LO corrections subtracted off.
First of all, we can see that the full numerical PDE solutions of the
$N=2$ baby Skyrmions (magenta pluses) match extremely well with the
exact energies from the ODE calculations (black solid line).
Second of all, we confirm that the two $N=1$ baby Skyrmions sitting
side-by-side always have a slightly higher energy than the $N=2$ one.
The energy difference, however, is very small and hence the
metastability is possible even with a small energy barrier between the 
two kinds of solution.

\begin{table}[!htp]
  \begin{center}
    \begin{tabular}{c|ccc}
      & BPS & Fig.~\ref{fig:N=2+2} & Fig.~\ref{fig:N=1+1+1+1_triangle}\\
      \hline
      $Q=4$ & 50.2655 & 50.2683 & 50.2685\\
      & BPS & Fig.~\ref{fig:N=1+1} & Fig.~\ref{fig:N=2}\\
      \hline
      $Q=2$ & 25.1327 & 25.1330 & 25.1357
    \end{tabular}
    \caption{Energies of BPS solutions (i.e.~with $\epsilon=0$)
      compared with the theoretical BPS mass for given $Q$.
      }
    \label{tab:BPSen}    
  \end{center}
\end{table}
In sharp contrast to the compacton case, we have in this model been
able to send $\epsilon$ all the way to zero, obtaining numerical BPS
solutions, 3 of which do not possess axial symmetry, see the last row
of figs.~\ref{fig:N=2+2}, \ref{fig:N=1+1+1+1_triangle},
\ref{fig:N=1+1} and \ref{fig:N=2}.
Since $\epsilon=0$ cannot be plotted on the logarithmic ordinate of
figs.~\ref{fig:NumEnQ4} and \ref{fig:NumEnQ2}, and they are all of the
same value (the BPS mass), we present the numerically calculated
energies in tab.~\ref{tab:BPSen} as a handle on the numerical accuracy
of our solutions.
For the $Q=4$ sector, the accuracy (discrepancy) of the numerically
calculated energy is about $6\times 10^{-5}$, whereas for the $Q=2$
sector it is $1\times 10^{-5}$ and $1\times 10^{-4}$, for the
two solutions. 

\begin{figure}[!thp]
\begin{center}
\mbox{\subfloat[$Q=4$]{\includegraphics[width=0.49\linewidth]{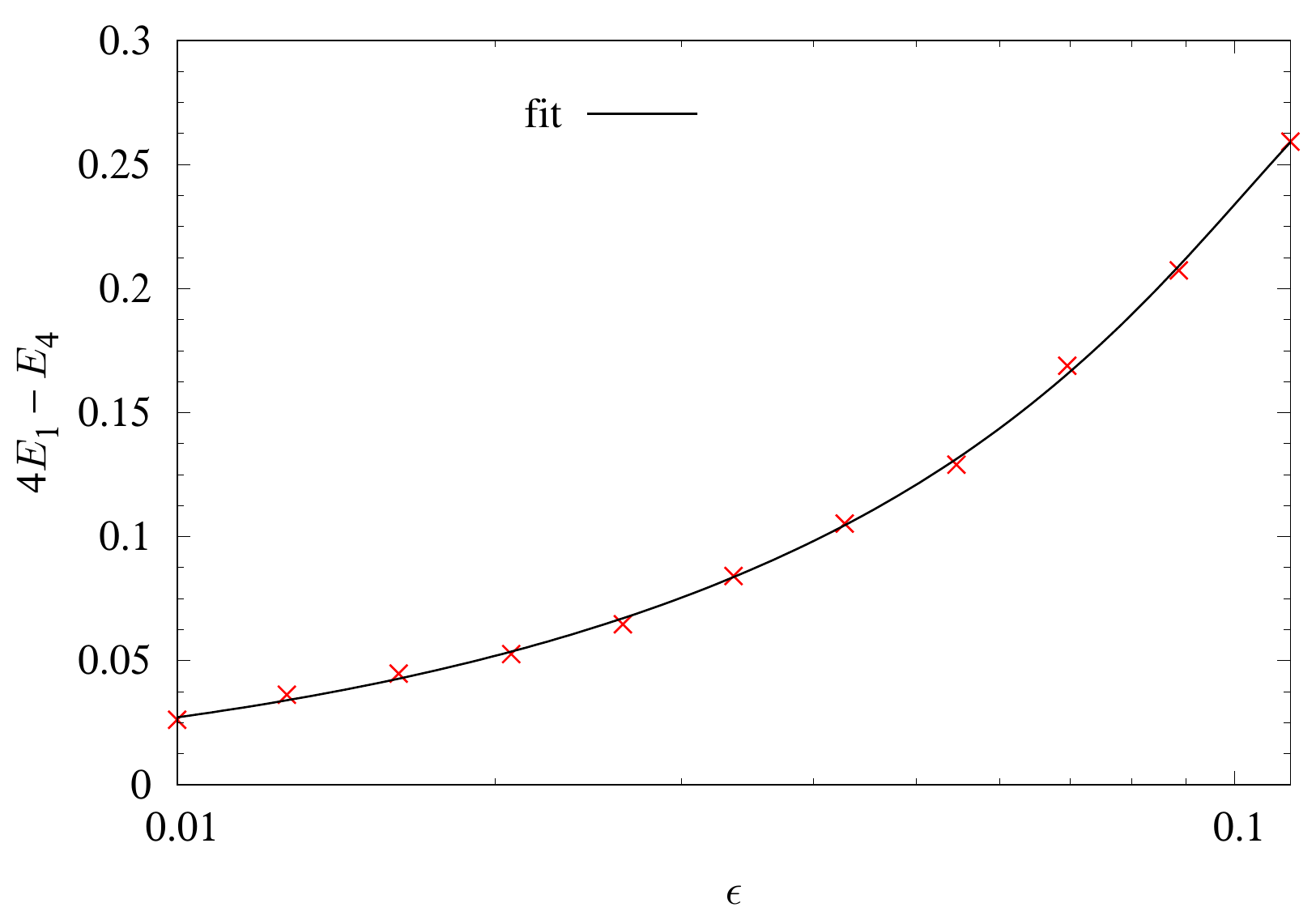}}
\subfloat[$Q=2$]{\includegraphics[width=0.49\linewidth]{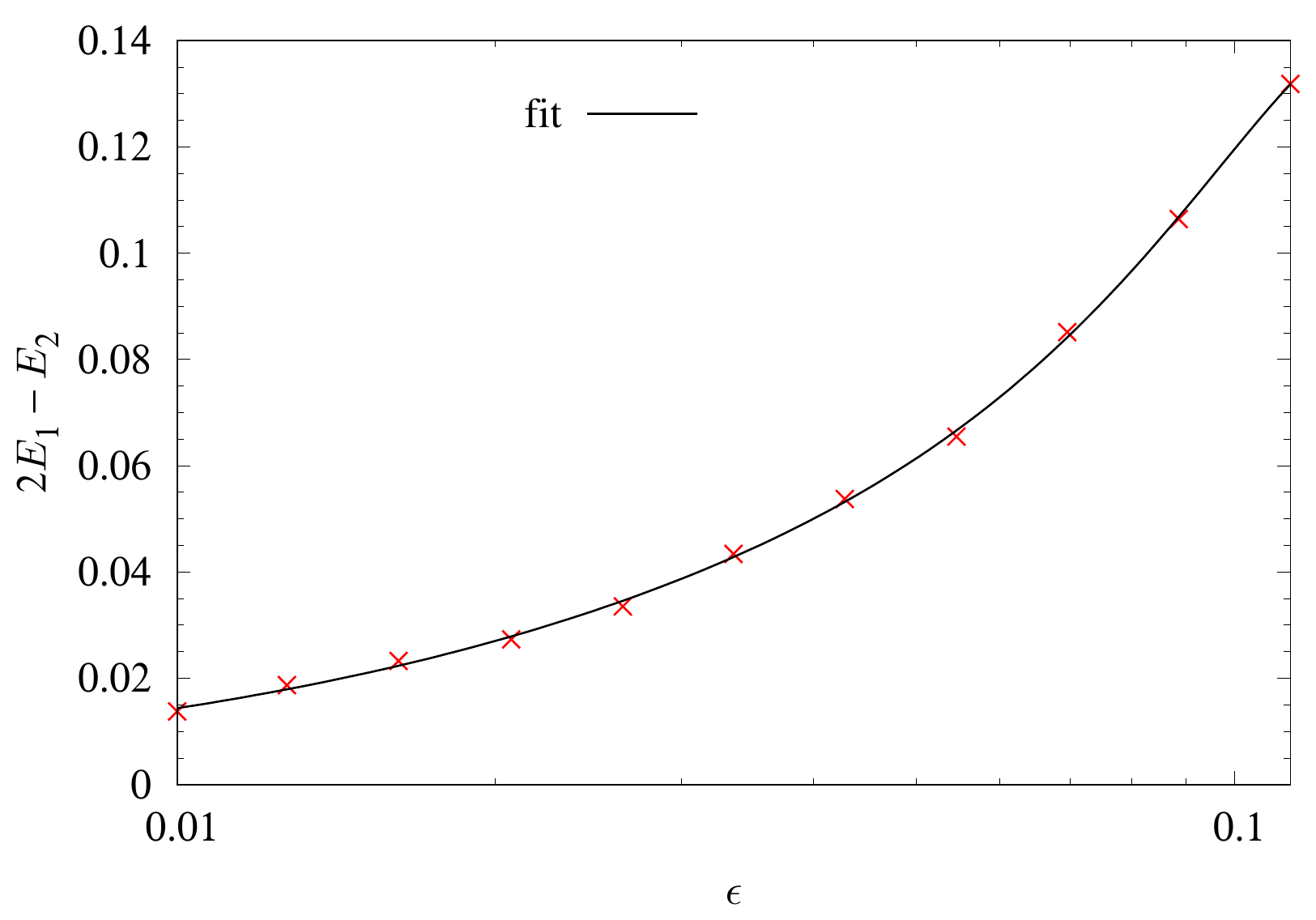}}}
\caption{Binding energies for (a) the $Q=2+2$ bound state solution of 
fig.~\ref{fig:N=2+2} and (b) of the $Q=1+1$ bound state solution of
fig.~\ref{fig:N=1+1}. 
The fits are shown in the text.
Here $m_1=0.5$.
}
\label{fig:ben_boundstate}
\end{center}
\end{figure}
We evaluate the binding energy
\beq
\BE_Q \equiv Q E_1 - E_Q,
\eeq
of the $Q=2+2$ bound state solution of
fig.~\ref{fig:N=2+2} in fig.~\ref{fig:ben_boundstate}(a) and of the
$Q=1+1$ bound state solution of fig.~\ref{fig:N=1+1} in
fig.~\ref{fig:ben_boundstate}(b) and the fits displayed on the figures
are given by
\begin{align}
\BE_4 &\simeq 2.860\epsilon - 16.99\epsilon^2 + 208.4\epsilon^3 - 907.4\epsilon^4,\label{eq:BE4}\\
\BE_2 &\simeq 1.552\epsilon - 13.15\epsilon^2 + 170.2\epsilon^3 - 743.7\epsilon^4,\label{eq:BE2}
\end{align}
and these binding energies are fitted for solutions with $m_1=0.5$.
The binding energies are tiny and the slight oscillation of the
numerically evaluated binding energies around the fit shows that we
are close to the edge of our numerical precision.
In principle, we could also numerically evaluate the energy of the
bond between two axially symmetric $N=2$ Skyrmions, that is
$2E_2-E_4$, but this quantity turns out to be at the level of
$10^{-4}$ and beyond our numerical precision, hence we do not attempt
at plotting it.

We notice that there seems to be a linear contribution to the binding
energy, in contradistinction with the compacton case of
ref.~\cite{Gudnason:2020tps}.
We will discuss this point in later sections.

\section{Long-range interaction}\label{sec:int}

In the case of the compactons, discussed in
ref.~\cite{Gudnason:2020tps}, the restricted harmonic condition was
not sufficient to fix the right background solution in the case of two
interacting baby Skyrmions. In particular, no restrictions in the
choice of the relative orientation or separation distance between the
non-overlapping compactons emerge from such a condition.

In this paper, we have turned on a more physical potential and now
have long-range forces. This means that at any separation distance,
two baby Skyrmions feel each other and the force between them is
dependent on the relative orientation, as was shown by
Piette-Schroers-Zakrzewski \cite{Piette:1994ug}.
Since we are only interested in baby Skyrmion bound states, among all
the possible GRH maps we must
choose among those that are in the attractive channel.
This uniquely fixes the relative orientation and hence the choice of
the GRH map. In the following,
we briefly review the 
calculation of ref.~\cite{Piette:1994ug}, adapting it to the case of
our Lagrangian \eqref{eq:L}.  

Let $u$ and $v$ be baby Skyrmion solutions of charge $N$ and $M$ and
let $\bphi^u$ and $\bphi^v$ be their representation in 3-vector
coordinates. Each field $\bphi$ is a solution of the full equations of
motion:
\begin{equation}
  \partial_i \mathbf{j}_i=\epsilon m_1^2\mathbf{n}\times \bphi+(\mathbf{n}\times \bphi)(1-\mathbf{n}\cdot \bphi),\label{eq:EulerL}
\end{equation}
where $\bn\equiv(0,0,1)$ is the vacuum of $\bphi$ and
\begin{equation}
  \mathbf{j}_i=\epsilon\bphi\times \partial_i \bphi+\partial_j\bphi(\partial_j\bphi\cdot\bphi\times \partial_i \bphi).
\end{equation}
When the two baby Skyrmions are well separated, the composite solution
of total charge $Q=N+M$ can be written as
$\bphi^w\equiv\bphi^{u+v}$. In order to evaluate the interaction
potential $V$ between the two solitons, it is necessary to decompose
the total energy in the form $E[\bphi^w]=E[\bphi^u]+E[\bphi^v]+V$. To
this end, it is useful to separate the coordinate space $\mathbb{R}^2$
into two regions such that $\bphi^u\approx\bn$ is close to the vacuum
$\bn$ in region $2$ and $\bphi^v\approx\bn$ is close to the vacuum
$\bn$ in region $1$. We can thus decompose the field into the vacuum
$\bn$ and an infinitesimal correction:
\begin{equation}
  \bphi^u\approx \mathbf{n}+\bdphi^u+\mathcal{O}(\bdphi^u\cdot\bdphi^u),
\end{equation}
where $\bdphi^u\cdot\bn=0$.
Since $\bphi^u$ solves the Euler-Lagrange equation \eqref{eq:EulerL},
$\bdphi^u$ satisfies the linearized equation 
\begin{equation}
  (\Delta-m_1^2)\,\bdphi^u=0,\label{eq:lineriz}
\end{equation}
and we have that in region $1$:
\beq
\bphi^w \approx \bphi^u + \bepsilon^v\times\bphi^u
+ \frac12\bepsilon^v\times(\bepsilon^v\times\bphi^u),
\label{eq:phi_w}
\eeq
where $\bepsilon^v$ is linear in $\bdphi^v$.
The expansion for the field $\bphi^w$ in terms of $\bdphi^u$ in the
region $2$ is analogous.
The asymptotic result \eqref{eq:lineriz} is independent of
$\epsilon$ as we expected and equivalent to that of
ref.~\cite{Piette:1994ug}. 
Once we have the form of $\bphi^w$ in the regions $1$ and $2$ (see
eq.~\eqref{eq:phi_w}), it is possible to evaluate the energy
$E[\bphi^w]$ as 
\begin{align}
\begin{split}
  E[\bphi^w]&\approx
  \int_1\d^2x\;\mathcal{E}(\bphi^w)+\int_2\d^2x\;\mathcal{E}(\bphi^w)\\
  &\approx E[\bphi^u]+E[\bphi^v]\\
  &\phantom{\approx\ }
  +\int_1\d^2x\left[\,\mathbf{j}^u_i\cdot\partial_i\bepsilon^v+ \epsilon m_1^2(\bepsilon^v\cdot\bn\times\bphi^u)+(\bepsilon^v\cdot\bn\times\bphi^u)(1-\bn\cdot\bphi^u)\right]\\
  &\phantom{\approx\ }
  +\int_2\d^2x\left[\,\mathbf{j}^v_i\cdot\partial_i \bepsilon^u
 + \epsilon m_1^2(\bepsilon^u\cdot\bn\times\bphi^v)+(\bepsilon^u\cdot\bn\times\bphi^v)(1-\bn\cdot\bphi^v)\right],
\end{split}
\end{align}
where $\bepsilon^u$ is linear in $\bdphi^u$ and is defined by
\begin{equation}\label{eq:epsilonu}
  \bepsilon^u=\frac12\bphi^v \times \big((1+\bphi^v\cdot\bn)\,\bdphi^u-(\bphi^v\cdot \bdphi^u)\,\bn\big),
\end{equation}
and $\bepsilon^v$ is equivalent to eq.~\eqref{eq:epsilonu} with the
superscripts $u$ and $v$ exchanged. In the end, using the equations of
motion \eqref{eq:EulerL} and Gauss's law, we can finally write the
long-range interacting potential $V$ as 
\begin{equation}\label{eq:attrctV}
V=\epsilon\int_{\Gamma}\left(\bdphi^v\cdot\partial_i\bdphi^u-\bdphi^u\cdot\partial_i\bdphi^v\right)\;\d S_i,
\end{equation}
where $\Gamma$ is a curve without self-intersections, separating region
$1$ from region $2$ and $\d S_i=\varepsilon_{ij}\Dot{\gamma}_j\d t$
for any parametrization $\gamma(t)$ of $\Gamma$. Given
eq.~\eqref{eq:lineriz}, the form of this potential is equivalent to
the one obtained in ref.~\cite{Piette:1994ug} albeit with an overall
factor of $\epsilon$.

With the result \eqref{eq:attrctV}, it is now possible to calculate
the long-range interaction for different multisoliton configurations
to determine the Skyrmions' relative orientation maximizing their
attraction. For the multisoliton case of charge $Q=1+1$, the potential
takes the form 
\begin{equation}
  V_{1+1}\propto\, \epsilon^3\cos(\alpha-\beta)\frac{e^{-m_1d}}{\sqrt{m_1d}},
  \label{eq:pot1+1}
\end{equation}
where $d$ is the relative distance, $\alpha$ and $\beta$ are the
two respective phases (orientations) of the baby
Skyrmions and we have assumed that $\bdphi\sim\mathcal{O}(\epsilon)$,
see the discussion in sec.~\ref{sec:NLO}. From this 
expression, we recognize that the maximally attractive channel is
obtained for $\alpha-\beta=\pi$, i.e.~when the two baby Skyrmions have
opposite orientation. Following this indication, we choose the
GRH map of this topological
sector with the solitons oppositely oriented. In the case of
the baby Skyrmion pair of charge $Q=2+2$, the asymptotic potential is
\begin{equation}
  V_{2+2}\propto\, -\epsilon^3\cos(\alpha-\beta)\frac{e^{-m_1d}}{\sqrt{m_1d}},
  \label{eq:pot2+2}
\end{equation}
in which case the maximum of the attraction requires the solitons to
have the same orientation, i.e.~$\alpha=\beta$ and we have
again assumed that $\bdphi\sim\mathcal{O}(\epsilon)$. Therefore, the proper
GRH map in this case is chosen to have the solitons
equally oriented. Our choices, based here on analytical
considerations, are confirmed by the numerical calculations in
figs.~\ref{fig:N=2+2} and \ref{fig:N=1+1} for small values of $\epsilon$.

\section{Binding energies in the perturbative scheme}\label{sec:binding}

\begin{figure}[!ht]
  \begin{center}
    \includegraphics[width=0.4\linewidth]{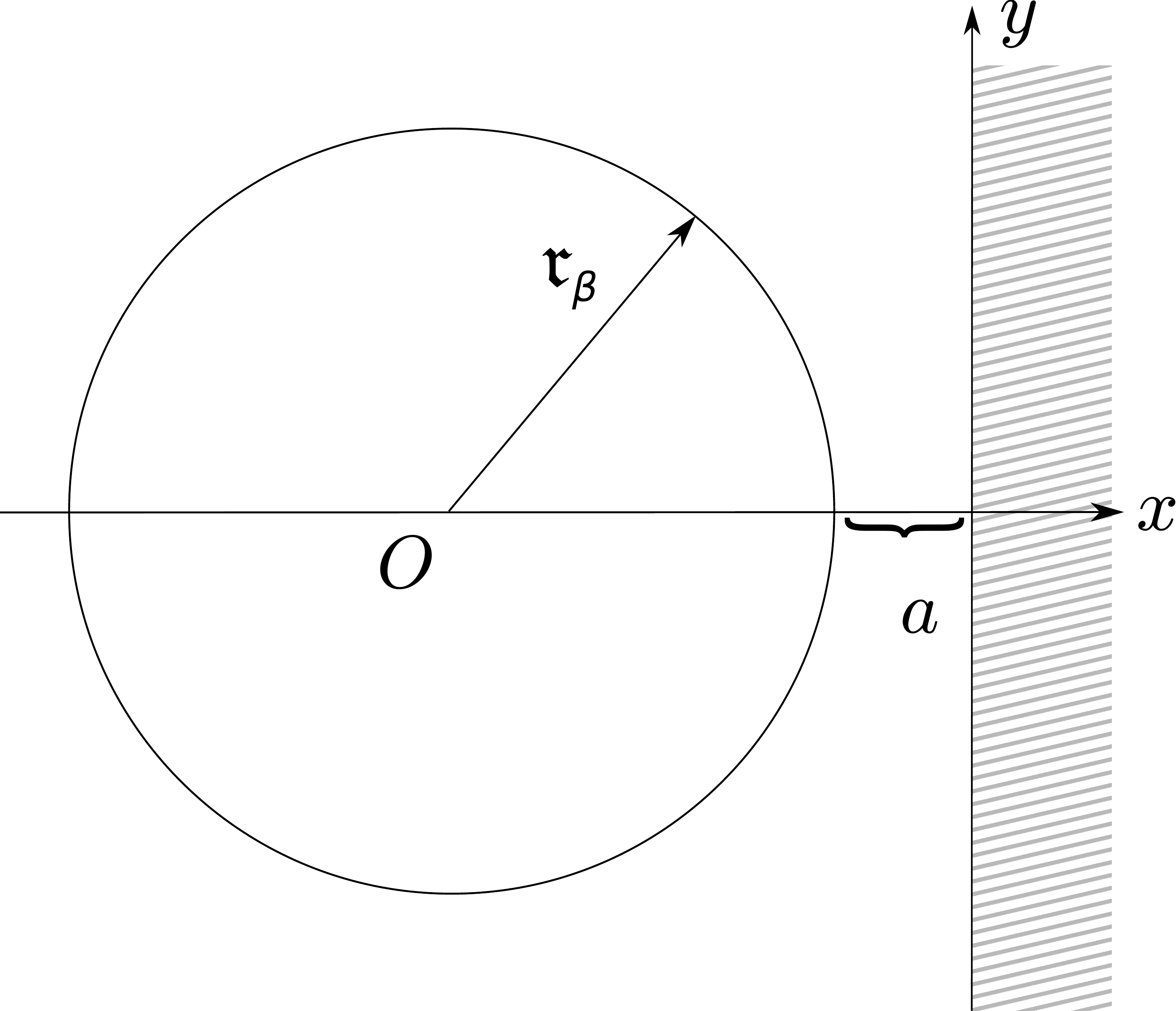}
    \caption{Sketch of the setup for calculating the binding energy
      between two axially symmetric $N$-Skyrmions.
      The radius containing most of the energy of the baby Skyrmion is
      $\mathfrak{r}_\beta$ with a suitable value for $\beta$
      (i.e.~the fraction of the BPS energy density contained within the
      radius); here we shall use
      $\mathfrak{r}_{0.99}\simeq 1.517\sqrt{2N}$.
      The separation between the two baby Skyrmions is $2a$ and the
      gluing boundary conditions in the problem are imposed at $x=x^1=0$
      (i.e.~the $y$-axis in the figure).
    }
    \label{fig:BC}
  \end{center}
\end{figure}

In the previous section, we have seen that two $N$-Skyrmions in the
attractive channel attract each other at asymptotic distances.
This attraction presumably breaks down due to nonlinearities at some
finite distance and for $\epsilon>0$ there will be a bound state with
two $N$-Skyrmions separated by a finite distance $2a$, see
fig.~\ref{fig:BC}.

In order to calculate the binding energies using the perturbative
method of the $\epsilon$-expansion that we have put forward in
ref.~\cite{Gudnason:2020tps} and this paper, we place two
$N$-Skyrmions side-by-side in the attractive channel, viz.~with the
pion field matching on the gluing line ($x^1=0$), see
fig.~\ref{fig:BC}.
In order to be able to glue the two baby Skyrmions together, we need
appropriate boundary conditions at $x^1=0$ (see fig.~\ref{fig:BC}),
which we shall call gluing conditions:
\begin{align}
  \p_x \phi^1(0,y) &= 0, \non
  \phi^2(0,y) &= 0, \label{eq:gluing_cond}\\
  \p_x\phi^3(0,y) &= 0. \nonumber
\end{align}
The reason for the odd condition on $\phi^2$, is that even boundary
conditions on all $\phi^a$ will turn the mirror Skyrmion into an
anti-Skyrmion for odd $N$, which is not what we want to glue the
solution with. 
The relatively simple-looking conditions above, become nonlinear
Robin-type boundary conditions for the fluctuation field $\bDelta$,
once we use that $\bphi=\bvarphi+\bdphi$ and $\bdphi$ from
eq.~\eqref{eq:Delta_form}: 
\begin{align}
  \p_x\left(
  \varphi^a
  +\varepsilon^{a b c}\Delta^b\varphi^c
  +\frac12(\bvarphi\cdot\bDelta)\Delta^a
  -\frac12(\bDelta\cdot\bDelta)\varphi^a
  \right)
  &= 0, \qquad a=1,3,\\
  \varphi^a
  +\varepsilon^{a b c}\Delta^b\varphi^c
  +\frac12(\bvarphi\cdot\bDelta)\Delta^a
  -\frac12(\bDelta\cdot\bDelta)\varphi^a
  &= 0, \qquad a=2.
\end{align}
It turns out to be a rather tricky boundary condition to implement
numerically.
For this reason we will use the Ansatz \eqref{eq:Delta_transverse} for
$\bDelta$ which reduces the above nonlinear Robin gluing conditions to
\begin{align}
  \p_x\left(\df\cos f\cos N\theta
  +\frac12\sin f\cos N\theta(2-\df^2-\dtheta^2)
  -\dtheta\sin N\theta\right) &= 0,\label{eq:gluing_cond_dfdt1}\\
  \dtheta\cos N\theta        
  +\df\cos f\sin N\theta
  +\frac12\sin f\sin N\theta(2-\df^2-\dtheta^2) &= 0.\label{eq:gluing_cond_dfdt2}
\end{align}
We will now use a finite difference approximation for the
$x$-derivative\footnote{Here for simplicity we use only a second-order 
  formula, although in the numerical code we use a fourth-order
  formula.} of $\df$ and $\dtheta$: 
\begin{equation}
\df_x = \frac{\df_{-2} - 4\df_{-1} + 3\df_0}{2h_x}, \qquad
\dtheta_x = \frac{\dtheta_{-2} - 4\dtheta_{-1} + 3\dtheta_0}{2h_x}, \qquad
\df = \df_0, \qquad \dtheta = \dtheta_0,
\end{equation}
where the subscript $0$ refers to the $x^1=0$ lattice point on the gluing
line, $-1$ to one lattice point to the left of the gluing line and so
on.
This reduces the eqs.~\eqref{eq:gluing_cond_dfdt1} and
\eqref{eq:gluing_cond_dfdt2} to two quadratic algebraic equations,
which in principle can be solved.
The issue is to let the algorithm choose the appropriate root for each
field, for which there are 4 possibilities.
For this reason, we solve this equation iteratively, by setting
\beq
\df_0 = \overline\df_0 + \ddf_0, \qquad
\dtheta_0 = \overline\dtheta_0 + \ddtheta_0,
\eeq
inserting these into the algebraic conditions and linearizing with
respect to $\ddf_0$ and $\ddtheta_0$.
This yields expressions for $\ddf_0$ and $\ddtheta_0$ which are not
particularly illuminating, so we will not display them here.
In principle the method is simple, $\overline\df_0$ is the previous
value of $\df_0$ and $\df_0$ is updated by adding the solution for
$\ddf_0$ to $\overline\df_0$ at each step of the algorithm (and
similarly for $\ddtheta_0$).
Notice that $\ddf_0$ vanishes when $\overline\df_0$ is a solution to
the full nonlinear Robin type gluing condition. 
Unfortunately, this simplest form of Newton iteration does not have
particularly good convergence properties \cite{Higham:2001}.
For this reason we had to implement a line search algorithm following
ref.~\cite{Higham:2001}, which updates $\df_0$ and $\dtheta_0$ using
\beq
\df_0 = \overline\df_0 + t\ddf_0, \qquad
\dtheta_0 = \overline\dtheta_0 + t\ddtheta_0, \qquad
t\in(0,2),
\eeq
where $t$ is a parameter that should be optimized for each step in the
iteration.
Note that $t=1$ is just the simplest version of Newton iteration.
We implement a rather crude line search algorithm that tries out 20
points of $t$ in its interval and refines the search one time.
It turns out that often the optimal value of $t$ is around $t\sim0.5$
or $t\sim0.6$.
Although this algorithm is rather computationally expensive,
ref.~\cite{Higham:2001} showed the line search is one of the cheapest
algorithms with improved convergence properties for this type of
algebraic Riccati equation.

We now turn to solving the coupled PDEs \eqref{eq:Xlineq} with the
boundary conditions
\beq
\lim_{|\bx|\to\infty}\bDelta = \mathbf{0}, \qquad
\bDelta(\bx_0) = \mathbf{0}, 
\eeq
for all $\bx_0$ being centers of $N$-Skyrmions, as well as the gluing
conditions described above.
The ``boundary condition'' at all $\bx_0$ is to prevent the
perturbation from trying to unwrap the soliton solution.
Since the calculation is computationally very expensive, we have
implemented the code in CUDA C and run it on an NVIDIA GPU cluster. 

\begin{figure}[!htp]
  \begin{center}
    \mbox{\subfloat[]{\includegraphics[width=0.49\linewidth]{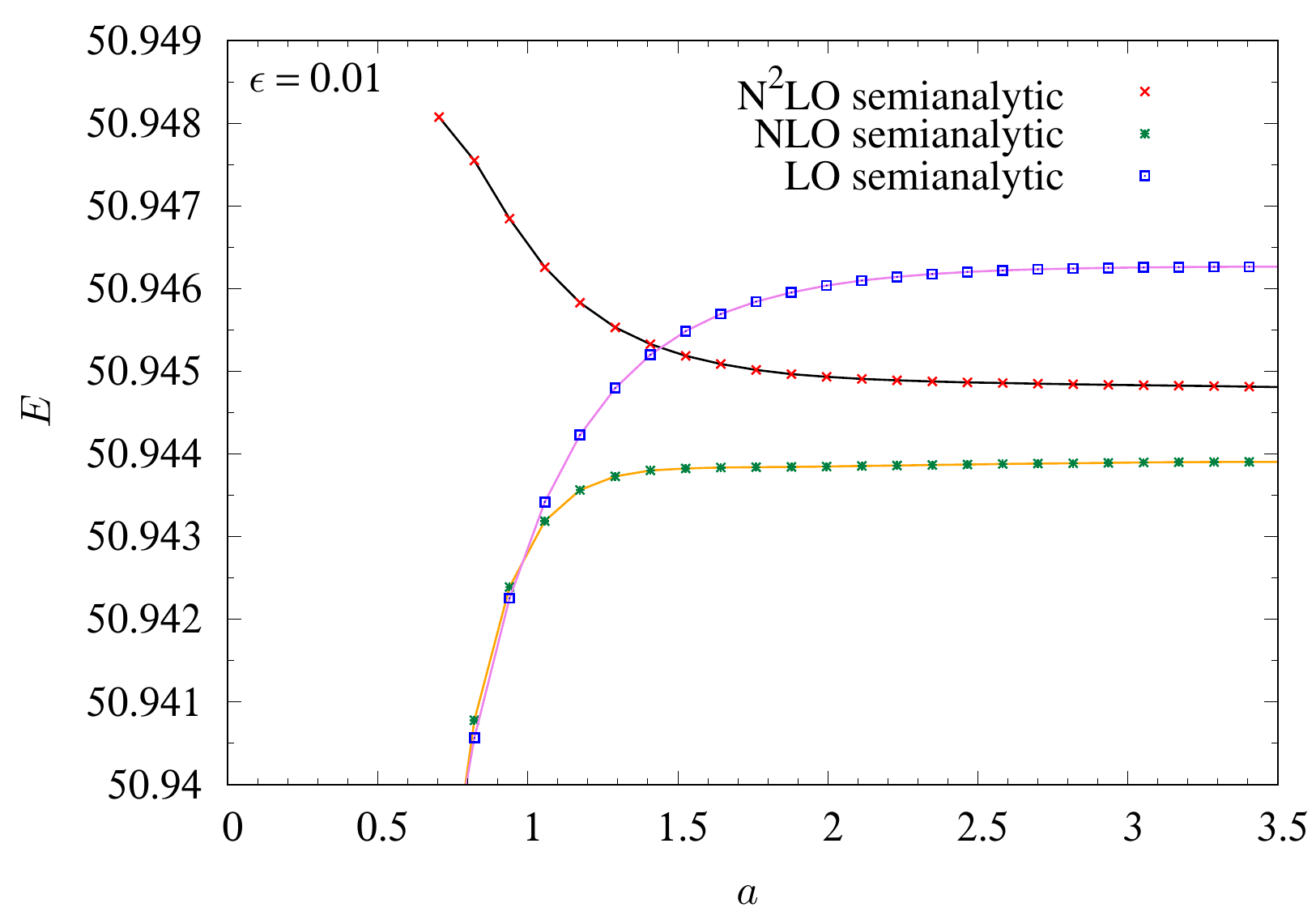}}
      \subfloat[]{\includegraphics[width=0.49\linewidth]{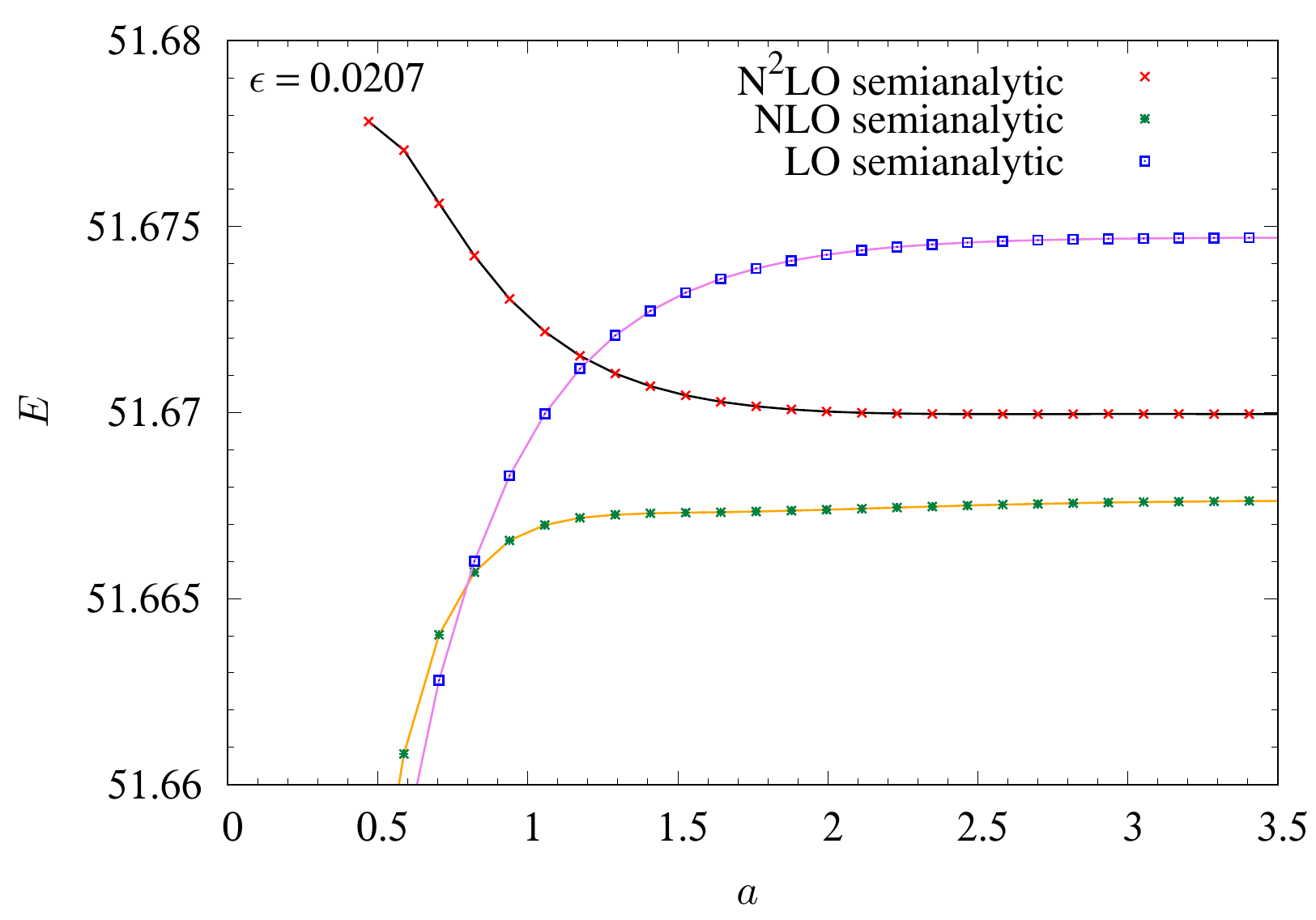}}}
    \mbox{\subfloat[]{\includegraphics[width=0.49\linewidth]{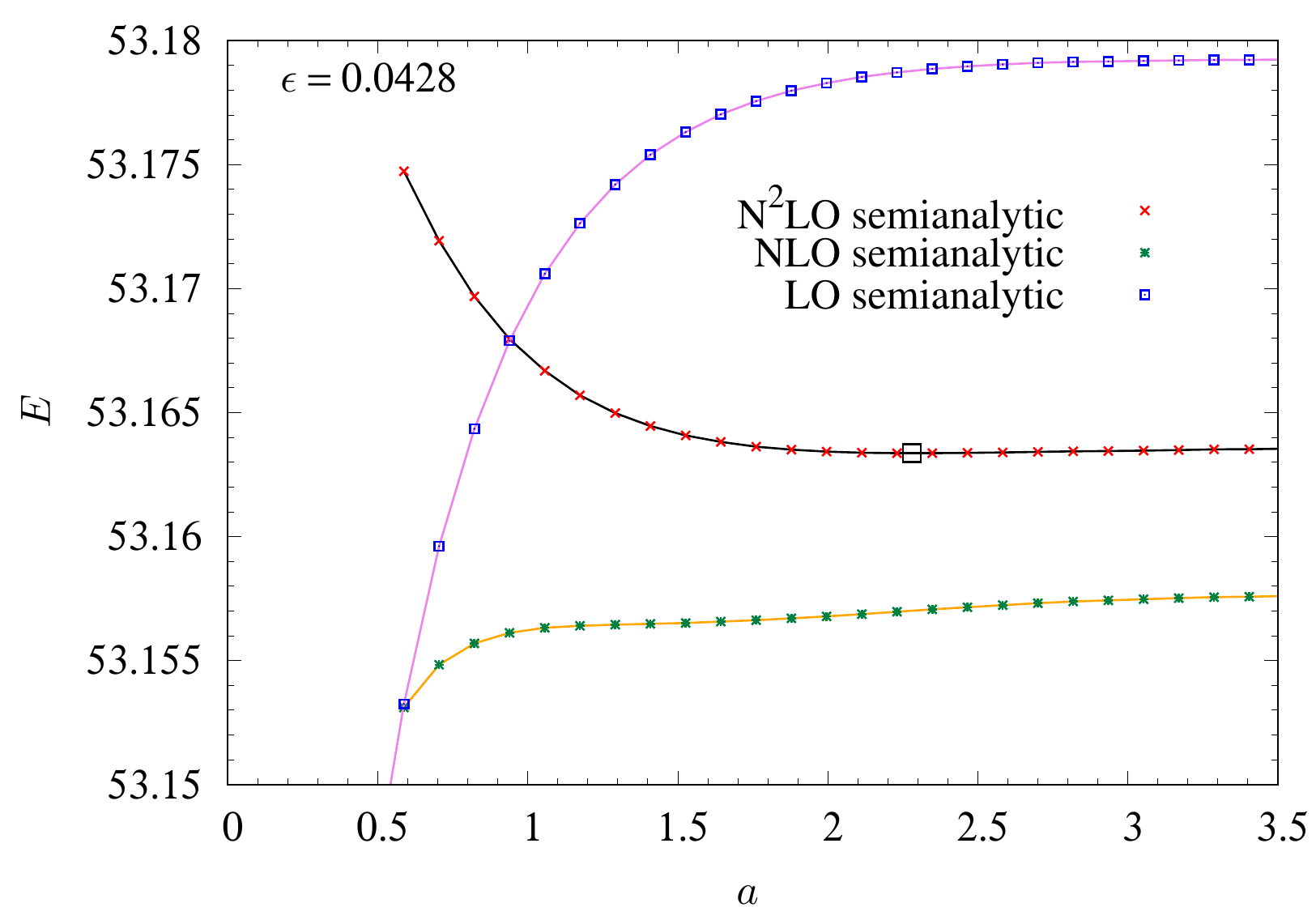}}
      \subfloat[]{\includegraphics[width=0.49\linewidth]{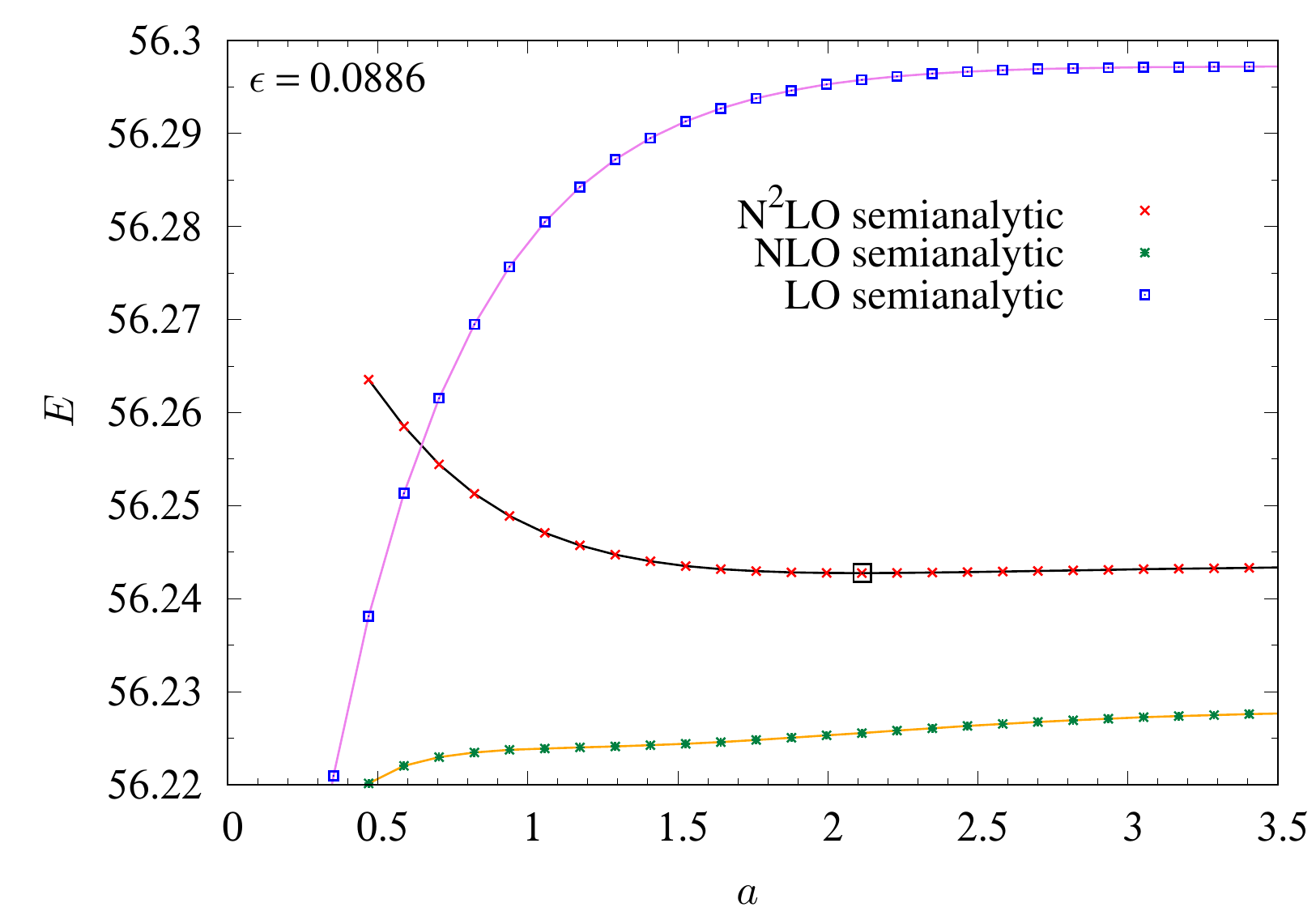}}}
    \caption{The LO (blue squares), NLO (green crosses) and N$^2$LO
      corrections (red crosses) to the energy of two $N=2$ baby
      Skyrmions as functions of the separation distance $2a$, see
      fig.~\ref{fig:BC}. The panels show different values of
      $\epsilon$: (a) $\epsilon=0.01$, (b) $\epsilon=0.0207$, (c)
      $\epsilon=0.0428$ and (d) $\epsilon=0.0886$.
      The LO energy is calculated by geometrically cutting off the
      BPS energy plus the LO correction at the gluing line ($x^1=0$) (and
      multiplying by 2).
      A large black square shows the minimum of the N$^2$LO energy in
      panels (c) and (d).
    }
    \label{fig:ben6}
  \end{center}
\end{figure}

In order to calculate the binding energy of two charge-$N$ baby
Skyrmions, we first need to determine their optimal separation
distance; that is the distance where the total energy of the bound
state is at a local or global minimum and then read off the energy at
that point.

In an attempt to understand which separation distance ($2a$) is
preferred by two $N=2$ baby Skyrmions in the attractive channel, we
calculate the energy density as a function of $a$ for various values
of $\epsilon$.
The result is shown for $\epsilon=0.01,0.0207,0.0428,0.0886$ in
fig.~\ref{fig:ben6}.
Let us first see what happens at each order in our $\epsilon$
expansion.
At the leading order (LO), there is no fluctuation field and the BPS
background solution $\bvarphi$ is used to calculate the energy from
the Lagrangian \eqref{eq:L}, including the BPS energy.
Of course, we have stitched together two baby Skyrmions side-by-side,
so the LO energy is evaluated without overlap: this means we integrate
the energy over the background BPS solution up to the gluing line
$(x^1=0)$ and then multiply by two.
Clearly, this order is flawed in the following sense: The further we
push the two baby Skyrmions together, the less the LO energy is.
This cannot be the correct physical picture.
For that, the perturbation field $\bDelta$ is necessary.
In particular, the gluing conditions are crucial as should become
clear momentarily.
For completeness, we show both the energy calculated to order
$\epsilon^2$ (NLO) and to order $\epsilon^3$ (N$^2$LO).
As we can see from the figure, what happens is that the LO energy goes
down as $a$ decreases, and so does the NLO energy, but the N$^2$LO
energy goes up. 
This is simply the gluing condition twitching the field at the gluing
line and building up energy localized around the gluing condition.
However, neither the N$^2$LO nor the NLO semi-analytic energies
capture a minimum at the separation distance $a\sim0.8$ that was found
in sec.~\ref{sec:numcalc}, and we conclude that the perturbative
method has failed to calculate the bound state for the baby-Skyrmions
with (Gaussian) tails, as opposed to the case of the compactons, for
which the method was very successful \cite{Gudnason:2020tps}.
We can also see the minimum (which only appears for the N$^2$LO
energies) is extremely shallow and orders of magnitude smaller than
expected. 

Before, we discuss the sources of the failure of the method to capture
the bound states, we consider 
the bound state of two $N=1$ baby Skyrmions, which are
only metastable (see fig.~\ref{fig:LON}), since the axially symmetric
$N=2$ solution has lower energy than the bound state of two $N=1$
Skyrmions.
The bound state does, nevertheless, exist for $\epsilon\lesssim0.15$,
see fig.~\ref{fig:N=1+1}.

\begin{figure}[!htp]
  \begin{center}
    \mbox{\subfloat[]{\includegraphics[width=0.49\linewidth]{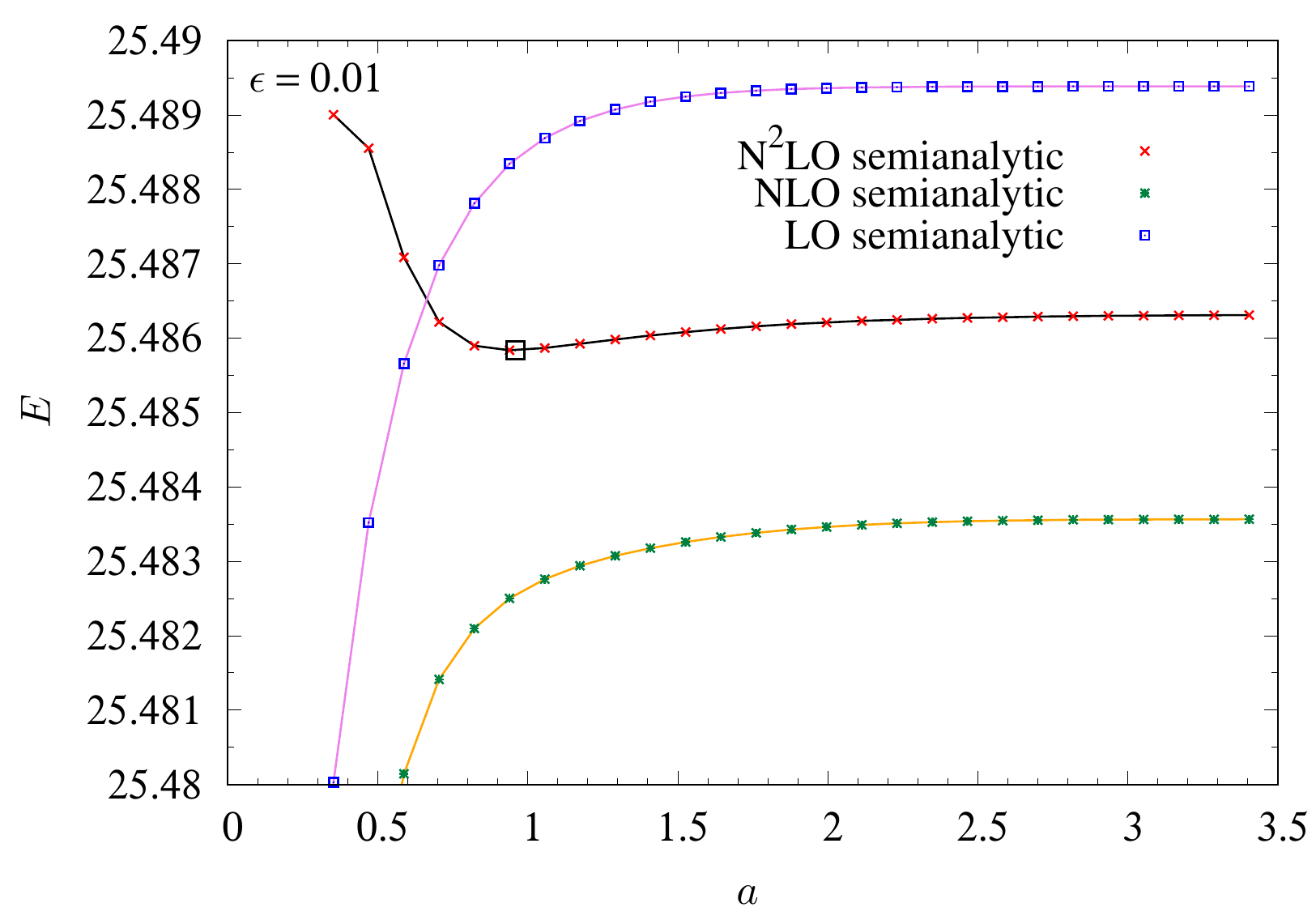}}
      \subfloat[]{\includegraphics[width=0.49\linewidth]{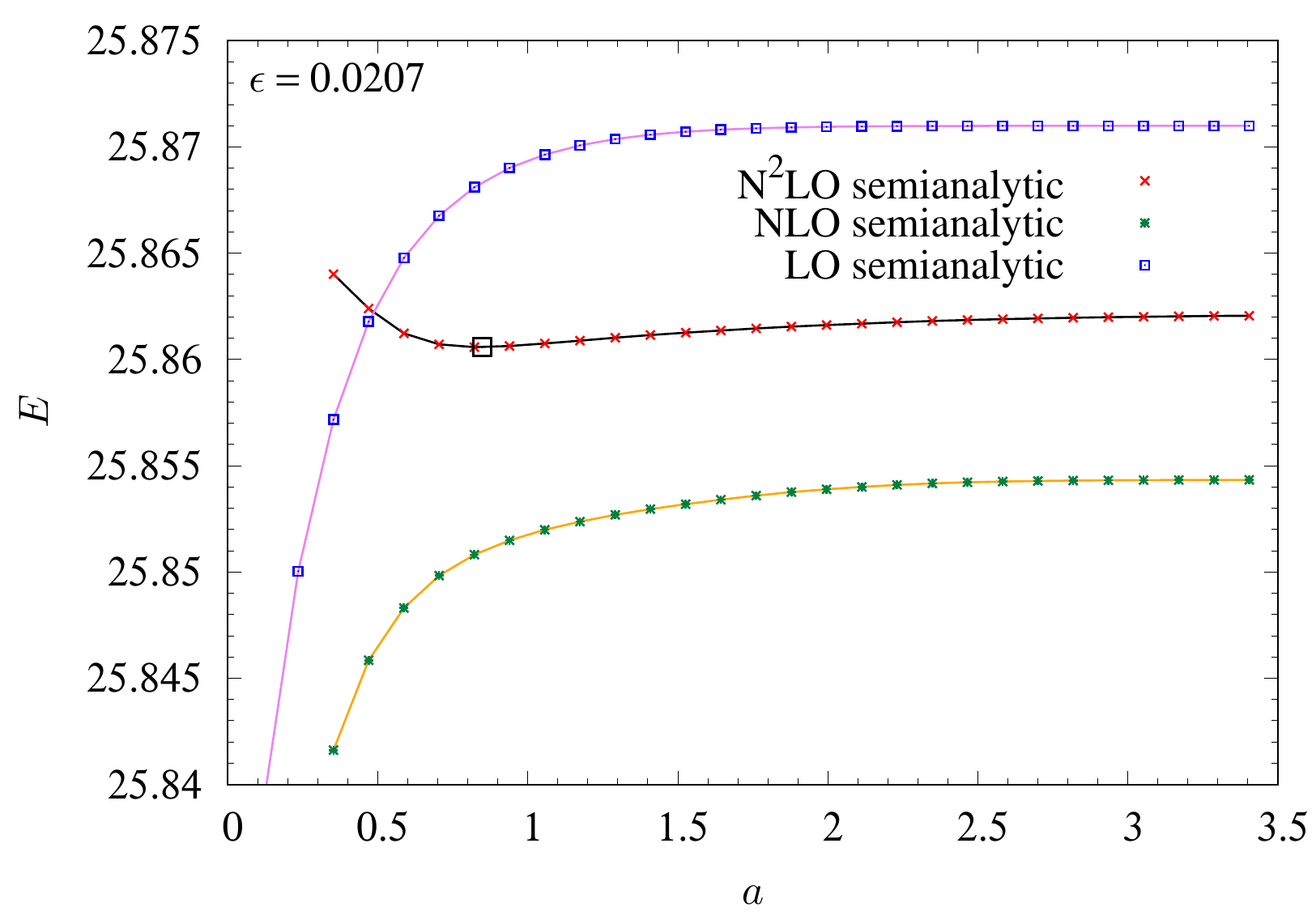}}}
    \mbox{\subfloat[]{\includegraphics[width=0.49\linewidth]{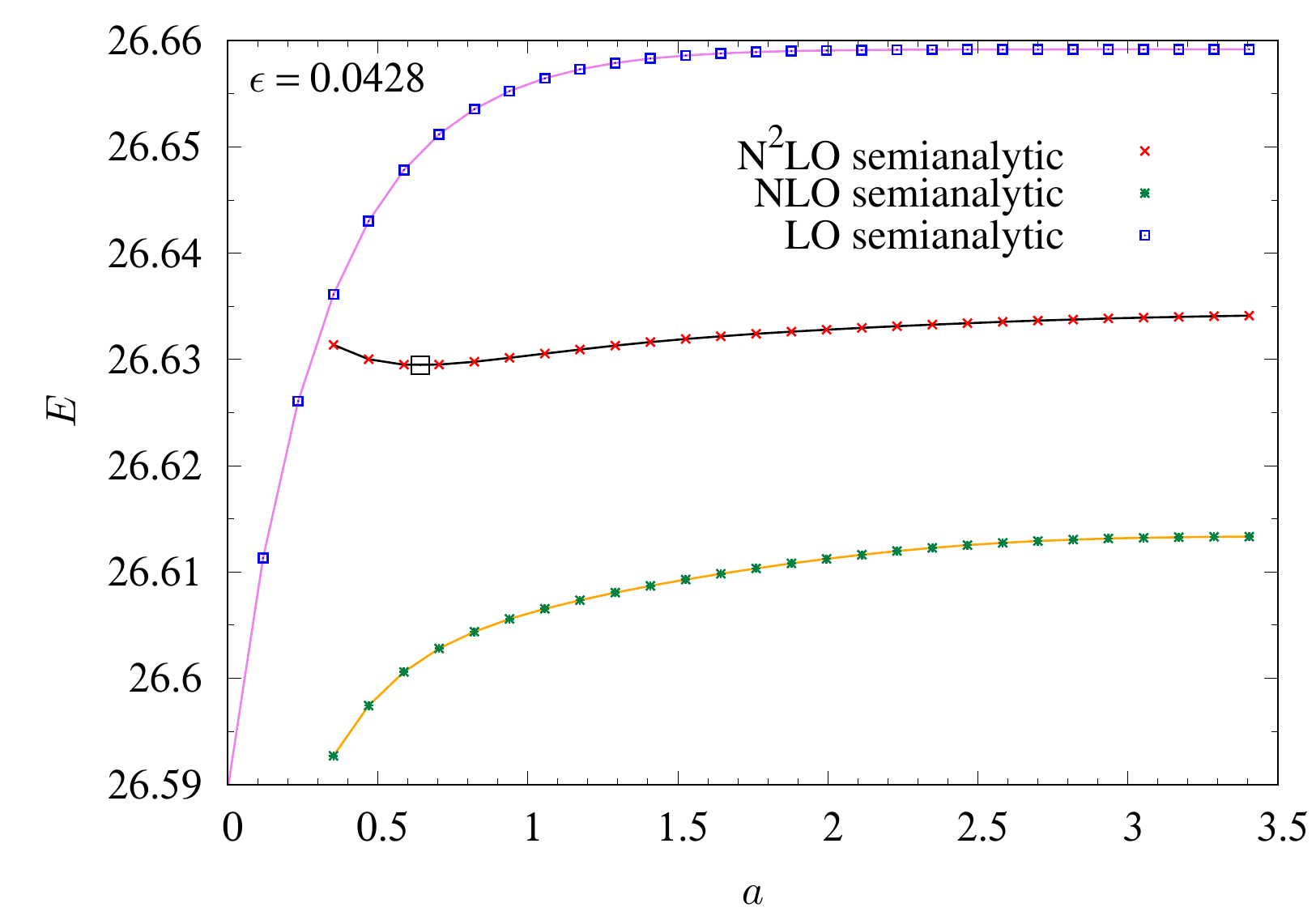}}
      \subfloat[]{\includegraphics[width=0.49\linewidth]{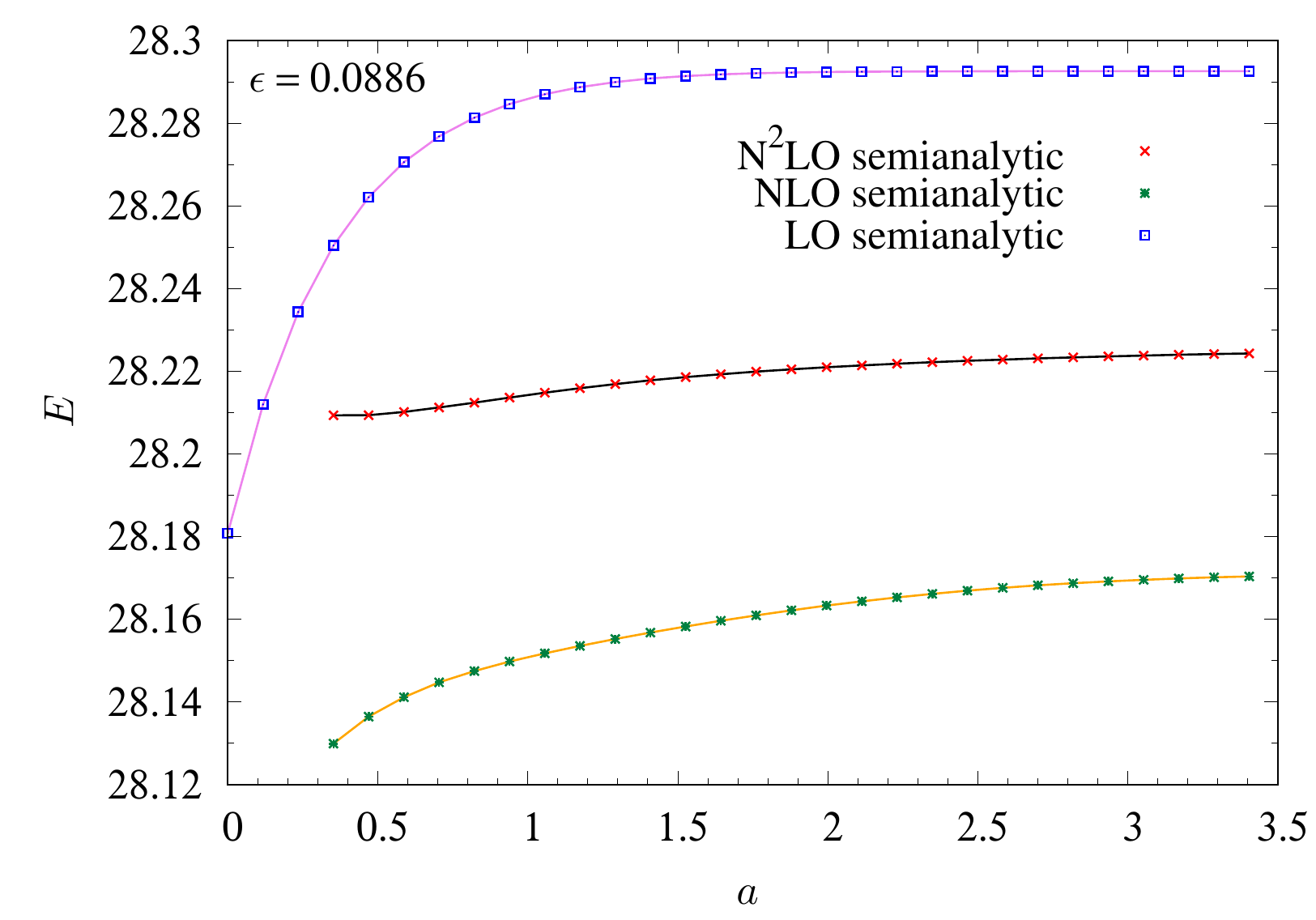}}}
    \caption{The LO (blue squares), NLO (green crosses) and N$^2$LO
      corrections (red crosses) to the energy of two $N=1$ baby
      Skyrmions as functions of the separation distance $2a$, see
      fig.~\ref{fig:BC}. The panels show different values of
      $\epsilon$: (a) $\epsilon=0.01$, (b) $\epsilon=0.0207$, (c)
      $\epsilon=0.0428$ and (d) $\epsilon=0.0886$.
      The LO energy is calculated by geometrically cutting off the
      BPS energy plus the LO correction at the gluing line ($x^1=0$)
      (and multiplying by 2).
      A large black square shows the minimum of the N$^2$LO energy.
    }
    \label{fig:ben3}
  \end{center}
\end{figure}

In order to attempt at calculating the binding energy of the bound
state in this case, we again perform a large number of PDE
calculations in this semianalytic approach for many values of the
separation distance $2a$, and $\epsilon=0.01,0.0207,0.0428,0.0886$. 
The result is shown in fig.~\ref{fig:ben3}.
Now one slight improvement over the previous case, is that we find a
minimum of the energy at order N$^2$LO for all values of $\epsilon$
smaller than $\sim0.08$.\footnote{
The minimum almost certainly
pertains for slightly larger values of $\epsilon$, but the convergence
of our calculations fails to capture it in fig.~\ref{fig:ben3}(d), but
not by much.
}
The minimum, however, again appears at quite large values of $a$ and
the perturbative method has thus failed also in the case of two $N=1$
baby Skyrmions.

\begin{figure}[!htp]
  \begin{center}
    \mbox{\subfloat[]{\includegraphics[width=0.49\linewidth]{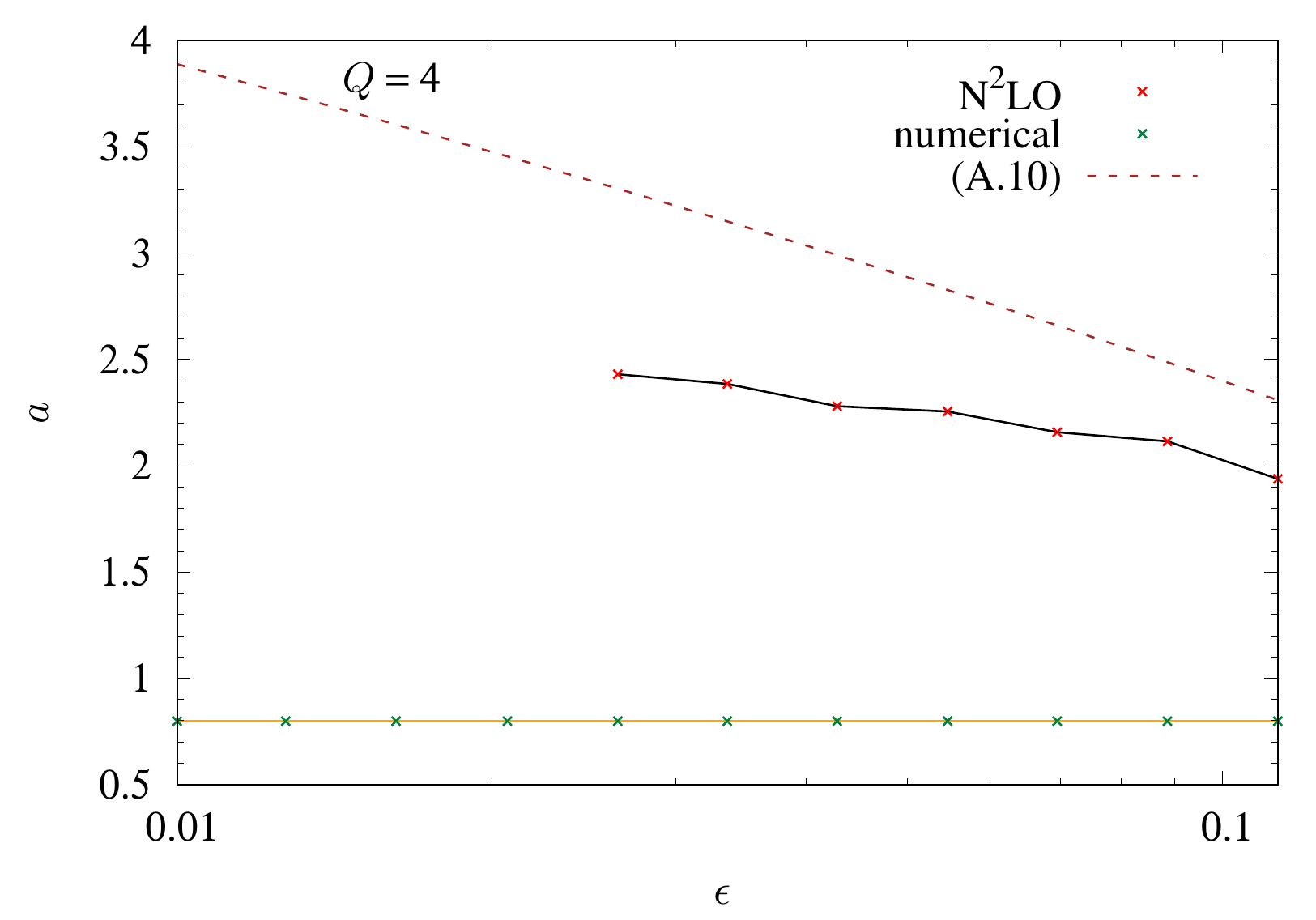}}
    \subfloat[]{\includegraphics[width=0.49\linewidth]{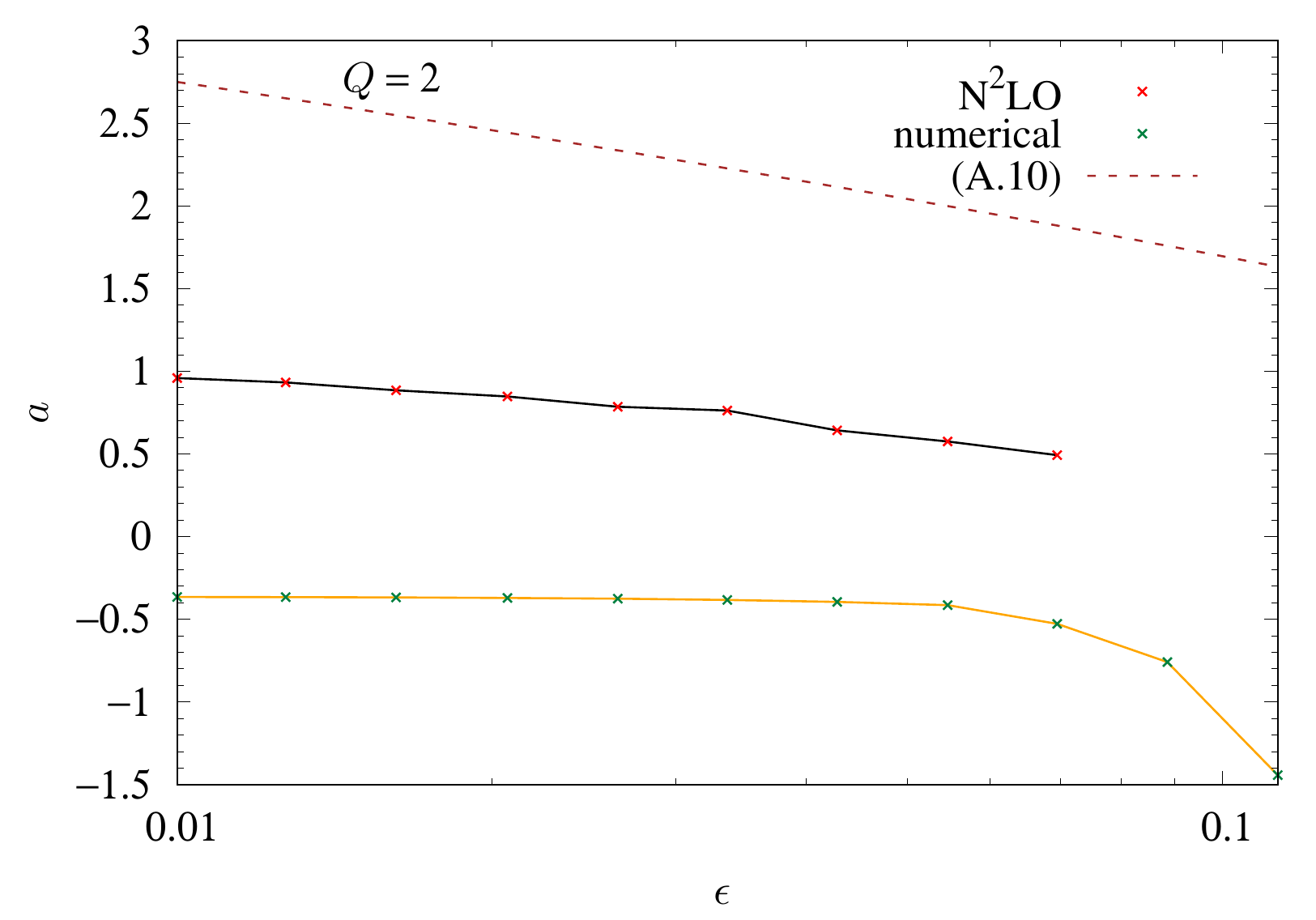}}}
    \caption{
      Separation distance $2a$ between (a) two $N=2$ and (b) two $N=1$
      baby Skyrmions as a function of $\epsilon$.
      The semianalytic method to order N$^2$LO
      is shown with red crosses and the full numerical brute-force
      computation of fig.~\ref{fig:N=2+2} is shown with green
      crosses.
      For comparison, the compacton radius $R_{\rm compacton}$ of
      eq.~\eqref{eq:R2compacton} is shown with a brown dashed line.
    }
    \label{fig:mina}
  \end{center}
\end{figure}

To summarize, we show in fig.~\ref{fig:mina} a comparison of the
separation distances found by the perturbative approach (green
crosses) and those found by precise full numerical calculations (red
crosses) of sec.~\ref{sec:numcalc}.
We also display the would-be compacton radius of a compacton that
corresponds to taking the limit $\epsilon\to0$ with
$\sqrt{\epsilon} m_1=:\tilde{m}_1$ held fixed.
What this limit amounts to, is to kill off the kinetic term
($\Lag_2$) while keeping the pion mass term $V_1$.
In this limit, there is a BPS solution with compacton radius
$R_{\rm compacton}$ given by eq.~\eqref{eq:R2compacton}.
None of the approaches agree and the only precise method used for
calculating the bound state is the numerical method used in
sec.~\ref{sec:numcalc}. 

The reasons why the semi-analytic or perturbative approach fails for
baby Skyrmions with tails, as compared to compactons, are:
\begin{itemize}
\item
The two axially symmetric BPS solutions glued together, used as a
background solution for the bound state calculation, is not a BPS
solution (although it was in the compacton
case \cite{Gudnason:2020tps}).
\item
The axially symmetric BPS solution is no longer close to the exact
solution at long distances (see fig.~\ref{fig:lpprofile}), as was the
case for the compactons \cite{Gudnason:2020tps}, which in turn has
implications for using it for the bound state calculation by gluing
two of them together. 
\item The nonlinearities become extremely important for the bound
states, because of the above reason.
\item In the case of the compactons, the cusp condition gave a crucial
and exact condition for the perturbation field at the boundary of the
compacton, outside of which the perturbation field was governed by a
free theory \cite{Gudnason:2020tps}. Without this crucial
nonperturbative ``bootstrap'', the perturbative method becomes
imprecise and fails to capture bound states.
\item Without the above-mentioned ``bootstrap'', the equations of
motions to be solved must include the nonlinearities, making the
problem as difficult as the original numerical problem -- the
perturbative approach thus fails in this case as it does not simplify
the problem to be solved.
\item Finally, the gluing condition in the case of the compactons
worked well because the perturbation fields were very small near the
gluing boundary \cite{Gudnason:2020tps}; this is not the case here for
the baby Skyrmions with (Gaussian) tails, in particular near the
optimal separation distance, and hence the gluing becomes unreliable
without taking into account nonlinearities. 
\end{itemize}

\section{Conclusion and discussion}\label{sec:discussion}

In this paper, we have studied the near-BPS regime of the baby Skyrme
model with a physical pion mass (i.e.~one that does not diverge in the
BPS limit) and a potential that is necessary for the BPS solution.
The latter potential is chosen such that the BPS solutions have
Gaussian tails (i.e.~$e^{-r^2}$), which means that we avoid a typical
complication of the BPS solutions being compactons -- as was the case
studied in ref.~\cite{Gudnason:2020tps}.
The problem with the compactons, is that the fields deviate from the
vacuum only on a compact region of space and the derivatives exhibit a
jump from a negative value to zero at the border of said region.
The true solution in the near-BPS regime does not possess such a
discontinuity in the first derivative of the fields and therefore we
had invented in ref.~\cite{Gudnason:2020tps} a cusp condition to make
the total field smooth, by inducing a counter-cusp to the perturbation
field. 
This has all been avoided in the present paper.
The motivation as stated above, is to get a pion mass that is of the
same order as the kinetic term and both are proportional to a small
parameter, $\epsilon$, hence giving rise to physical pions in 
the theory and also to avoid the technical difficulties coming along
with the mentioned cusps.
The leading order correction to the energy in $\epsilon$ of an
axially symmetric charge-$N$ baby Skyrmion is also in this case
obtained by evaluating the Dirichlet (kinetic) energy of the BPS solution,
which luckily is finite. This correction is linear in $\epsilon$. 
Bound states, however, cannot be studied unless we go to a higher
order in $\epsilon$ and introduce a perturbation field along with
suitable boundary conditions.
This is because the tails of the BPS solutions fall off to spatial
infinity and would just overlap each other if nothing extra is
introduced in the theory. 
The key to the perturbation field was again a transverse field used in
ref.~\cite{Gudnason:2020tps}, inspired by the work in
ref.~\cite{Piette:1994ug}.
The perturbation field in this paper does not require a so-called cusp
condition, mentioned above, which is a relieving simplification.
First we studied the energies of the baby Skyrmions in the
axially symmetric case up to order $\mathcal{O}(\epsilon^3)$ which we
call N$^2$LO and find excellent agreement between the perturbative
scheme and full numerical ODE calculations.

Then we move on to performing large-scale full numerical brute-force
computations of Skyrmions with charges $Q=4$ and $Q=2$, which include
the simplest bound states of the most stable baby Skyrmions in the
theory. 
For the $Q=4$ sector, the most stable solution is a bound state of two
$N=2$ approximately axially symmetric baby Skyrmions quite well
separated, but still having their tales intertwined in a bound state.
The axially symmetric $Q=N=4$ solution is unstable for small
$\epsilon$, but another metastable solution composed of four $N=1$
baby Skyrmion in a nearly triangular arrangement was found.
Interestingly, this solution was completely triangularly symmetric in
the compacton case studied in ref.~\cite{Gudnason:2020tps}, whereas in
this case two of the solitons move close together and repel the other
two, one more than the other. This is probably explained by a
quadrupole force, which is incompatible with the triangular
symmetry. 
In the $Q=2$ topological sector, the most stable solution is simply
the $Q=N=2$ axially symmetric solution, but for small $\epsilon$,
there is also a bound state of two $N=1$ baby Skyrmions.
Finally, we would like to stress that we have been able to find
numerical BPS solutions for all the mentioned cases with $\epsilon=0$
set strictly to zero.
That is, these solutions are nontrivial BPS solutions for which we do
not yet know a suitable analytic Ansatz.

The last part of the paper is an attempt at calculating the binding
energies perturbatively in our semianalytic $\epsilon$ expansion
scheme, which was very successful for the compacton case of
ref.~\cite{Gudnason:2020tps}.
Unfortunately, it turns out that this
perturbative scheme is unreliable for baby Skyrmions with (Gaussian)
tails, basically because the background field configuration
is no longer close to the true solution and nonlinearities become
crucial, thus invalidating the linearized approach used in the
perturbative scheme. 
For more details, see the list at the end of the previous section.

Interestingly, it appears that the binding energy in this model, where
the BPS solutions have a Gaussian tail, has a leading term which is
linear in $\epsilon$ -- in stark contradistinction to the case of the
compactons of ref.~\cite{Gudnason:2020tps}, where the binding energy
appears only at the order $\mathcal{O}(\epsilon^2)$. 

Although the compactons require painful cusp conditions for allowing
one to study their near-BPS limit, the solitons with tails turn out to
be even more difficult and can probably be well understood only by using 
nonlinear techniques or simply full numerical computations.

It would be interesting to consider other potentials and as we found
in this paper, compared with ref.~\cite{Gudnason:2020tps}, everything
depends strongly on the choice of the potential.
Potentials in the BPS sector, giving rise to a power-law
tail \cite{Adam:2010jr}, would probably provide very different
properties with respect to the $\epsilon$ expansion as well as to the
solutions in general. We will leave the investigation of such cases to
future studies.

\subsection*{Acknowledgments}

S.~B.~G.~thanks the Outstanding Talent Program of Henan University for
partial support.
The work of S.~B.~G.~is supported by the National Natural Science
Foundation of China (Grants No.~11675223 and No.~12071111).
The work of M.~B.~and S.~B.~is supported by the INFN special project 
grant ``GAST (Gauge and String Theories)''.

\appendix

\section{BPS solution with potential \texorpdfstring{$\tilde{m}_1^2V_1+V_2$}{m1tilde**2*V1+V2}}\label{app:would-be_BPS_sol}

Writing the BPS equation for the Skyrme term, but including both
$\tilde{m}_1^2V_1+V_2$ instead of only $V_2$ (as in
eq.~\eqref{eq:Bogomolnyi_trick}), we have
\beq
\mathcal{Q}_{12} = -\bphi\cdot\p_1\bphi\times\p_2\bphi=
\mp\sqrt{(1-\phi_3)^2 + 2\tilde{m}_1^2(1-\phi_3)}.
\eeq
Switching to stereographic coordinates yields
\beq
\frac{\epsilon^{ij}\p_i\omega\p_j\bar\omega}{(1+|\omega|^2)^2}
=\mp\i\frac{|\omega|\sqrt{\tilde{m}_1^2+(1+\tilde{m}_1^2)|\omega|^2}}{1+|\omega|^2}.
\eeq
Inserting the axially symmetric Ansatz $\omega=e^{\i N\theta}\zeta(r)$, 
we obtain
\beq
\frac{\p_r\zeta}{r}
=-\frac{(1+\zeta^2)\sqrt{\tilde{m}_1^2+(1+\tilde{m}_1^2)\zeta^2}}{2N},
\eeq
where we have chosen the lower sign.
Switching to the $\gamma$ \eqref{eq:gamma_var} and $y$ variables
\eqref{eq:y_var}, the differential equation reduces to
\beq
\frac{\d\gamma}{\d y} = -\frac{\sqrt{\gamma(\tilde{m}_1^2+\gamma)}}{N}.
\eeq
Neatly, this differential equation reduces exactly to
eq.~\eqref{eq:diff_gamma} in the limit of $\tilde{m}_1\to0$. 
Integrating the above differential equation, we obtain
\beq
2\log\left(\sqrt{\gamma} + \sqrt{\gamma+\tilde{m}_1^2}\right)
=-\frac{y}{N} - \kappa,
\eeq
where $\kappa$ is an integration constant.
Solving for $\gamma$ yields
\beq
\gamma = \frac14 e^{-\frac{y}{N} - \kappa}
\left(1 - \tilde{m}_1^2 e^{\frac{y}{N}+\kappa}\right)^2.
\eeq
The boundary condition corresponding to $\omega$ being singular at
$y\to0$ are $\gamma=1$ at $y=0$, which determines $\kappa$:
\beq
\kappa = \log\frac{2 + \tilde{m}_1^2 - 2\sqrt{1 + \tilde{m}_1^2}}{\tilde{m}_1^4}.
\eeq
Inserting the integration constant $\kappa$ into the solution, we find
\beq
\gamma = \frac{e^{\frac{-y}{N}}\left(\tilde{m}_1^2 - e^{\frac{y}{N}}\left(2 +
  \tilde{m}_1^2 - 2\sqrt{1 - \tilde{m}_1^2}\right)\right)^2}{4\left(2 + \tilde{m}_1^2 -
  2\sqrt{1+\tilde{m}_1^2}\right)}.
\eeq
In terms of $\zeta$ and $r$, we have
\begin{equation}
\zeta = \sqrt{\frac{2}{2+\tilde{m}_1^2 - (2+\tilde{m}_1^2)\cosh\left(\frac{r^2}{2N}\right)
    + 2\sqrt{1+\tilde{m}_1^2}\sinh\left(\frac{r^2}{2N}\right)} - 1}, \qquad
r \in [0,R_{\rm compacton}],
\label{eq:zeta_m1_sol}
\end{equation}
with $R_{\rm compacton}$ being the compacton radius and is given by
\beq
R_{\rm compacton} = \sqrt{2N}\sqrt{\log\frac{\tilde{m}_1^2}{2+\tilde{m}_1^2-2\sqrt{1+\tilde{m}_1^2}}}. 
\label{eq:R2compacton}
\eeq
Notice that the solution \eqref{eq:zeta_m1_sol} reduces to
eq.~\eqref{eq:BPSsol} in the limit of $\tilde{m}_1\to0$.
The compacton radius is finite for any finite value of $\tilde{m}_1>0$, but in
the limit of $\tilde{m}_1\to0$ the compacton radius $R_{\rm compacton}$ tends
to infinity.
This is consistent with the solution tending to eq.~\eqref{eq:BPSsol},
which indeed has a tail that falls off exponentially (or rather like a
Gaussian). 
Finally, the BPS mass of the solution \eqref{eq:zeta_m1_sol} is given
in eq.~\eqref{eq:would-be_BPS_bound}.

\end{document}